\documentclass[useAMS,usenatbib]{mn2e}
\usepackage[totalwidth=515pt,totalheight=680pt]{geometry}
\usepackage{graphicx,amssymb,amsbsy}
\input psfig

\title[4-body Disruption of Primordial Planet Signatures]
{Disrupting Primordial Planet Signatures:  The Close Encounter of Two Single-Planet Exosystems in the Galactic Disc}
\author[Veras \& Moeckel]{Dimitri Veras$^{1}$\thanks{E-mail:veras@ast.cam.ac.uk}, 
Nickolas Moeckel$^{1}$\thanks{E-mail:nickolas1@gmail.com}  \\
$^{1}$Institute of Astronomy, University of Cambridge, Madingley Road, Cambridge CB3 0HA}

\begin{document}

\date{Accepted 2012 June 20.  Received 2012 June 4; in original form 2012 May 8}

\pagerange{\pageref{firstpage}--23} \pubyear{XXXX} 

\maketitle

\label{firstpage}

\begin{abstract}
During their main sequence lifetimes, the majority of all Galactic Disc field stars must endure at least one stellar intruder passing within a few hundred AU.  Mounting observations of planet-star separations near or beyond this distance suggest that these close encounters may fundamentally shape currently-observed orbital architectures and hence obscure primordial orbital features.  We consider the commonly-occurring fast close encounters of two single-planet systems in the Galactic Disc, and investigate the resulting change in the planetary eccentricity and semimajor axis.  We derive explicit 4-body analytical limits for these variations and present numerical cross-sections which can be applied to localized regions of the Galaxy.  We find that each wide-orbit planet has a few percent chance of escape and an eccentricity that will typically change by at least $0.1$ due to these encounters.  The orbital properties established at formation of millions of tight-orbit Milky Way exoplanets are likely to be disrupted.
\end{abstract}


\section{Introduction}

After leaving their birth clusters, most stars 
undertake a potentially harrowing multi-Gyr journey 
through the Galactic Disc.
The stars are continuously perturbed by global Galactic phenomena
and are periodically nudged by individual stellar encounters.
Occasionally, an encounter is close enough to cause 
major disruption to any planets orbiting in
the approaching systems.  The currently observed exoplanet
population may be shaped in part by these encounters.

\subsection{Typical Closest Encounter Distances}

\begin{figure}
\centerline{
\psfig{figure=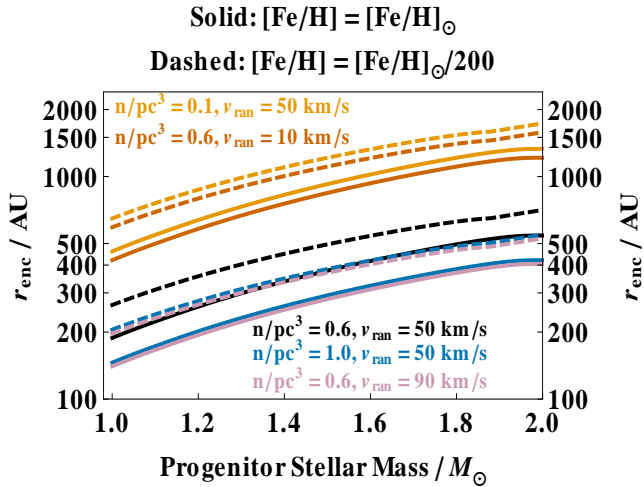,width=8.5cm,height=6.5cm} 
}
\caption{
Upper bound estimates for typical main-sequence closest encounter 
distances, $r_{enc}$, between 
exosystems in the field of the Galactic Disc.
Solar metallicity stars and very low metallicity stars
are represented by solid and dashed lines, respectively.  
The random velocity of stars is $v_{\rm ran}$
and the space density of stars is $n$.  Fiducial values of $n$
($0.6 {\rm pc}^{-3}$) and $v_{\rm ran}$ ($50 {\rm km/s}$) are represented 
by the black curves.  Because the majority
of Galactic Disc stars are less massive than $1 {\rm M}_{\odot}$, 
they will have $r_{enc}$ values less than hundreds of AU.
}
\label{closeenc}
\end{figure}

Using simple arguments \citep[e.g. from Pg. 3 of][]{bintre2008},
one can crudely estimate an upper bound for the typical 
encounter distance, $r_{enc}$, 
over a main sequence lifetime.  If $n$ denotes the space
density of stars in the Galactic Disc, and $v_{\rm ran}$ is the random
velocity of stars, then 
$r_{enc} \approx (4 \pi n v_{\rm ran } t_{\rm MS})^{-1/2}$, where $t_{\rm MS}$ is the main
sequence lifetime.  This estimate is conservatively large 
because gravitational
focusing is not included.  We can estimate $t_{MS}$ through
simulations from the SSE stellar evolution code \citep{huretal2000}.
Doing so yields Fig. \ref{closeenc}, which plots the closest
encounter main sequence distance as a function of progenitor 
mass from $1 {\rm M}_{\odot} - 2 {\rm M}_{\odot}$, which represents a common
range of exoplanet host masses. The majority of stars drawn from a
standard stellar initial 
mass function \citep[see e.g.][]{paretal2011} will have masses under 1 ${\rm M}_{\odot}$, further 
suggesting that the typical encounter separations in Fig. \ref{closeenc}
represent overestimates.  The solid and dashed lines represent
Solar and very low (1/200th of Solar) metallicities, respectively.
The metallicity of a star helps dictate its main sequence lifetime,
and hence the expected close encounter distance.  The plot
partially illustrates that differences in the metallicity
of stars have little (indirect) effect on the close encounter distance.

The figure
demonstrates that the majority of all stars will suffer a close
encounter of just a few hundred AU for a reasonable range
of $n$ and $v_{\rm ran}$ values.  Even in sparse environments, like 
the Solar neighborhood (with $n \approx 0.1$ pc$^{-3}$), Sun-like 
stars will approach one another at least once within a few hundred
AU.  This estimate corroborates the rough estimate of $500$ AU 
given by \cite{zaktre2004}, who consider only a 5 Gyr encounter 
timescale.

\subsection{Stellar Encounter Orientations with Respect to Galactic Centre}

Given that close encounters within hundreds of AU will 
typically occur, we can now attempt to characterize the 
orientations of the collisions with respect to the Galactic
Centre.  As outlined
by \cite{quietal2011}, the distribution of velocities
in the Galactic Disc is affected by a multitude of factors.
Potential perturbers include Galactic Lindblad resonances
\citep[e.g.][]{yuakuo1997,lepetal2011}, stellar streams
from past mergers and interactions with satellite subhaloes
\citep[e.g.][]{bekfre2003,gozetal2010}, and transient
spiral density waves \citep[e.g.][]{deetal2004}.
Stellar velocities may also be highly dependent on the phase 
and pattern speed of the Milky Way's spiral arms \citep{antetal2011},
suggesting drastic differences in the velocity distribution
in different regions.  These factors might help explain
why the velocity components of the stars in the Solar
neighborhood are neither isotropic nor Gaussian
\citep{binetal2000,naketal2010}.
Generally, the orbits of Disc stars
are modulated vertically and 
epicyclically \citep[e.g. pgs. 164-166,][]{bintre2008}, 
and may undergo significant radial migration \citep[e.g.][]{schoenrich2011}.
Further, the amplitude of the epicyclic and vertical
oscillations are of the same order of magnitude 
\citep[e.g., Pg. 18,][]{bintre2008}, and are orders
of magnitude longer than then the physical radii of 
the stars themselves.  Therefore, we should  
expect that stars suffer close encounters with each
other at random orientations with respect to the
Galactic Centre.

\subsection{Planetary Orbit Orientations with Respect to the Galactic Disc}

Now we assess whether the planes of the planetary
orbits should have a preferential orientation to 
the Galactic Disc.  The severe misalignment of the Solar System's
invariable plane with the Galactic plane at $\approx 60^{\circ}$ 
\citep[][]{huawad1966,dunetal1987}
foreshadows the likely answer.
Observations constrain the 
distribution of exoplanetary orbital planes
poorly because most extrasolar planets
have been discovered by the Doppler radial velocity
technique, which alone does not provide any information
about line-of-sight inclinations.  Similarly,
the stellar rotational axis orientation 
-- which is suggestive of planetary orbit orientation -- 
of the vast majority of non-exoplanet host stars is unknown.  
However, in cases where this information has been obtained, 
\cite{abt2001} and \cite{howcla2009} find that the
these axes are orientated randomly.  For exoplanet-host
stars that harbour transiting planets, we {\it do} have 
line-of-sight inclination information.
According to the Exoplanet Data 
Explorer\footnote{See the Exoplanet Data Explorer at http://exoplanets.org/}, as of
15 January, 2012, there are 141 transiting exoplanets.
At most, the orbital plane of any of these planets
is misaligned with our line-of-sight by $\approx 13.4^{\circ}$.
However, the median misalignment angle is just $\approx 2.79^{\circ}$ and
the standard deviation is $\approx 2.73^{\circ}$.

\begin{figure}
\centerline{
\psfig{figure=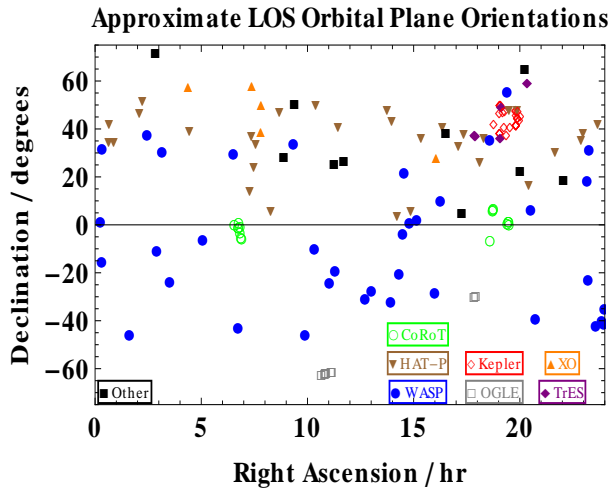,width=8.5cm,height=6.5cm} 
}
\caption{
Approximate line-of-sight exoplanetary orbital plane orientations.
Plotted are the spatial coordinates of stars which host transiting
planets.  All data is taken from the Exoplanet Data 
Explorer, as of 15 January, 2012.  Plot markers are determined based
on whether the orbiting planet's name includes 
``WASP'' (blue filled circles), 
``HAT-P'' (downward-pointing brown filled triangles),
``OGLE'' (hollow gray squares)
``Kepler'' or ``KOI'' (hollow red diamonds),
``TrES'' (purple filled diamonds) or
``XO'' (upward-pointing orange filled triangles).
Other transiting planets are given by filled black
squares.  The plot demonstrates that planetary 
orbital planes are known to encompass a wide range
of alignments with respect to the Galactic Disc.
}
\label{transit}
\end{figure}

Therefore, effectively we observe transiting planets edge-on,
and the locations of these planets on the sky might 
suggest a relation between the Galactic plane and planetary
orbital planes.  In Fig. \ref{transit}, we plot the declination
versus right ascension of the host stars of these 141 transiting planets
from this database.  In order to help assuage the strong 
observational bias in the
plot, the plot markers are colored and shaped according to the planet
names, which are often indicative of the  
program or collaboration who first discovered the planet.  For example, 
the planets with names containing {\it ``Kepler''} or 
{\it ``KOI''} (Kepler Object of Interest) are all clustered in the
same region on the plot.  This is due to the the fixed field
the {\it Kepler} space mission is observing.  The plot definitively
illustrates that observed planetary orbital planes are 
known to encompass a wide range of orientations with respect
to the Galactic Disc.

These considerations lead us to treat
close encounters between two planetary systems 
in arbitrary directions with orbital planes
that are arbitrarily aligned with each other.
However, we must sensibly restrict the vast
phase space of these encounters.  We do so
first by reviewing some published literature
related to this topic.

\subsection{Extending Previous Scattering Studies}

The three-body problem which includes a star-planet
pair experiencing a perturbation from an
intruder star has been the subject of several 
studies, and is well-characterized in many regions
of phase space.  Most studies, however, treat these
interactions in the context of cluster encounters
\citep{hegras1996,davsig2001,freetal2006,spuetal2009},  
which typically have a higher $n$, smaller $v_{\rm ran}$,
and much shorter lifetime (tens of Myr) than
in the field.  An
exception is \cite{zaktre2004}, who do
consider perturbations in the field from
a stellar intruder, but on a multi-planet system.
They treat the perturbation as a superposition
of three-body interactions, neglecting the contribution
to the potential from the planets.  Also, they
treat the velocity vector of the intruder
and the orbital plane as coplanar.  Among their several
useful results are i) about $10\%$ of all stars
experience close encounters within 200 AU, ii)
planetary eccentricities may be excited up
to $0.1$ in the field, and iii) the extent of the excitation
is strongly dependent on system size and phase.

Here, we provide a multi-tiered extension to that
work.  First, we consider the potential of all four
bodies in the close encounter of two one-planet systems,
as most Milky Way stars are now thought to have planets 
\citep{casetal2012}.  Previous studies of the 
4-body problem often consider the more general
case of the interaction of two
stellar binaries 
\citep{mikkola1984,hut1993,bacetal1996,heggie2000,giespu2003,freetal2004,pfamut2006,sweatman2007}
or a planet-less intruder perturbing a multi-planet exosystem \citep{maletal2011,boletal2012}. 
However, none of these studies consider the close encounter of two
single-planet systems.

Second, because field encounters are fast, we develop an 
analytical method based on impulses that can determine
the change in orbital parameters without resorting to numerical
simulations.  We consider two extremes in phase for our analysis, 
although the method can in principle be generalized to arbitrary 
phases, and even arbitrary numbers of planets.

Third, we do perform numerical simulations, here specifically
for the purpose of obtaining normalized cross sections.  These
quantities then enable one to determine
the overall rate of encounters and eccentricity excitations
over a main sequence lifetime in localized patches of
the Milky Way.  As already argued earlier, we consider 
encounters of all mutual orientations, independent of their
locations with respect to the Galactic Centre.

\begin{table*}
 \centering
 \begin{minipage}{180mm}
  \caption{Variables Used in this Paper}
  \begin{tabular}{@{}ll@{}}
  \hline
   Variable & Explanation \\
 \hline
 $a_h$ & Hyperbolic semimajor axis for a star \\
 $a_{k0}^{(*)}$ & Initial semimajor axis for planet $k$ in the {\tt far} ($*=f$) and {\tt close} ($*=c$) cases \\[4pt]
 $a_{kf}^{(*)}$ & Final semimajor axis for planet $k$ in the {\tt far} ($*=f$) and {\tt close} ($*=c$) cases \\[4pt]
 $a_{\chi}^{(*)}$ & Contribution to Planet \#2's semimajor axis variation due to Planet \#1 alone in the {\tt far} ($*=f$) and {\tt close} ($*=c$) cases \\[4pt]
 $\alpha$ & Number of planetary orbital periods to numerically integrate before the close encounter \\ 
 $b$ & Impact parameter of both stars \\
 $b_{{\rm eje}}^{(f)}$ & {\tt far} case impact parameter value at which a planet escapes \\[3pt]
 $b_{{\rm eje,1}}^{(c)}$ & Maximum {\tt close} case impact parameter value separating planetary escape from boundedness  \\[3pt]
 $b_{{\rm eje,2}}^{(c)}$ & Middle {\tt close} case impact parameter value separating planetary escape from boundedness  \\[3pt]
 $b_{{\rm eje,3}}^{(c)}$ & Minimum {\tt close} case impact parameter value separating planetary escape from boundedness \\[3pt]
 $b_{\rm max}$ & Maximum impact parameter used in the numerical simulations \\
 $b_{\rm min}$ & Impact parameter which causes a planet-planet collision \\
 $b_{p1p2}$ & Impact parameter of both planets \\[2pt]
 $b_{p1s2}$ & Impact parameter of Planet \#1 and Star \#2 \\
 $b_{s1p2}$ & Impact parameter of Star \#1 and Planet \#2 \\
 $b_{{\rm stat,<}}^{(c)}$ & {\tt close} case lower impact parameter value at which there is no net perturbation on the planets \\[3pt]
 $b_{{\rm stat,>}}^{(c)}$ & {\tt close} case upper impact parameter value at which there is no net perturbation on the planets \\[3pt]
 $\beta$ & Factor by which $(a_{10} + a_{20})$ is multiplied to obtain $q$ for the numerical simulations \\
 $\gamma$ & Fraction of the innermost planetary orbit used as a numerical integration timestep bound \\
 $\delta$ & Dimensionless planet/star mass ratio for each system when both are physically equivalent \\
 $\delta_k$ & Dimensionless planet/star mass ratio for system $k$ \\
 $e_{\rm ext,max}^{(c)}$ & {\tt close} case local eccentricity maximum, for $ (b_{{\rm stat,>}}^{(c)}) < b $ \\[3pt]
 $e_{\rm ext,min}^{(c)}$ & {\tt close} case local eccentricity minimum, for $ b_{{\rm eje,1}}^{(c)} < b < b_{{\rm eje,2}}^{(c)}$ \\[3pt]
 $e_h$ & Hyperbolic eccentricity of a star \\
 $e_{kf}^{(*)}$ & Final eccentricity for planet $k$ in the {\tt far} ($*=f$) and {\tt close} ($*=c$) cases \\[3pt]
 $e_{\chi}^{(*)}$ & Contribution to Planet \#2's eccentricity variation due to Planet \#1 alone in the {\tt far} ($*=f$) and {\tt close} ($*=c$) cases \\[4pt]
 $E_h$ & Hyperbolic anomaly of a star \\
 $\epsilon$ & Dimensionless ratio equal to $a_{10}/a_{20}$  \\ 
 $\eta$ & Dimensionless ratio equal to $V_{\infty}/V_{\rm crit}$ \\
 $M_{pk}$ & Mass of planet $k$ \\
 $M_{sk}$ & Mass of star $k$ \\
 $M_{\rm tot}$ & Total mass of the 4-body system \\
 $\mu$ & Sum of both stellar masses, times the Gravitational Constant \\
 $n$ & Space density of stars \\
 $N$ & Number of experiments \\
 $\mathcal{N}$ & Number of times over a main sequence lifetime that $\left|\Delta e_1 \right| > \Upsilon$ occurs \\
 $q$ & Pericenter of the star-star hyperbolic orbit \\
 $r_{\rm enc}$ & Typical closest encounter distance for two stars in the Galactic Disc \\
 $r_{\rm start}$ & Separation used to initialize numerical integrations \\ 
 ${\rm RAND}$ & Low-discrepancy quasi-random Niederreiter number between 0 and 1 \\
 $\sigma$ & Cross section \\
 $\sigma_{\rm norm}$ & Normalized cross section \\ 
 $t_{\rm enc}$ & Timescale of close encounter between both planetary systems \\
 $t_{\rm integrate}$ & Numerical integration timescale \\
 $t_{\rm MS}$ & Main Sequence lifetime \\
 $T_k$ & Orbital period of planet $k$ about star $k$ \\
 $\Upsilon$ & Given extent of an eccentricity perturbation \\
 $v_{\rm ran}$ & Random stellar velocity \\
 $V_{\infty}$ & Velocity of Star \#1 with respect to Star \#2 at an infinite separation  \\
 $V_{{\rm circ},k}$ & Circular velocity of planet $k$ about star $k$ \\
 $V_{{\rm circ},k0}$ & Circular velocity of planet $k$ about star $k$ assuming $M_{pk}=0$ \\
 $V_{{\rm circ},0}$ & Circular velocity of either planet for equal planetary masses and semimajor axes \\
 $V_{\rm crit}$ & Velocity at which the total energy of the 4-body system equals zero \\
 $V_{\rm peri}$ & Pericenter velocity of the star-star hyperbolic orbit \\ 
 $|\Delta \vec{V}_{\bot}|$ & Magnitude of the velocity kick perpendicular to the direction of motion \\
\hline
\end{tabular}
\end{minipage}
\end{table*}

\subsection{Plan for Paper}

We outline some of the key quantities
in the hyperbolic 4-body problem in Section 2
before our analytical (Section 3 and the Appendix) and 
numerical (Section 4) explorations.  
Of particular note are the two specific orientations we model without
numerical integrations (Sections
\ref{farsec} and  \ref{closesec}) and the 
eccentricity excitation 
frequencies arising from our numerical integrations
(Section \ref{exfreq}).  In Section 5,
we interpret the results.  Section 6 discusses
related topics, and Section 7 provides a short conclusion.

\subsubsection{Variables used}

Table 1 delineates the variables applied throughout this paper. 
The subscript $k$ takes the values ``1'' and ``2'' and
is used to describe the planet and star belonging to the different planetary systems taking part in the encounter.
Primed and double-primed values are explained in the text
where necessary.

\section{4-body Problem Setup}

Consider a planet with mass $M_{p1}$ orbiting a
Galactic Disc star with mass $M_{s1}$, and an independently
evolving planet with mass $M_{p2}$ orbiting a different Disc
star with mass $M_{s2}$.  Initially, assume
the distance between the systems (denoted ``1'' and ``2'') is infinity.
Each planetary orbit is described by the planet's semimajor axis, $a_k$, 
and eccentricity, $e_k$, where $k=1$ or $2$
depending on the planet.  At $t = 0$, the orbital parameters are denoted by an 
additional subscript,``$0$''.

As argued in Section 1, the systems may approach each other
at any orientation, and the relative orientation of the planetary
orbital planes is also unconstrained. 
Now consider the plane in which the stars approach each other,
and fix the reference frame on $M_{s2}$.  System \#1 will approach
System \#2 such that $M_{s1}$ will be traveling at a velocity $V_{\infty}$
with an impact parameter $b$.  Because $V_{\infty} > 0$ 
the stars will approach each other on approximate hyperbolic orbits,
approximate because of the presence of the planets.  Denote
the reduced mass of the 2-body hyperbolic system as 
$\mu \equiv G \left( M_{s1} + M_{s2} \right)$.
We treat values of $V_{\infty}$, $b$ and $\mu$, as well as
all four individual masses and $a_{k0}$,
for $k=1,2$, as given, known quantities throughout the paper.
Further, $e_{k0} = 0$ always.

\subsection{Key Orbital Parameters}

The total energy of the system is equal to
$V_{\infty}^2/2 = -\mu/2a_h$, where $a_h$ is the (negative) hyperbolic semimajor axis.   Hence

\begin{equation}
a_h = -\frac{\mu}{V_{\infty}^2}.
\label{semihyp}
\end{equation}

\noindent{The} total angular momentum of the system is equal to $b V_{\infty}$, which
can be related to the hyperbolic eccentricity, $e_h > 1$, such that

\begin{equation}
e_{h}^2 = 1 + \frac{b^2 V_{\infty}^4}{\mu^2}.
\label{ecchyp}
\end{equation}

The pericenter, $q > 0$,
of the star-star hyperbolic orbit is

\begin{equation}
q \equiv |a_h| \left(e_h - 1 \right)
=
\frac{\mu}{V_{\infty}^2} 
\left[
\sqrt{1 + \frac{b^2 V_{\infty}^4}{\mu^2}} - 1
\right]
.
\label{peri}
\end{equation}

\subsection{Velocity Comparisons}

The total energy of a 2-body system with a nonzero
relative velocity is positive.  The critical 
velocity of the four body system, $V_{\rm crit}$, for which the total system energy
is zero and ionization is possible is \citep{freetal2004}:

\begin{equation}
V_{\rm crit} \equiv 
\sqrt{
\frac{G M_{\rm tot}}
{\left(M_{s1} + M_{p1} \right) \left(M_{s2} + M_{p2} \right)}
\left(
\frac{M_{s1} M_{p1}}{a_{10}} + \frac{M_{s2} M_{p2}}{a_{20}}
\right)
}
\end{equation}

\begin{figure*}
\centerline{
\psfig{figure=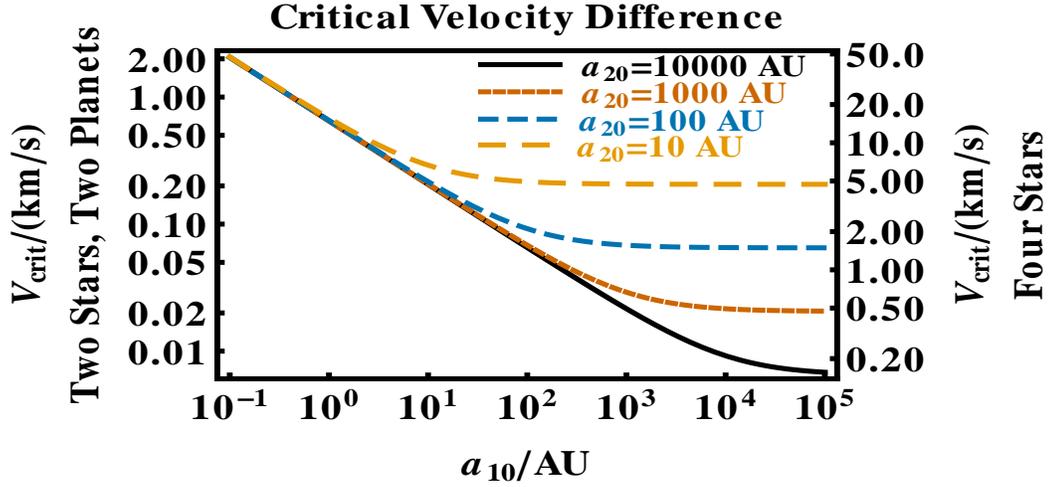,height=6.5cm,width=14cm} 
}
\caption{
The critical velocity as a function of semimajor axes for 
two $1 {\rm M}_{\odot}$ stars and two $1 {\rm M}_{\rm J}$ planets
(left axis) and four $1 {\rm M}_{\odot}$ stars (right axis).
Hence, given typical field velocities, unlike pairs of binary stars
in the Galactic Disc, almost never will two single-planet
exosystems be completely ionized.
}
\label{Vcrit}
\end{figure*}

\noindent{where} $M_{\rm tot} = M_{s1} + M_{p1} + M_{s2} + M_{p2}$. We plot 
typical values of $V_{\rm crit}$ in Fig. \ref{Vcrit}, showing that $V_{\rm crit}$ 
is nearly 23 times lower for
two ${\rm M}_{\rm J}$ planets and two ${\rm M}_{\odot}$ stars
than for four ${\rm M}_{\odot}$ stars, where ${\rm M}_{\rm J}$ is the mass
of Jupiter.  Hence,
comparison of typical stellar velocities in the Galactic
Disc ($\approx 10$km/s - $100$km/s) implies 
that one-planet systems are moving
too fast to ionize all four bodies through encounters
regardless of the values of $a_{10}$ and $a_{20}$.

Now we can compare the circular velocity of a planet with respect to its
parent star, $V_{{\rm circ},k}$, to typical values of $V_{\infty}$.  We have

\begin{equation}
V_{{\rm circ},k} = 29.79 \frac{\rm km}{\rm s} 
          \sqrt{\left( \frac{M_{sk} + M_{pk}}{{\rm M}_{\odot}} \right) 
                \left( \frac{1 {\rm AU}}{a_{k0}}  \right)  }
\label{Vcirc}
\end{equation}

\noindent{Therefore}, for wide orbit planets and typical Disc velocities,
$V_{\infty} \gg V_{{\rm circ},k}$.  However, for planets on tight orbits,
the velocities are comparable.  Further, we denote $V_{{\rm circ},k0}$ as the circular
velocity of planet $k$ when $M_{pk} = 0$ (such that $V_{{\rm circ},k} \approx V_{{\rm circ},k0}$).

The fastest velocity achieved in a hyperbolic orbit is at the pericenter 
of that orbit.  The pericenter velocity $V_{\rm peri}$, is related to $V_{\infty}$
through

\begin{equation}
V_{\rm peri} = 
\sqrt{
\frac{\mu}{\left|a_h\right|}  
\left[
\frac
{e_h + 1}
{e_h - 1}
\right]
}
= V_{\infty}
\left[
\frac
{\sqrt{1 + \frac{b^2 V_{\infty}^4}{\mu^2}} + 1}
{\sqrt{1 + \frac{b^2 V_{\infty}^4}{\mu^2}} - 1}
\right]^{\frac{1}{2}}
\end{equation}

\noindent{which} is always greater than $V_{\infty}$ and becomes infinite as $b \rightarrow 0$.
$V_{\infty}$ represents the minimum velocity of the orbit.

\section{Impulse Analytics} \label{impulse}

Although we must resort to numerical simulations to fully explore
the 4-body problem consisting of two planet-star systems, 
here we investigate how this cases of this problem may be solved analytically
in the impulse regime.  As suggested by Eq. (\ref{Vcirc}),
perturbations on wide orbit planets due to passing planetary systems
may be treated in the impulse approximation.  \cite{zaktre2004}
claim that this assumption holds for their planetless intruder if 
the stellar perturber is fast and if the
planetary period is much longer
than the characteristic timescale of the
encounter, $t_{\rm enc} \approx b/V_{\rm peri}$.  This condition is analogous here to

\begin{equation}
\frac{T_{k''}}{t_{\rm enc}} \approx  2\pi \frac{a_{k''}}{b} \frac{V_{\rm peri}}{V_{{\rm circ},k''}}
\gg 1,
\label{impcri}
\end{equation}

\noindent{where} $k''$ indicates the planet with the smaller orbital period.  
As demonstrated by Eq. (\ref{impcri}), the impulse approximation is well-suited 
for wide orbits due to the resulting low value of $V_{{\rm circ},k''}$.
In the impulse regime, the planets do not progress in their orbits around
their parent stars during the encounter (i.e., the mean anomaly is approximated as stationary).

The impulse approximation allows us to isolate and estimate analytically the 
planets' mutual perturbations during the encounter.
For simplicity, let us treat both planets on circular orbits.  By symmetry, in the 
impulse approximation the only net perturbation is perpendicular to the velocity 
vector of the perturber. For ease of reference to \cite{zaktre2004}, 
we also take both planetary systems to be coplanar with each
other and with the perturber's velocity vector.
We will be estimating
the perturbations on Planet \#2.  By symmetry, the perturbations on Planet \#1
will yield the same change in orbital parameters.

\subsection{General Case}

\subsubsection{Total Perturbations on the Passing Star}

First, let us estimate the perturbations on Star \#2 due to Star \#1.
Pg. 422 of \cite{bintre1987} shows that the imparted velocity kick is

\begin{equation}
|\Delta \vec{V}_{\bot}|_{s1s2} =
\frac{2 b V_{\infty}^3}{\mu}
e_{h}^{-2}
.
\label{stanb}
\end{equation}

Planet \#1 will also kick Star \#2.
The effective impact parameter between Planet \#1 and Star \#2, 
$b_{p1s2}$, will depend on the planet's position during the encounter.
We have,

\begin{equation}
|\Delta \vec{V}_{\bot}|_{p1s2} =
\frac{2 M_{p1} b_{p1s2} V_{\infty}^3}{G\left(M_{p1} + M_{s2} \right)^2}
\left(1+\frac{b_{p1s2}^2 V_{\infty}^4}{G^2\left(M_{p1} + M_{s2} \right)^2}\right)^{-1}
.
\end{equation}

\subsubsection{Total Perturbations on the Passing Planet}

Similarly to the impulse imparted on Star \#2 by Planet \#1, the impulse imparted by Star \#1 on Planet \#2 is:

\begin{equation}
|\Delta \vec{V}_{\bot}|_{s1p2} =
\frac{2 b_{s1p2} V_{\infty}^3}{G\left(M_{s1} + M_{p2} \right)}
\left(1+\frac{b_{s1p2}^2 V_{\infty}^4}{G^2\left(M_{s1} + M_{p2} \right)^2}\right)^{-1}
.
\end{equation}

\noindent{The} impulse on Planet \#2 from Planet \#1 is:

\begin{equation}
|\Delta \vec{V}_{\bot}|_{p1p2} =
\frac{2 b_{p1p2} V_{\infty}^3}{G\left(M_{p1} + M_{p2} \right)}
\left(1+\frac{b_{p1p2}^2 V_{\infty}^4}{G^2\left(M_{p1} + M_{p2} \right)^2}\right)^{-1}
.
\label{bp1p2}
\end{equation}

\subsubsection{Net Perturbations on the Passing Planet}

Therefore, Planet \#2 experiences a net velocity kick relative to its
parent star of

\begin{eqnarray}
|\Delta \vec{V}_{\bot}|_{p2} &=& |\Delta \vec{V}_{\bot}|_{s1s2} + |\Delta \vec{V}_{\bot}|_{p1s2}
\nonumber
\\                               
&-& \left( |\Delta \vec{V}_{\bot}|_{s1p2} + |\Delta \vec{V}_{\bot}|_{p1p2} \right)
\label{Vperp}
\end{eqnarray}

Equipped with Eqs. (\ref{stanb})-(\ref{Vperp}), we can insert these velocity kicks into the formalism of
\cite{jacwya2012} [see their Eqs. 1-2].\footnote{The assumption under which these equations
are derived is that the impulse is instantaneous, which is equivalent to our Eq. (\ref{impcri}).} 
in order to determine the resulting variation in $a_2$ and $e_2$.
Let us denote the post-encounter values of $a_2$ and $e_2$ as $a_{2f}$ and $e_{2f}$ (recall $e_{20}=0$).  Then

\begin{eqnarray}
\frac{a_{20}}{a_{2f}} &=& 1 - \left(\frac{|\Delta \vec{V}_{\bot}|_{p2}}{V_{{\rm circ},2}}\right)^2 
,
\label{af}
\\
e_{2f}  &=& \left|\frac{|\Delta \vec{V}_{\bot}|_{p2}}{V_{{\rm circ},2}}\right|
.
\label{ef}
\end{eqnarray}

\subsection{Specific Example}

\begin{figure}
\centerline{
\psfig{figure=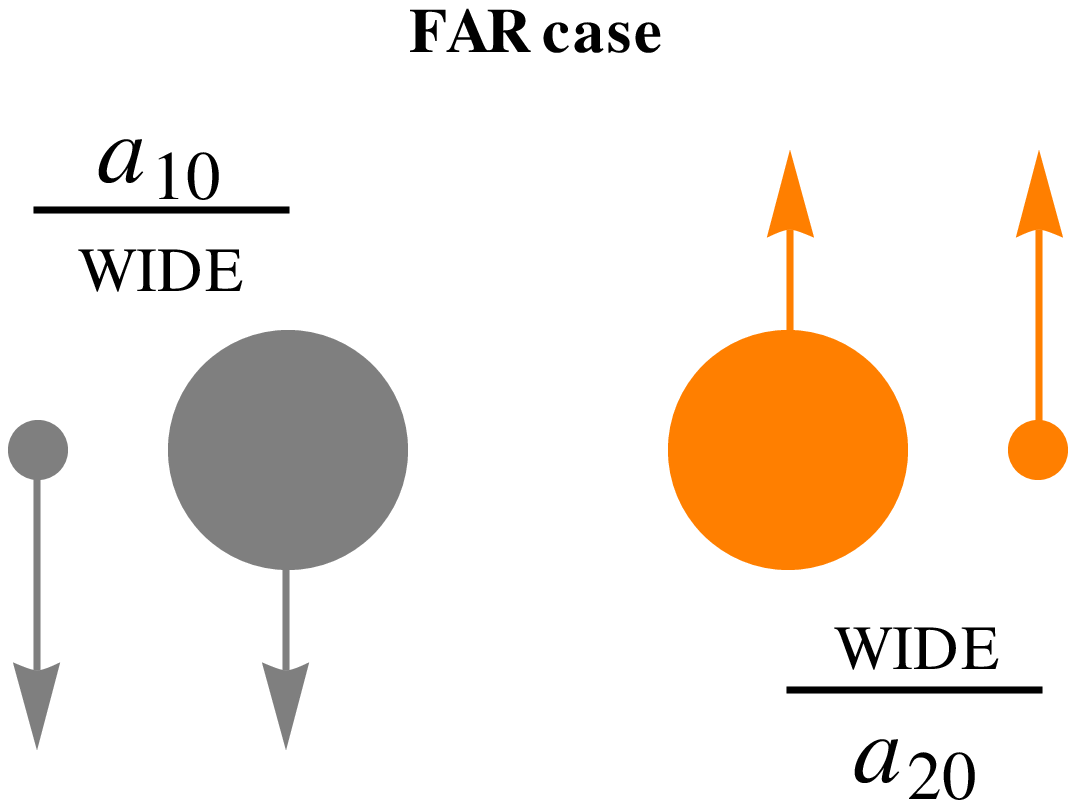,height=6cm,width=6.2cm}
}
\centerline{ 
\psfig{figure=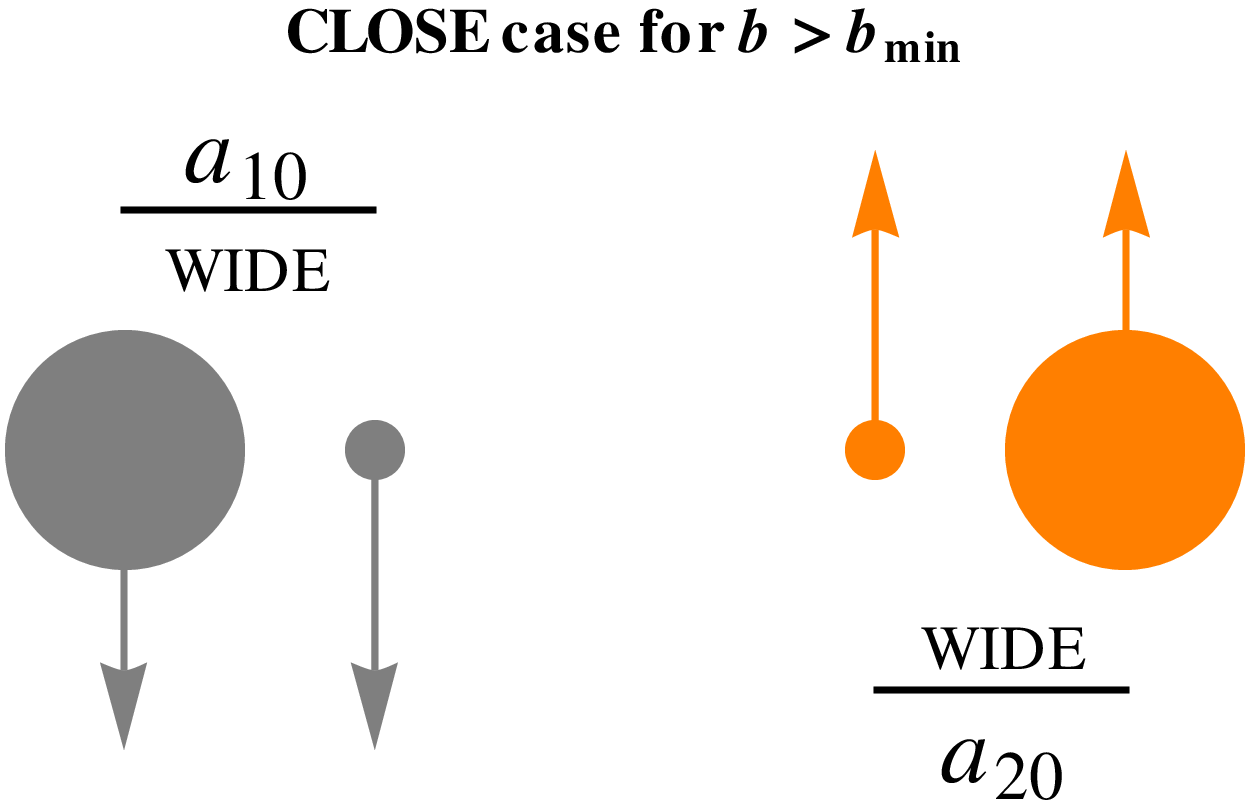,height=6cm,width=6.2cm}
}
\centerline{
\psfig{figure=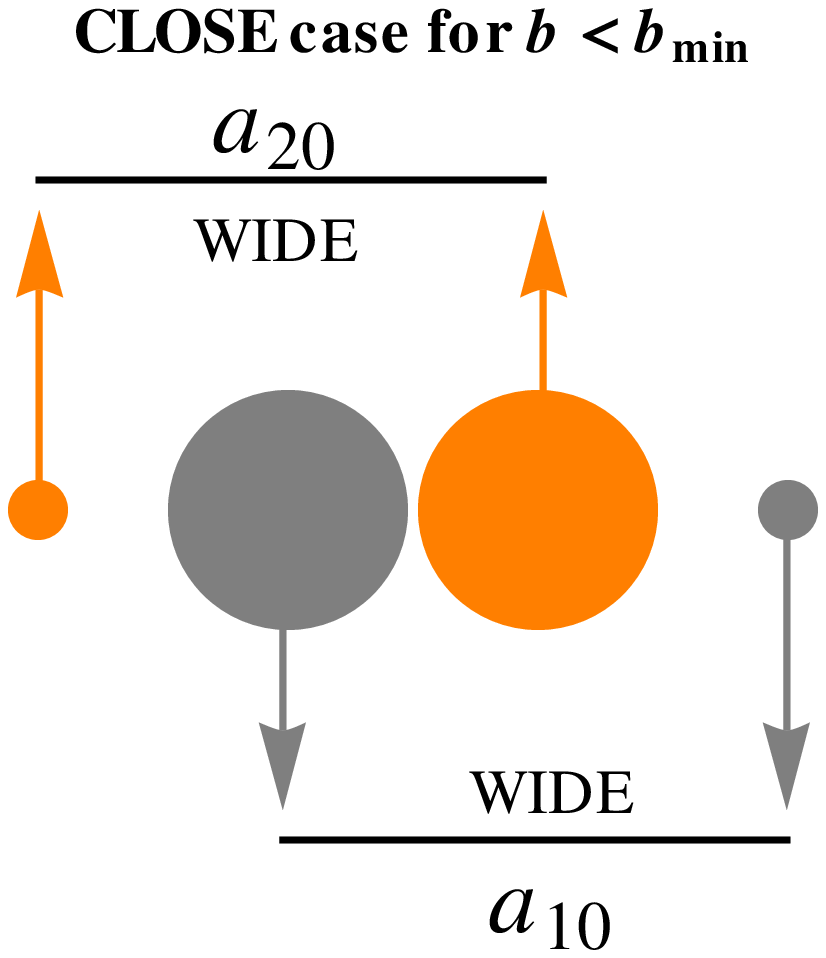,height=6cm,width=6.2cm}
}
\caption{
Cartoon of different close approach configurations
modelled by impulses.  The larger objects are stars
and the smaller objects are planets.  Different colors
denote the two different systems.
}
\label{cartoon}
\end{figure}

We wish to relate $a_{2f}$ and $e_{2f}$ to $b$, $V_{\infty}$
and $M_{s1}$, $M_{s2}$, $M_{p1}$, and $M_{p2}$ in an analytically tractable manner.  Thus,
we will focus on two specific cases of interest, as illustrated in Fig. \ref{cartoon}.
In the first case, which we denote by {\tt far}, both planets are the furthest
possible distance from each other as the systems pass each other; both stars
are in-between the planets.  Here, the value of $b$ may be any value from $0$ to $\infty$.
In the second case, which we denote by {\tt close}, the direction of the vectors
from each star to its child planet are pointing towards each other.
Here, a value of $b$ that we denote $b_{\rm min}$ will cause
both planets to collide.  For $b < b_{\rm min}$, the orbits will overlap.
Figure \ref{cartoon} 
shows a cartoon of the encounters at pericenter for different cases.

We have derived analytical formulae for the critical points of the motion,
asymptotic limits, and the individual contribution to the perturbations from
the planets alone.  All these formulae are presented and explained in the Appendix
in order to help retain the focus of the reader here.  In this section, we provide
just the most important results.

\subsubsection{Analytic Simplification} \label{ansimp}

In order to obtain compact, understandable formulae,
for the remainder of Section 3, we assume $a_{10} = a_{20}$, $M_{p1} = M_{p2}$ and
$M_{s1} = M_{s2}$ such that both systems are equivalent except for their labels.  
Define  $\delta \equiv M_{p2}/M_{s2}$ and $V_{{\rm circ},0}$ as the circular
velocity of either planet assuming the planet mass is zero.  In other specific cases
of interest, these assumptions may be lifted and the more general results rederived
in a similar manner as below.

\subsubsection{Fiducial Sample} \label{fidsimp}

In order to provide tangible numbers that accompany the analytics
and resulting plots, we concurrently consider fiducial 
values of $M_{s2} = {\rm M}_{\odot}$, $M_{p2} = {\rm M}_{\rm J}$, $a_{20} = 1000$ AU 
and $V_{\infty} = 30 {\rm km \ s}^{-1} \approx 1000 V_{\rm crit}$ unless otherwise 
indicated\footnote{Our choice of fiducial semimajor axis helps us demonstrate all of the
regimes of interest for realistic close encounter distances (Fig. \ref{closeenc})
and known exoplanet separations \citep[e.g.][]{goletal2010,kuzetal2011,luhetal2011}.}.
These values give $V_{{\rm circ},2} \approx 1.3 {\rm km \ s}^{-1}$ such that the impulse
approximation is valid as long as $b \ll 1.45 \times 10^5$ AU (Eq. \ref{impcri}).

\begin{figure}
\leftline{
\psfig{figure=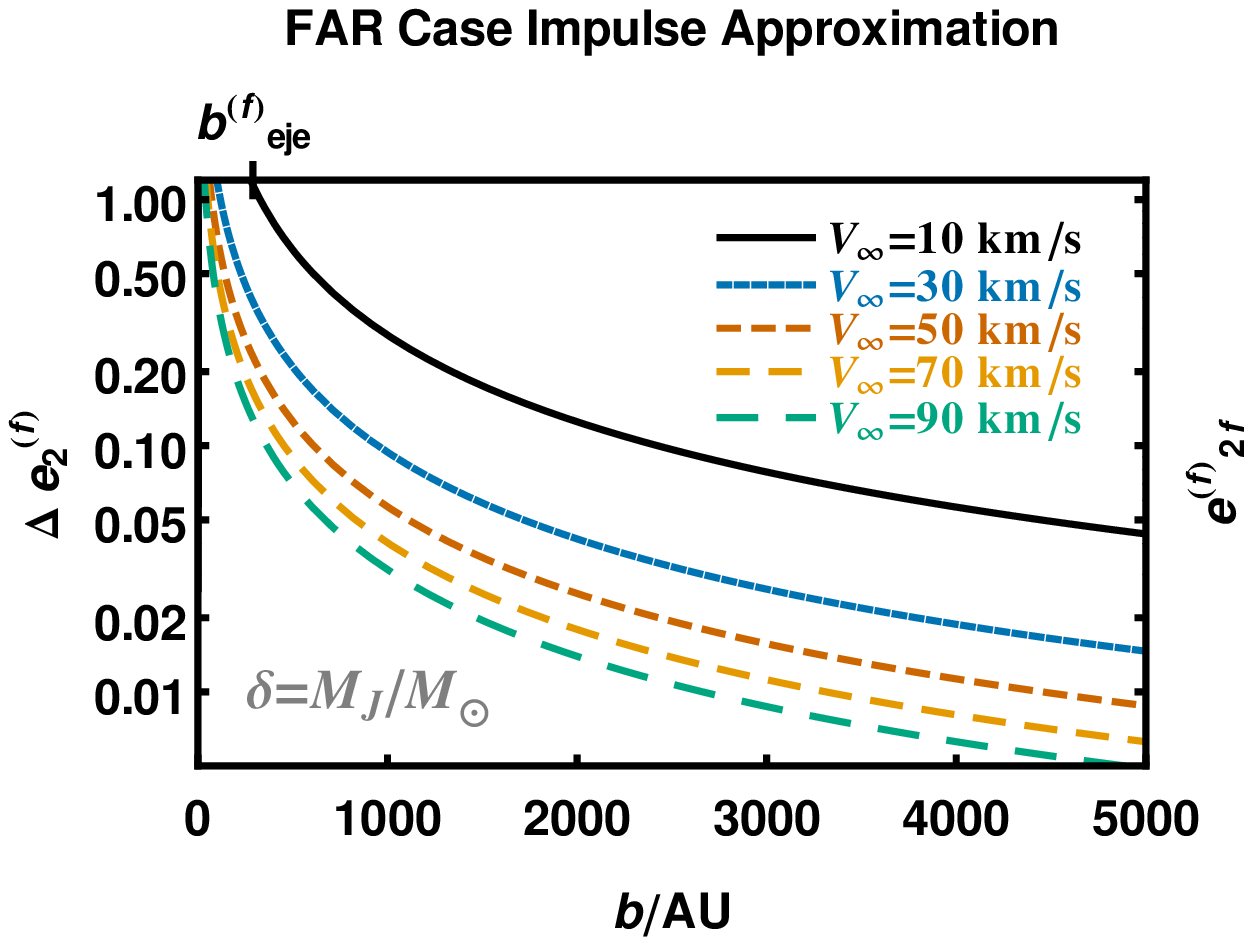,height=6cm,width=7cm} 
}
\centerline{}
\leftline{
\hspace{0.1mm}
\psfig{figure=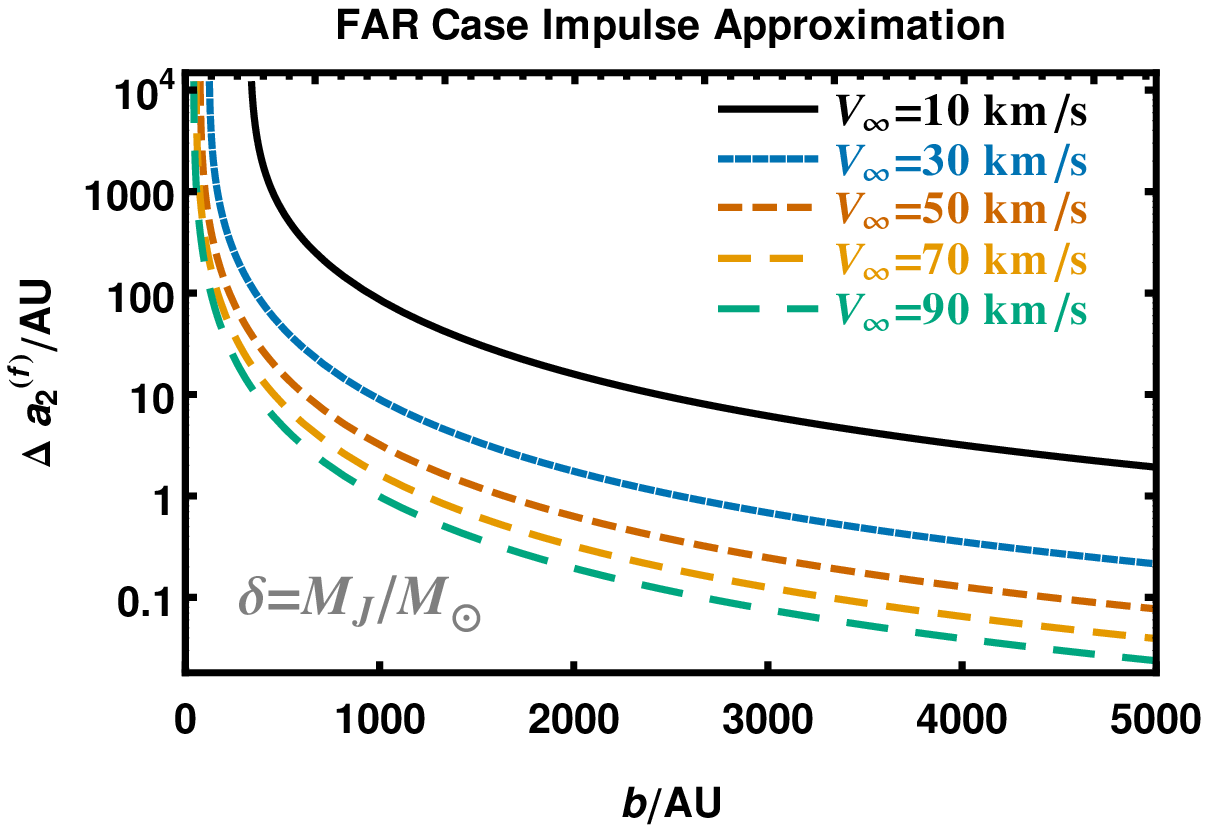,height=6cm,width=6.8cm}
}
\caption{
Eccentricity and semimajor axis variation of Planet \#2 in the {\tt far} case.
The variation monotonically increases as $b$ decreases from infinity
to $b_{{\rm eje}}^{(f)}$, when the planet is ejected.  The eccentricity
and semimajor axis always increase due to interactions in the {\tt Far} case.
}
\label{aefar}
\end{figure}

\subsubsection{The {\tt far} Case}  \label{farsec}

Even though the planets are at opposition to each other in the 
{\tt far} case, planetary ejection will occur when the stars have a
close-enough encounter, when $b \le b_{{\rm eje}}^{(f)}$.  Alternatively, for
$b > b_{{\rm eje}}^{(f)}$, the orbital parameter evolution is

\begin{eqnarray}
e_{2f}^{(f)} &\approx& \left[ \frac{2 \left(2 a_{20} + b \right)}{a_{20} + b} \right]
              \left[ \frac{\sqrt{ \frac{G M_{s1} a_{20}}{b^2}  }  }{V_{\infty}} \right]
,
\label{far2e}
\\
a_{2f}^{(f)} &=& \frac{a_{20}}{1 - e_{2f}^{{(f)}^2}}.
\label{far2a}
\end{eqnarray}

\noindent{Note} that $e_{2f}^{(f)} \rightarrow 0$ and $a_{2f}^{(f)} \rightarrow a_{20}$ as 
$b \rightarrow \infty$, as expected.  Also, $a_{2f}^{(f)}$ cannot decrease due
to the close encounter.  

Equations (\ref{far2e})-(\ref{far2a}) show that
as long as the planetary mass is neligible compared to the stellar mass,
the planetary contribution is also negligible everywhere in the {\tt far} case
parameter space.  Nevertheless, we quantify this contribution in the Appendix.
Note that $e_{2f}^{(f)} \rightarrow 0$ and $a_{2f}^{(f)} \rightarrow a_{20}$ as 
$b \rightarrow \infty$, as expected.  Also, $a_{2f}^{(f)}$ cannot decrease due
to the close encounter.

Figure \ref{aefar} illustrates these properties.  Depending on $V_{\infty}$, Planet \#2
will be ejected when $b$ is within a few tens or hundreds of AU; $b_{{\rm eje}}^{(f)}$ is marked
on the upper axis of the left panel for the slowest $V_{\infty}$.  Planets which 
remain bound after surviving a passing star at $b \approx 500$ AU expand their orbits by tens to hundreds
of AU and stretch their orbits through eccentricity increases of at least $0.1$.

\subsubsection{The {\tt close} Case}  \label{closesec}

Now let us consider the opposite limit, where the position vectors
from each star to their orbiting planet point towards each other.
The resulting orbital parameter
evolution is a more complicated function of $b$.  

In particular, there are two separate ranges of $b$ in which
a planet will escape: i) $0 < b < b_{{\rm eje,3}}^{(c)}$, when both
stars are in between both planets and the stars are close
to each other, and ii) $b_{{\rm eje,2}}^{(c)} < b < b_{{\rm eje,1}}^{(c)}$,
when one star nearly collides with one planet.
Further, there is one region,
$(b_{{\rm stat,<}}^{(c)}) < b < (b_{{\rm stat,>}}^{(c)})$ , where the
planet-planet interaction becomes important.
Additionally, there are two local extrema:
i) in between the escape regions, the perturbations are minimized
at $e_{\rm ext,min}^{(c)}$, and ii) for $b$ well beyond $b_{\rm min}$, 
the perturbations are maximized at $e_{\rm ext,max}^{(c)}$.

All of the physical features mentioned above and illustrated in
both Figs. \ref{aenear} and \ref{aeneardelta} (which can be
used as guides for the location of the critical points)
can be reproduced with a compact
analytical form for $e_{2f}^{(c)}$ as a function of impact parameter.
We remind the reader that this quantity, among several others,
are derived in the Appendix:

\begin{figure*}
\leftline{
\psfig{figure=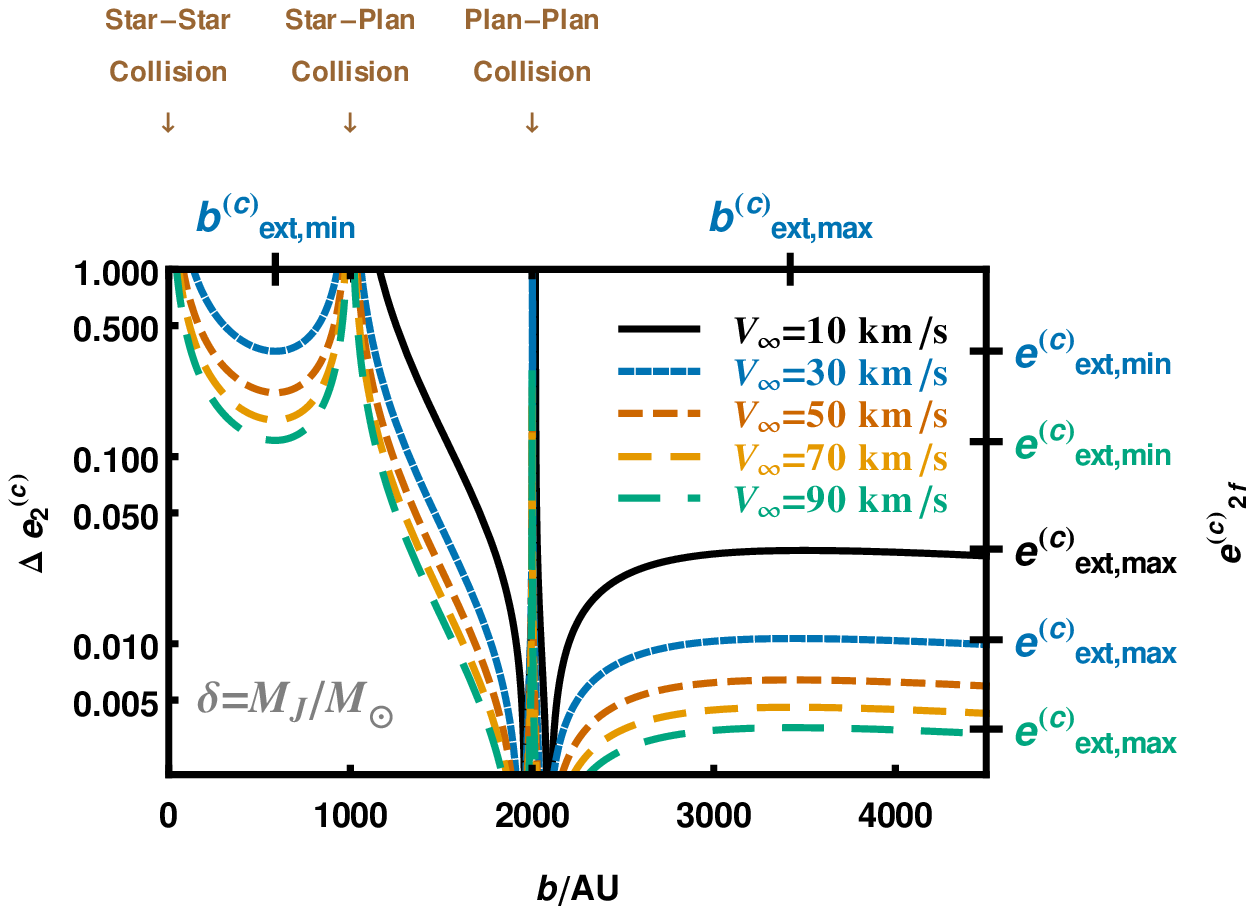,height=10cm,width=17cm} 
}
\leftline{
\hspace{4.3 mm}
\psfig{figure=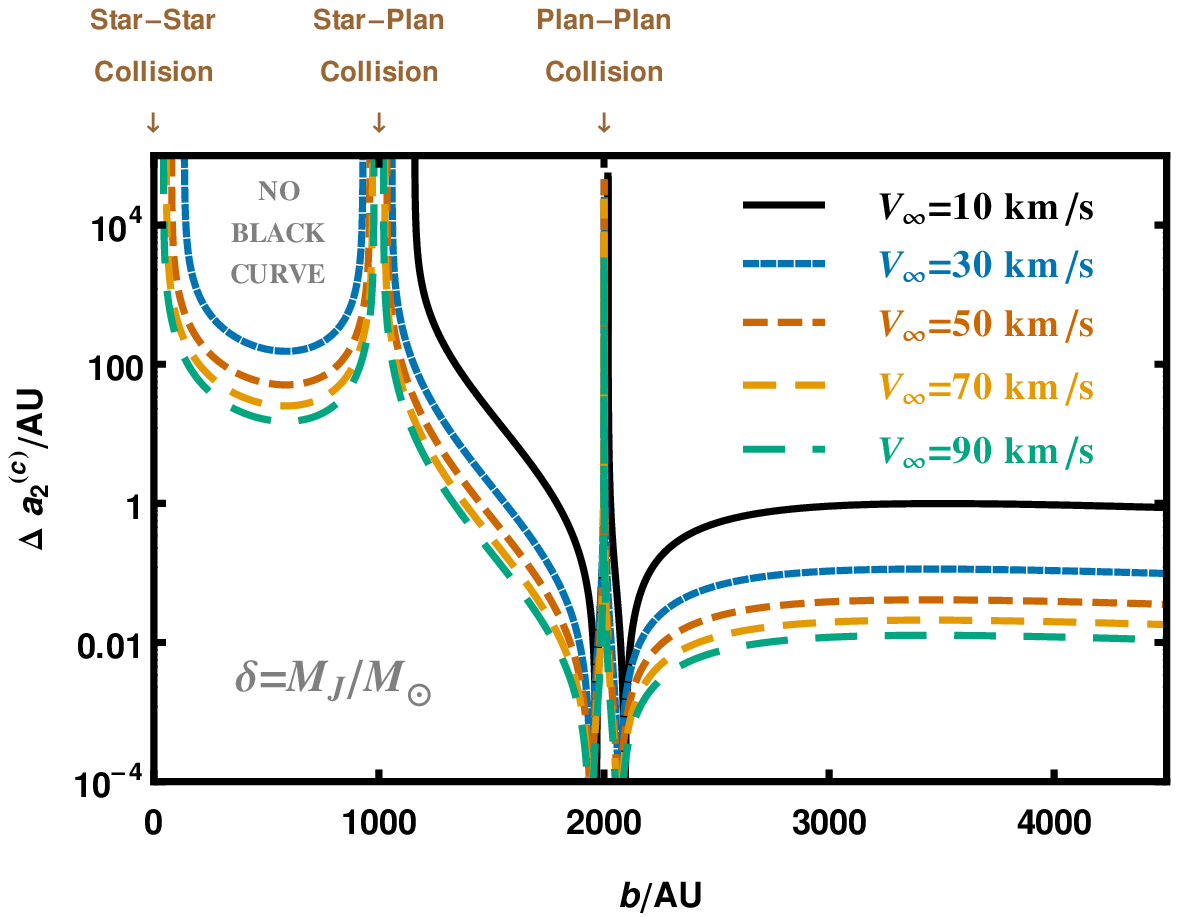,height=8.7cm,width=13.85cm}
}
\caption{
Eccentricity and semimajor axis variation of Planet \#2 
in the {\tt close} case.  At $b=0$ AU, the stars collide.
At $b \approx 1000$ AU, Planet \#2 collides with Star \#1.
At $b \approx 2000$ AU, Planet \#2 collides with Planet \#1.
The extreme points present in the panels are 
explained in Eqs. (\ref{bstatless})-(\ref{aextmax}).  The explicit functional
form of $e_{2}^{(c)}$ is given in Eqs. (\ref{e2close})-(\ref{Z7}).  Both
panels demonstrate that planets will experience major disruption 
and potentially ejection if their orbits cross.  
}
\label{aenear}
\end{figure*}

\begin{equation}
e_{2f}^{(c)} \approx
\left|
2 \left(\frac{a_{20}}{b}\right) \left(\frac{V_{{\rm circ},0}}{V_{\infty}} \right)
\frac
{
8 G^2 M_{s1}^2 + Z_4 + Z_5
}
{
Z_6 Z_7
}
\right|
\label{e2close}
\end{equation}

\noindent{such that}

\begin{eqnarray}
Z_4 &=& 2 G M_{s1} V_{\infty}^2 \left(6 a_{20} - b\left(3 - 2 \delta  \right)  \right)
\\  
Z_5 &=& V_{\infty}^4 \left(4 a_{20}^2 + b^2 \left(1 - 2 \delta \right) - 2 b a_{20} \left(2 - \delta \right)  \right)
\\
Z_6 &=& 2 G M_{s1} + V_{\infty}^2 \left(a_{20} - b \right)
\\
Z_7 &=& 2 G M_{s1} + V_{\infty}^2 \left(2 a_{20} - b \right)
\label{Z7}
\end{eqnarray}

\noindent{where} $a_{2f}^{(c)}$ is derived from $e_{2f}^{(c)}$ in the usual way (Eq. \ref{far2a}).

\subsection{Consequences}

The analytics show that a planet's eccentricity can be raised
to any value due to a realistic close encounter.  Even in
the limiting case where both planets are furthest from each other
during the encounter, if the stars endure a close enough approach, 
then the planets will be ejected.  In the other extreme, measurable 
eccentricity excitation can occur over a wide, realistic range of 
impact parameters.  When a star crosses in between another star and 
planet, the eccentricity excitation is at least a tenth, 
but is likely many tenths.  When any two bodies narrowly miss each other,
planetary escape may occur.  However, there are locations at which
this net perturbation is zero; this range of locations
increases along with planetary mass. 

This analytic exploration helps us to gauge expectations for
the outcomes of numerical simulations, and perhaps more importantly,
provides an explanation for some of the trends seen in the outputs
of our numerical simulations.  We describe these simulations 
in the following section.

\section{Numerical Cross Sections}
We now calculate cross sections of various encounter outcomes via numerical 
scattering experiments. Cross sections of this type, introduced in stellar dynamical 
research by \cite{hutbah1983}, represent the effective surface area for some 
outcome of a scattering event between involving stellar or planetary systems. Coupled with 
a velocity distribution and density of systems, the cross section yields a total
outcome frequency. By suitably setting up random stellar encounters and 
performing many experiments, the probabilistic outcome of these potentially 
chaotic encounters can be obtained.

Here, we are interested in the frequency of planetary systems
whose planets have eccentricities that are perturbed by a particular
amount, $\Upsilon$.  The cross section is a function of $\Upsilon$,
$\epsilon \equiv a_{10}/a_{20}$, and $\eta \equiv V_{\infty}/V_{\rm crit}$
such that one example is:

\begin{equation}
\sigma(\left|\Delta e_1 \right| > \Upsilon,\epsilon,\eta) = \pi b_{\rm max}^2 
\frac{N(\left|\Delta e_1 \right| > \Upsilon,\epsilon,\eta)}{N_{\rm total}}
\label{sig}
\end{equation}

\noindent{where} $N_{\rm total}$ represents the total number of experiments
and $N(\left|\Delta e_1 \right| > \Upsilon,\epsilon,\eta)$ represents
the number of experiments with a given $\epsilon$ and $\eta$ that
yield $\left|\Delta e_1 \right| > \Upsilon$.
One may compute errors in $\sigma(\left|\Delta e_1 \right| > \Upsilon,\epsilon,\eta)$
by using Gaussian counting statistics \citep{hutbah1983}.  Doing
so gives error bars which are equal to the RHS of Eq. (\ref{sig}) divided
by $\sqrt{N(\left|\Delta e_1 \right| > \Upsilon,\epsilon,\eta)}$.
In order to create a scale-free cross section -- for wider
applications -- $\sigma$ can be 
normalized as:

\begin{eqnarray}
\sigma_{\rm norm}(\left|\Delta e_1 \right| > \Upsilon,\epsilon,\eta) 
&=& 
\frac{\sigma(\left|\Delta e_1 \right| > \Upsilon,\epsilon,\eta)}
{\pi \left(a_{10} + a_{20}\right)^2  }
\nonumber
\\
&=&
\frac{\sigma(\left|\Delta e_1 \right| > \Upsilon,\epsilon,\eta)}
{\pi a_{10}^2}
\left(\frac{\epsilon}{1 + \epsilon}\right)^2
\end{eqnarray}

\noindent{Our} goal is to compute values of both 
$\sigma_{\rm norm}(\left|\Delta e_1 \right| > \Upsilon,\epsilon,\eta)$
and $\sigma_{\rm norm}(\left|\Delta e_2 \right| > \Upsilon,\epsilon,\eta)$
for different values of $\Upsilon$, $\epsilon$ and $\eta$.
Doing so requires a careful numerical setup.

\subsection{Numerical Simulation Setup}

First, we must
set up initial conditions such that the initial
separation of the stars is finite, and then select
this finite separation.  We also must
chose a sufficiently representative range of $b$
small enough to not be computationally prohibitive
but large enough to encompass all the regimes
in, for example, Fig. \ref{aenear}.
Finally, we must choose values of $V_{\infty}$ that
encompass a wide range of possible physical values.

\subsubsection{Characterizing Finite Separations}

Strictly, our numerical simulations cannot treat infinite distances.
Therefore, we must propagate forward the 2-body star-star hyperbolic solution 
until the mutual distance between the stars, $r(t)$, reaches 
a specified distance $r(t=t_{\rm start}) \equiv |r_{\rm start}|$.  At this separation, the initial
conditions for the numerical simulations are established.  
For any finite $|r_{\rm start}|$, the components of the 
stars' velocity both parallel and perpendicular to the
impact parameter segment will be different than their initial
values at an infinite mutual separation.  Denote the position
and velocity components parallel to $b$ by $x$ and $\dot{x}$,
and the perpendicular components by $y$ and $\dot{y}$.
Given values of $\mu, V_{\infty}, b,$ and $|r_{\rm start}|$ of the relative orbit, we seek
to derive $x_{\rm start}$, $y_{\rm start}$, $\dot{x}_{\rm start}$ and
$\dot{y}_{\rm start}$.

Denote the hyperbolic anomaly as $E_h$.  Then we adopt
the same definition of $r$ as in Pg. 45 of \cite{taff1985} and Pg. 85
of \cite{roy2005} such that

\begin{equation}
\cosh{E_h} = \frac{1 + \frac{r_{\rm start}}{a_h}}{e_h}.
\label{coshf}
\end{equation}

\noindent{Note} that by this convention, $r_{\rm start}$ is negative.
Through geometry, we have:

\begin{eqnarray}
x_{\rm start} &=& a_h \left(e_h - \cosh{E_h} \right),
\label{x}
\\
y_{\rm start} &=& a_h \sqrt{e_{h}^2-1} \sqrt{\cosh^2{E_h} - 1}
.
\label{y}
\end{eqnarray}

\noindent{Differentiating} these gives

\begin{eqnarray}
\dot{x}_{\rm start} &=& -a_h \sqrt{\cosh^2{E_h} - 1} \frac{dE_h}{dt}
,
\label{u}
\\
\dot{y}_{\rm start} &=& a_h \sqrt{e_{h}^2 - 1} \cosh{E_h} \frac{dE_h}{dt}
\label{v}
\end{eqnarray}

\noindent{yielding} a total velocity of

\begin{equation}
\frac{dr_{\rm start}}{dt} = \pm a_h \frac{dE_h}{dt} \sqrt{e_{h}^2 \cosh^2{E_h} - 1} 
= \sqrt{\mu \left( \frac{2}{r_{\rm start}} - \frac{1}{a_h} \right)}
\end{equation}

\noindent{where} the RHS is due to the properties of a two-body hyperbolic orbit.
Thus,

\begin{equation}
\frac{dE_h}{dt} = \frac{\pm \sqrt{\mu \left( \frac{2}{r_{\rm start}} - \frac{1}{a_h} \right)}}
{a_h \sqrt{e_{h}^2 \cosh^2{E_h} - 1}},
\label{coshft}
\end{equation}

\noindent{where} the signs indicate the two possible velocity
vectors along the orbit.  Because we model the systems approaching
one another, we adopt the upper sign.  Subsequently, we can insert
Eq. (\ref{coshft}) into Eqs. (\ref{u}) and (\ref{v}) and use 
Eqs. (\ref{semihyp}), (\ref{ecchyp}), (\ref{coshf}) and (\ref{coshft})
to express the starting Cartesian elements in terms of given parameters:

\begin{eqnarray}
x_{\rm start} &=& -\frac{b^2 V_{\infty}^2 + r_{\rm start} \mu}
                      {\sqrt{b^2 V_{\infty}^4 + \mu^2}} 
,
\label{xstart}
\\
y_{\rm start} &=& -b V_{\infty} 
                \sqrt{  
                \frac{-b^2 V_{\infty}^2 + r_{\rm start} \left(r_{\rm start}V_{\infty}^2 - 2 \mu \right)}
                     {b^2 V_{\infty}^4 + \mu^2 } 
                 }
,
\\
\dot{x}_{\rm start} &=& \frac{\mu}{r_{\rm start}}
                \sqrt{ \frac{r_{\rm start}V_{\infty}^2 + 2 \mu}{r_{\rm start}V_{\infty}^2 - 2 \mu}}
\nonumber
\\
            &\times&\sqrt{ \frac{-b^2 V_{\infty}^2 + r_{\rm start} \left(r_{\rm start}V_{\infty}^2 - 2 \mu \right)}
                     {b^2 V_{\infty}^4 + \mu^2 }}
,
\\
\dot{y}_{\rm start} &=& \frac{b V_{\infty} \left( - r_{\rm start}V_{\infty} + \mu  \right) \sqrt{V_{\infty}^2 + \frac{2\mu}{r_{\rm start}}}}
                     {\sqrt{r_{\rm start} \left(r_{\rm start} V_{\infty}^2 - 2 \mu  \right)  \left(b^2 V_{\infty}^4 + \mu^2 \right)}}
.
\label{xyuv}
\end{eqnarray}

\noindent{Finally}, we convert these elements into the center of mass frame
for the numerical simulation initial conditions. 

We wish to i) model an approximately
equal approach and retreat for each simulation,
and ii) sufficiently sample both the approach and retreat.
Regarding i), in the reduced two-body hyperbolic problem, we need 
the time the approaching system takes to change 
$|E_h|$  to $-|E_h|$, or instead $|\sin{E_h}|$ to $-|\sin{E_h}|$:

\begin{eqnarray}
t_{\rm integrate} &=& 2 \frac{\left|\sinh{E_h}\right|}{\left|\frac{d\sinh{E_h}}{dt}\right|}
=
2\frac{\left|\tanh{E_h}\right|  }{ \left|\frac{dE_h}{dt}\right|  }
\nonumber
\\
&&
\nonumber
\end{eqnarray}
\begin{eqnarray}
=
2
\left[
\left( 
   \frac{r_{\rm start} \sqrt{2\mu - r_{\rm start} V_{\infty}^2}}
        {r_{\rm start} V_{\infty}^2 - \mu} 
\right)
\sqrt{   
   \frac{\left| \left(b^2 - r_{\rm start}^2\right)V_{\infty}^2 + 2 r_{\rm start} \mu \right|}
        {\left| r_{\rm start} V_{\infty}^2 + 2 \mu \right|  }
}
\right]
\nonumber
&&
\\
&&
\nonumber
\end{eqnarray}
\begin{equation}
\approx -2 \frac{r_{\rm start}}{V_{\infty}}
\label{tint0}
\end{equation}

\noindent{where} $r_{\rm start} < 0$.

Regarding ii), suppose the longer of the planetary orbital
periods is denoted by $T_{k'}$.  One wishes to sample $\alpha$ of these
periods {\it before the close encounter}.
Then, $t_{\rm integrate}/T_{k'} \ge 2\alpha$, or

\begin{equation}
r_{\rm start} = -2 \pi \alpha \sqrt{\frac{a_{k',0}^3}{\mu}} V_{\infty}
\label{ralph}
\end{equation}

\noindent{We} set $r_{\rm start}$ from Eq. (\ref{ralph}), and 
set $\alpha = 1.2$, in all of our numerical simulations
in order to sample at least one planetary orbit both
before and after the encounter.

\subsubsection{System Orientations}

As argued in Section 1, there is no apparent
preferred direction for planetary system close encounters with respect to the
Galactic Centre nor with one another.  Therefore, we randomly orient
both planets with respect to their parent stars and randomly orient
both planetary systems with respect to one another.

\subsubsection{Impact Parameter Range}

\noindent{Following} previous work \citep[e.g.][]{freetal2004}, we express $q$ as a multiple of the sum of
the initial separations of both planet systems, so that $q = \beta \left(a_{10} + a_{20}\right)$,
where $\beta$ is a constant.  Using this form of the pericenter, we obtain
from Eq. (\ref{peri}):

\begin{equation}
b_{\rm max} = 
\sqrt{
\left[\beta \left(a_{10} + a_{20}\right)\right]^2
+
\frac{2 \mu}{V_{\infty}^2}
\left[\beta \left(a_{10} + a_{20}\right)\right]
}
\label{bmax}
,
\end{equation}

\noindent{which} is bounded from below as

\begin{equation}
\min{(b_{\rm max})} = \lim_{V_{\infty} \to \infty} b = \beta \left(a_{10} + a_{20}\right)
.
\end{equation}

\noindent{The} general expression for the upper bound is long, but may
be simplified in specific cases.  If we assume
$M_{s1} = M_{s2}$,  $M_{p1} = M_{p2}$ and $a_{10} = a_{20}$, which are the same assumptions
adopted in our analytical cases, then 

\begin{equation}
b' = 2 a_{20} \sqrt{\beta \left[\beta + 2 \left( \frac{V_{{\rm circ},20}}{V_{\infty}} \right)^2  \right] }
\end{equation}

\noindent{where} $b'$ denotes the value of $b$ under the above assumptions. 
Hence,

\begin{eqnarray}
\max{(b'_{\rm max})} &=& \lim_{V_{\infty} \to V_{\rm crit}} b' 
\nonumber
\\
&=& 2 a_{10} 
\sqrt{\beta \left(\beta + \frac{M_{s2} + M_{p2}}{2M_{p2}} \right) }
\end{eqnarray}

\noindent{which} shows that the impact parameter may be arbitrarily
large for a small enough planetary mass.

For each set of simulations, we wish to sample a representative
range of impact parameters.  We chose $\beta = 2.5$ and select
values of $b$ from 0 out to $b_{\rm max}$ according to 
$b = \sqrt{{\rm RAND} \times b_{\rm max}^2}$,
where RAND is a low-discrepancy quasi-random Niederreiter number 
between zero and unity. The impact parameter is thus sampled according to its probability, and no weighting of the
scattering experiment outcomes is necessary to account for the larger frequency of wide encounters compared to nearly head-on encounters.

\subsubsection{$\epsilon$ and $\eta$ Range}

We choose three values of the planetary semimajor axis ratio
($\equiv \epsilon \equiv a_{10}/a_{20} = 1,10,100$), which represents
a wide variety of both already observed planetary systems and systems
with wide-orbit planets which have not yet been observed.  
In order to determine a range of plausible $\eta$ values,
reconsider Fig. \ref{Vcrit}.  For $a_{10} = 1000$ AU, our
three values of $\epsilon$, and plausible velocity values
in the field (10 km/s $\le V_{\infty} \le 100$ km/s),
we obtain $\eta$ ranges of roughly
$[330-3300]$ , $[120-1200]$, and $[50-500]$, respectively.
However, we need not restrict our $\eta$ ranges to field
values.  We can also include typically slow cluster velocities
of $\approx 1$ km/s (by reducing the lower bounds on the ranges by an order
of magnitude) and hyper-velocity stars (by increasing the upper bounds
by a factor of a few).  Therefore,
for each value of $\epsilon$, we choose between 15-20 values of $\eta$
based on these broad ranges.

\subsubsection{Numerical Code}

We use a modified version of Piet Hut's fourth-order Hermite
integrator\footnote{That code is available at http://www.artcompsci.org},
which we call {\tt SuperHermite}.  {\tt SuperHermite} was introduced 
in \cite{moever2012}, where the details of the implementation and code verification can be found. The {\tt SuperHermite} code
utilizes a P(EC)$^n$ method \citep{koketal1998} to achieve implicit time-symmetry. There is no preferred dominant force or geometry (such as one central
star or a circumbinary system).  {\tt SuperHermite} also contains collision detection;
in all simulations, we set each star's radius to be a Solar radius and
each planet's radius to be Jupiter's radius.  Although {\tt SuperHermite}
can accurately determine eccentricity variations many orders of magnitude
smaller than observationally detectable values, in this study we
consider only $\Upsilon \ge 10^{-4}$.

As a safety measure, we choose a maximum allowable timestep for our integrations:

\begin{equation}
t_{\rm step,max} = 2 \pi \gamma \sqrt{\frac{a_{k'',0}^3}{\mu}}
.
\end{equation}

\noindent{Here,} $\gamma$ is the fraction of the innermost orbit that 
the simulation is allowed to use as a timestep.  We use $\gamma = 1/20$.
However, {\tt SuperHermite}'s timestep choice will almost certainly be more 
conservative than this in all of our single-encounter simulations.

\subsubsection{Other Considerations}

For each pair $(\epsilon,\eta)$, we ran $N_{\rm total} = 10^4$
scattering experiments.  Strictly, the imposition of a 
finite separation means that 
due to the center of mass frame shift,
numerically $e_{k0}(t = t_{\rm start})$ and $a_{k0}(t = t_{\rm start})$ deviate 
from their given initial values at infinite separations by a factor of roughly a 
few $\delta$.  For example, $e_{k0}(t = t_{\rm start}) \approx 0.003$.  As we are primarily
concerned with the {\it change} in eccentricity, this initial small nonzero eccentricity
is not of concern.  Regarding the change of orbital parameters,
$a_{k}(t > t_{\rm start})$ and $e_{k}(t > t_{\rm start})$ do vary slightly as the 
stars approach each other,
well before the close encounter.  This variation, which increases with
decreasing $|r_{\rm start}|$, is natural and unavoidable, and is orders of magnitude less 
than the variation due to the close encounter.

\begin{figure*}
\centerline{
\psfig{figure=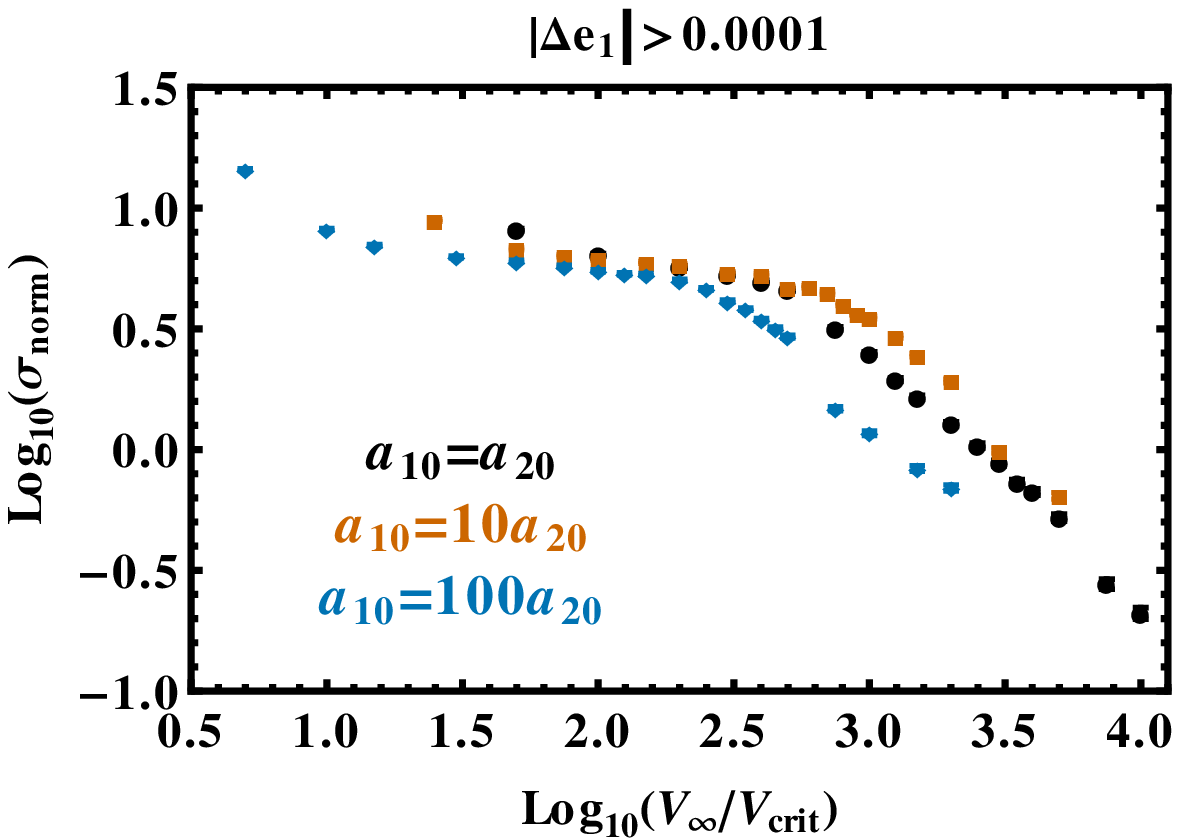,height=6cm,width=8cm} 
\psfig{figure=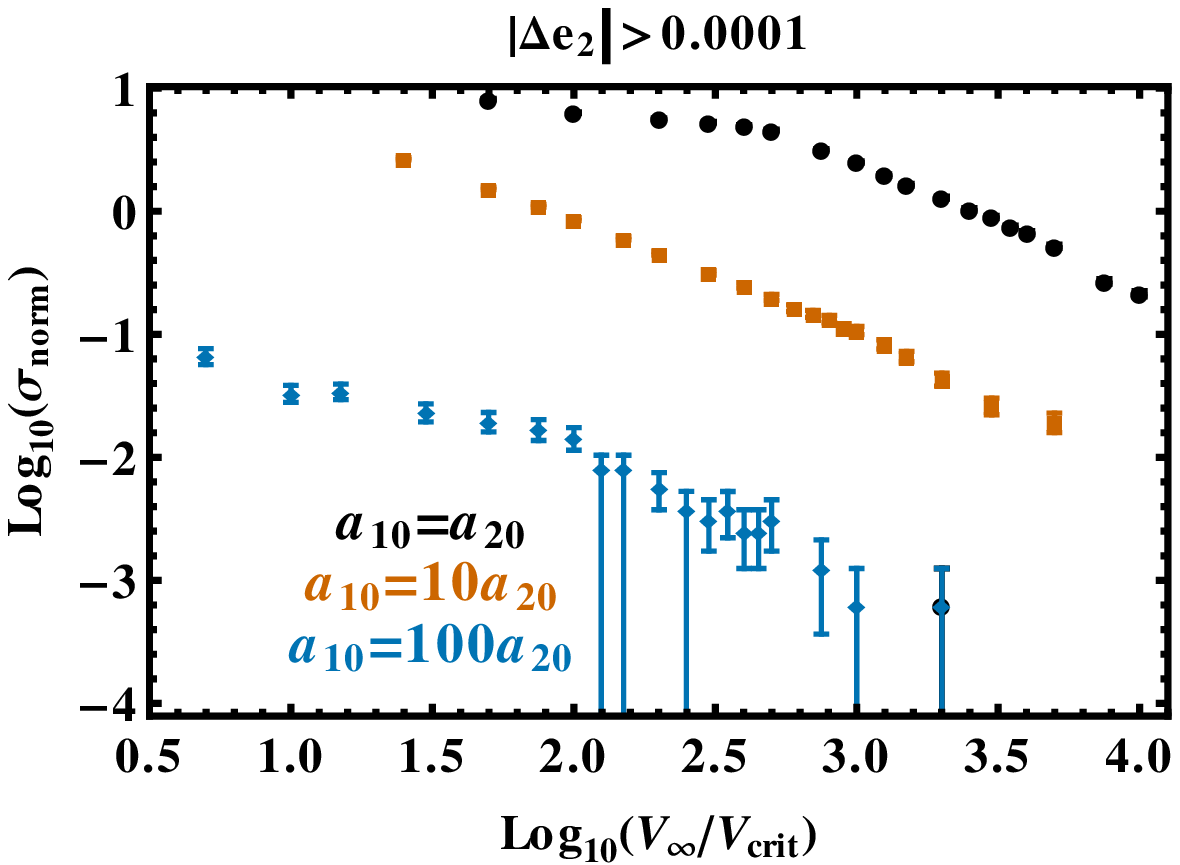,height=6cm,width=8cm}
}
\caption{
Normalized cross sections for outcomes corresponding
to $|\Delta e_1| > 10^{-4}$ (left panel)
and $|\Delta e_2| > 10^{-4}$ (right panel).
}
\label{cse0001}
\end{figure*}

\begin{figure*}
\centerline{
\psfig{figure=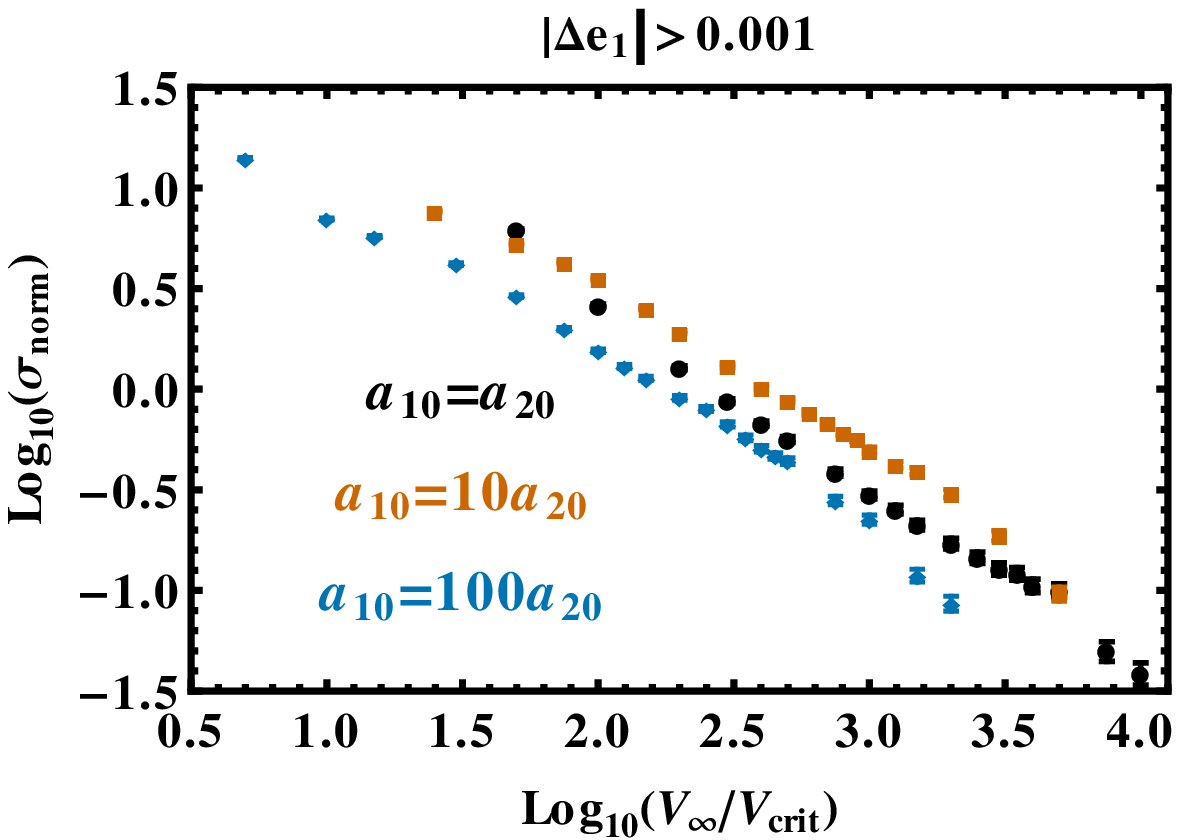,height=6cm,width=8cm} 
\psfig{figure=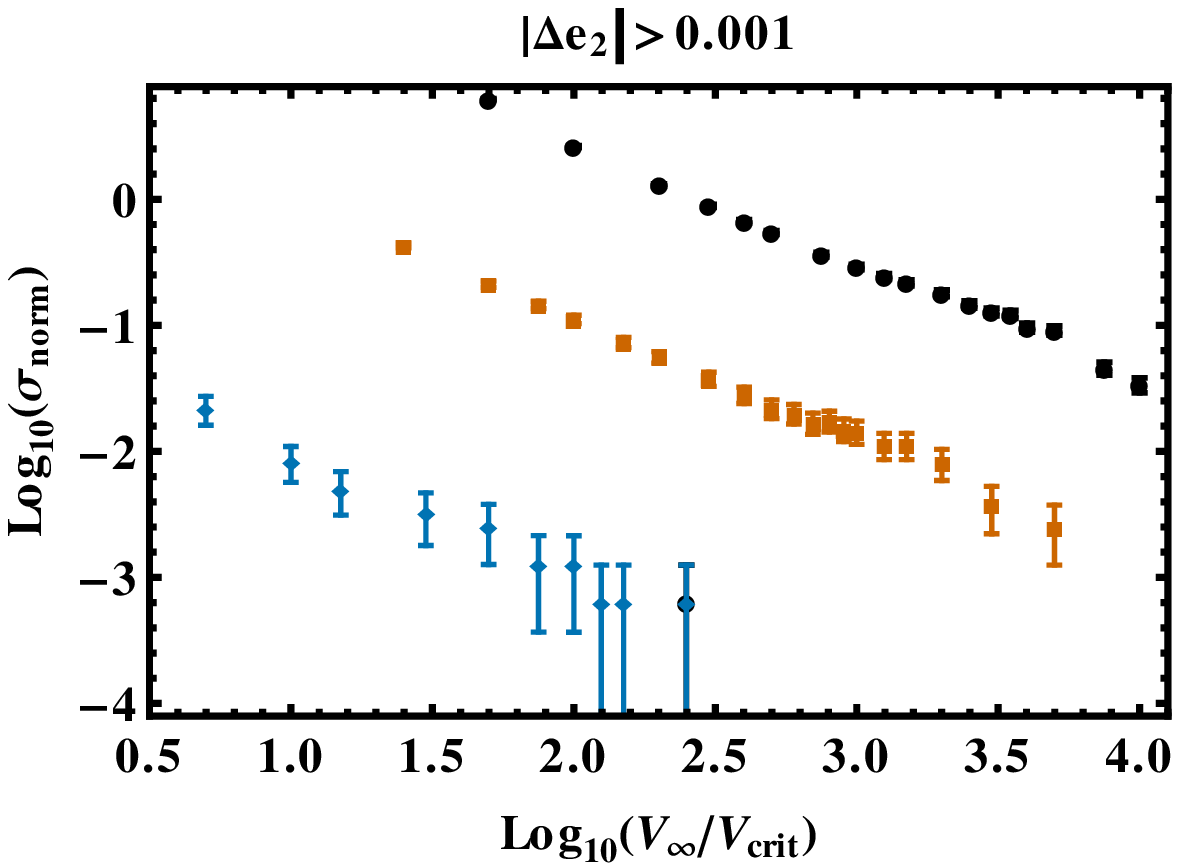,height=6cm,width=8cm}
}
\caption{
Normalized cross sections for outcomes corresponding
to $|\Delta e_1| > 10^{-3}$ (left panel)
and $|\Delta e_2| > 10^{-3}$ (right panel).
}
\label{cse001}
\end{figure*}

\begin{figure*}
\centerline{
\psfig{figure=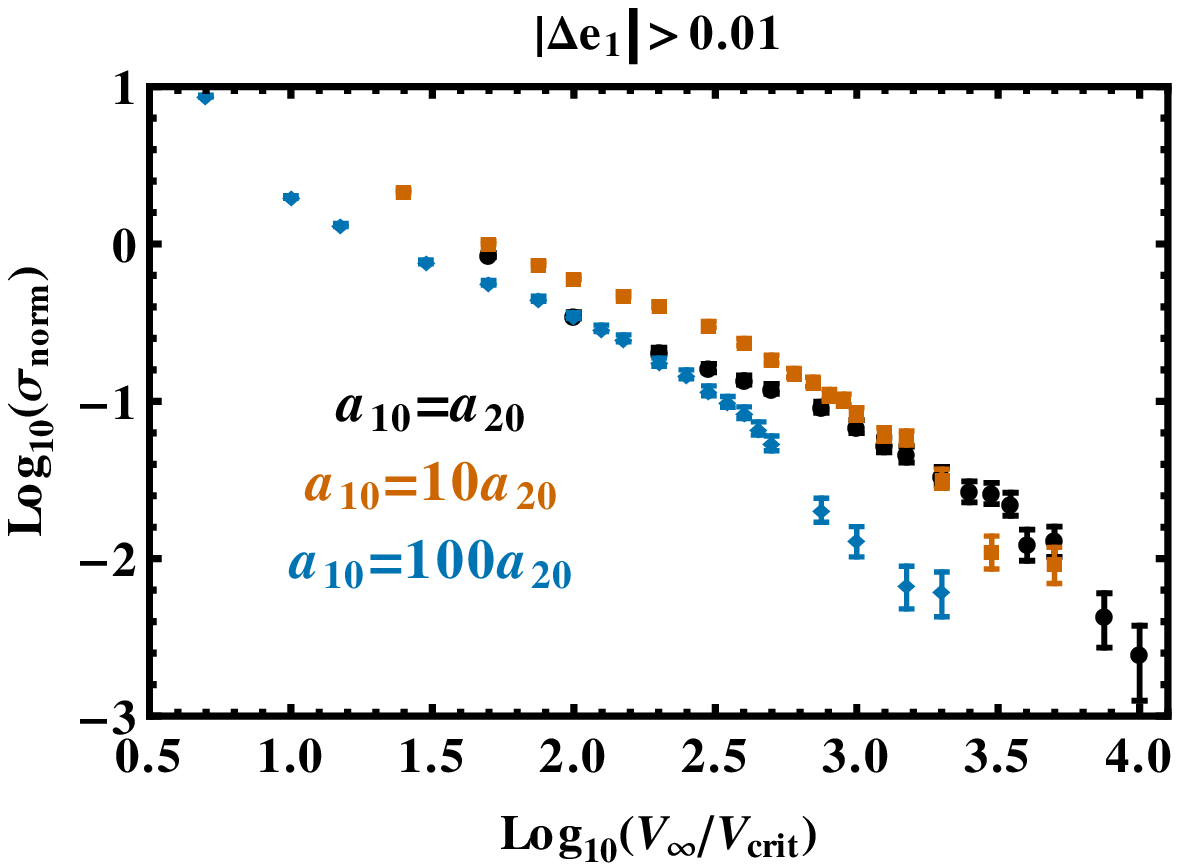,height=6cm,width=8cm} 
\psfig{figure=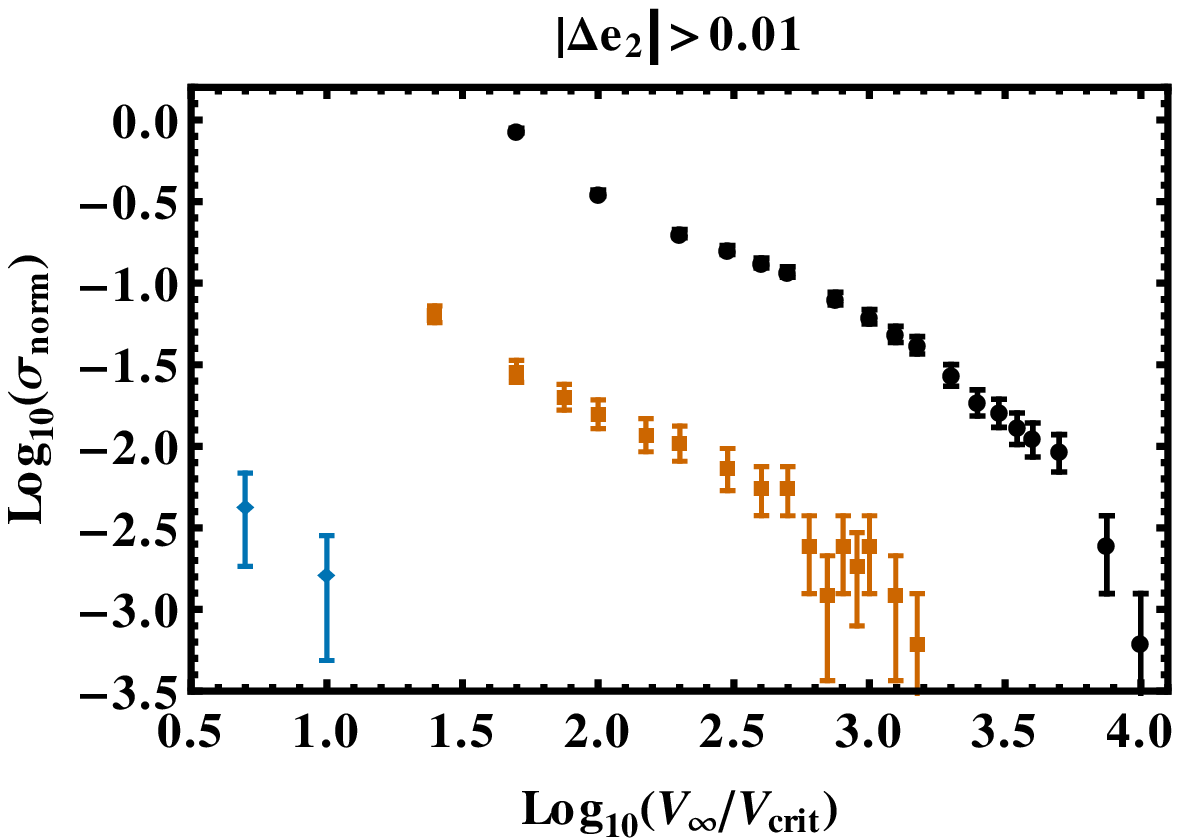,height=6cm,width=8cm}
}
\caption{
Normalized cross sections for outcomes corresponding
to $|\Delta e_1| > 10^{-2}$ (left panel)
and $|\Delta e_2| > 10^{-2}$ (right panel).
}
\label{cse01}
\end{figure*}

\begin{figure*}
\centerline{
\psfig{figure=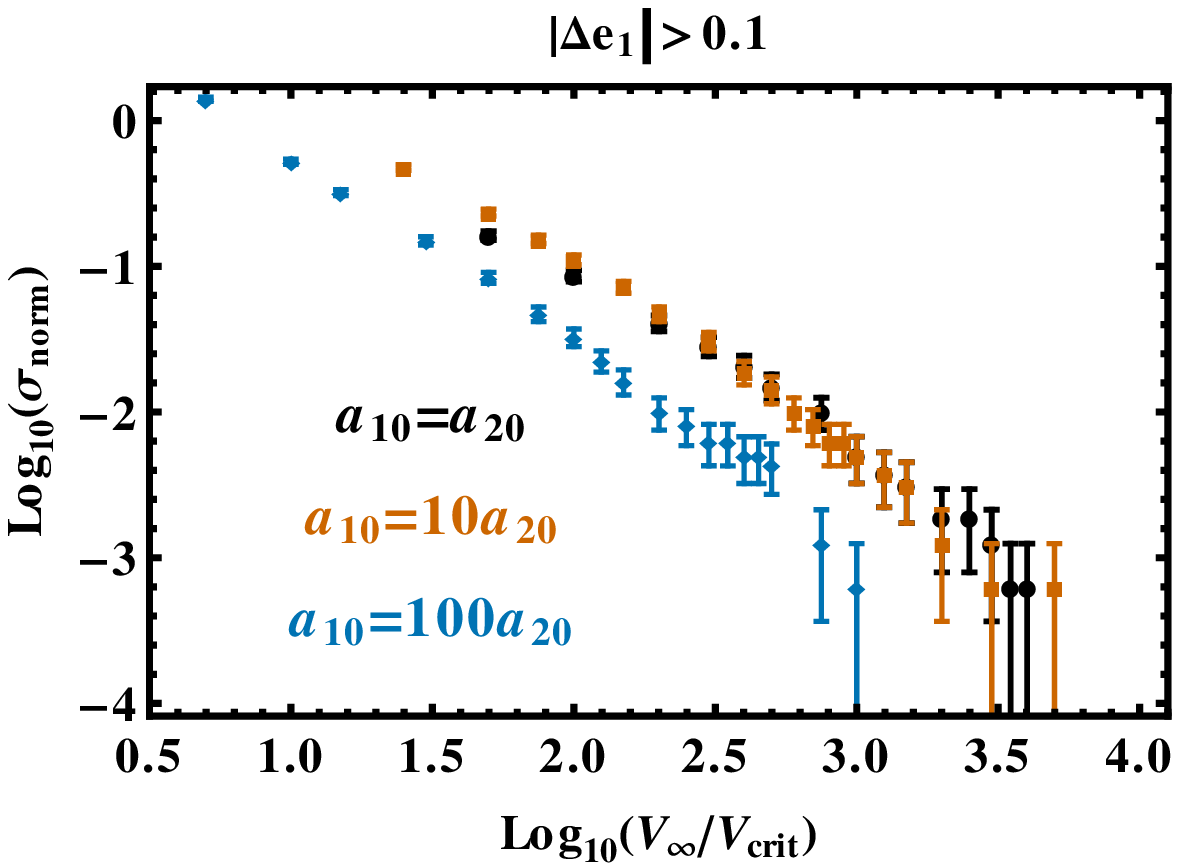,height=6cm,width=8cm} 
\psfig{figure=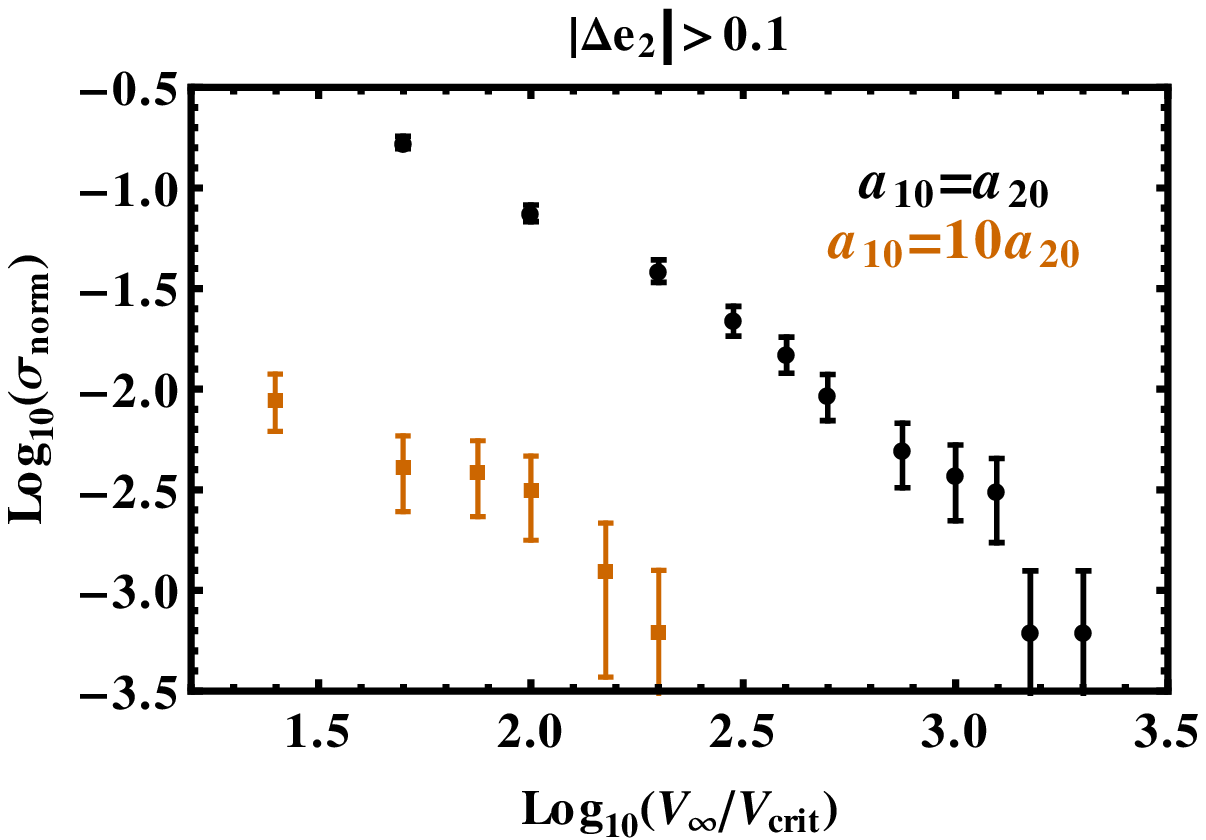,height=6cm,width=8cm}
}
\caption{
Normalized cross sections for outcomes corresponding
to $|\Delta e_1| > 10^{-1}$ (left panel)
and $|\Delta e_2| > 10^{-1}$ (right panel).
}
\label{cse1}
\end{figure*}

\begin{figure*}
\centerline{
\psfig{figure=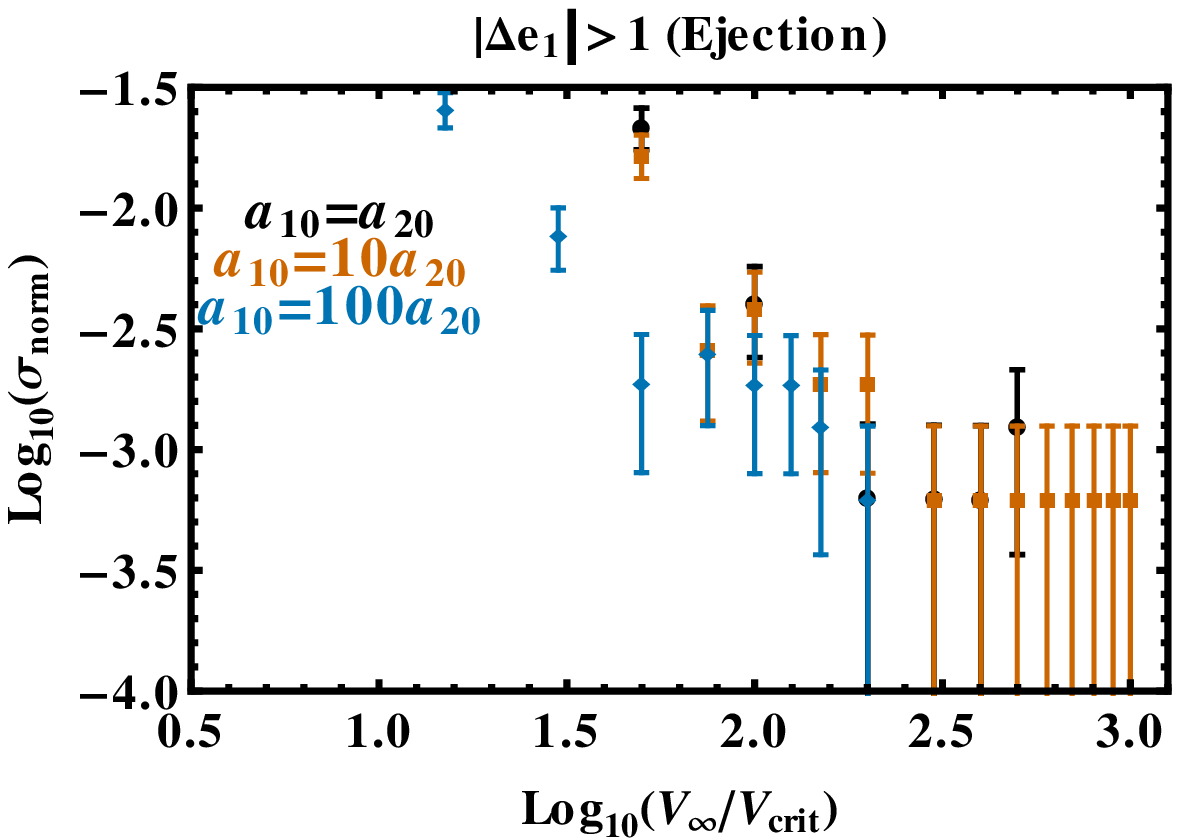,height=6cm,width=8cm} 
\psfig{figure=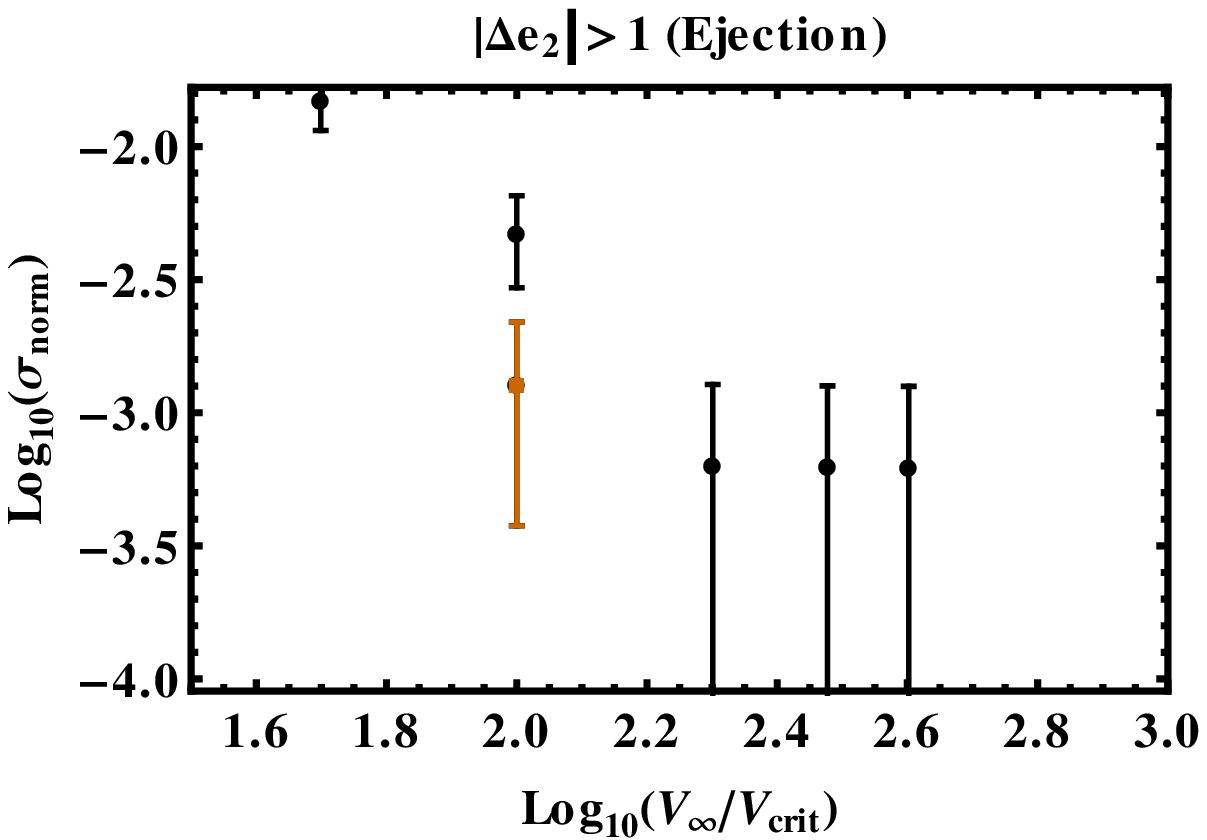,height=6cm,width=8cm}
}
\caption{
Normalized cross sections for outcomes corresponding
to ejection for $e_1$ (left panel) and 
$e_2$ (right panel).
}
\label{cseje}
\end{figure*}

\subsection{Simulation Results}

We present our cross-sections in
Figs. \ref{cse0001}-\ref{cseje}.
Each successive figure shows a higher
value of $\Upsilon$, culminating
with ejection ($\Upsilon = 1$).
Each figure contains two panels; the left
is for $|\Delta e_1|$ and the right is 
for $|\Delta e_2|$.  The black circles,
orange squares and blue diamonds respectively
show the cases $\epsilon = \lbrace{1,10,100\rbrace}$.
Each data point has vertical error bars; in
some cases these are so small that they are
not discernible.  

Due to symmetry, the black circles on
both panels in each plot should be, and are, roughly
equivalent.  For most of the cross sections in the 
ejection figure (Fig. \ref{cseje}), just one
data point was obtained for a particular $(\eta, \epsilon)$ pair. 
Nevertheless, the plot demonstrates that ejection can occur,
and predominantly for the widest orbit planets.

One perhaps surprising trend that is apparent in the left
panels of Figs. \ref{cse0001}-\ref{cse001} is that
the normalized cross sections do not appear
to be monotonic functions of $\epsilon$.  Now we
show how this trend indeed may arise naturally
through analytic considerations.

\subsubsection{Semimajor Axis Dependence Explanation}

We reconsider
the impulse approximation and the {\tt close} configuration.
Now remove the assumption $a_{10} = a_{20}$.  We seek
the perturbation on Planet \#1, whose initial semimajor axis
is fixed, while the initial $\epsilon$ is allowed to vary across
simulations.  This perturbation should
be equal to

\begin{eqnarray}
|\Delta \vec{V}_{\bot}|_{p1} &=& |\Delta \vec{V}_{\bot}|_{s1s2} + |\Delta \vec{V}_{\bot}|_{p2s1}
\nonumber
\\                               
&-& \left( |\Delta \vec{V}_{\bot}|_{s2p1} + |\Delta \vec{V}_{\bot}|_{p1p2} \right)
.
\label{perton1}
\end{eqnarray}

\noindent{A} similar analysis to that from Section 
\ref{impulse} and the Appendix yields
more complex formulae because here $\epsilon \ne 1$.
We find that the resulting eccentricity excitation can be well
approximated by:

\begin{eqnarray}
e_{1f}^{(c)} \approx 
\left|
\frac{\sqrt{2} a_{10} \left(2 a_{10} + \epsilon \left(b - 2q\right)  \right) }
{b \eta \left(a_{10} - \epsilon q \right) \sqrt{\delta \left(1 + \epsilon \right)   }   }
\right|
&&
\nonumber
\\
&&
\nonumber
\end{eqnarray}

\begin{equation}
\approx 
\left|
\frac
{\sqrt{2} a_{10} 
\left[         
\epsilon \left(1 + \epsilon \right) b \delta \eta^2 - 2 a_{10}
\left( \delta \eta^2 + \epsilon \left( 1 + \delta + \delta \epsilon^2 \right) \right)
\right]
}
{
b \eta \sqrt{\delta \left(1 + \epsilon \right)}
\left[
b \delta \epsilon \eta^2 \left(1 + \epsilon \right)
- a_{10} \left(\delta \eta^2 + \epsilon \left(1 + \delta + \delta \epsilon^2 \right)   \right)
\right]
}
\right|
\label{epecc}
\end{equation}

\begin{figure}
\centerline{
\psfig{figure=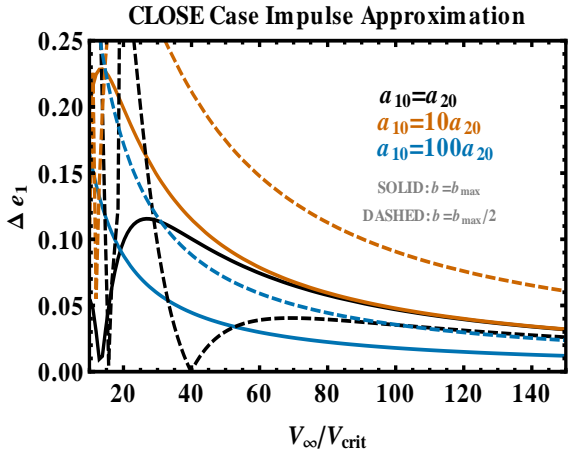,height=6cm,width=8cm} 
}
\caption{
The change in $e_1$ as a function of $\eta$ for a few different
curves of constant $\epsilon \equiv a_{10}/a_{20}$ in the {\tt close} case.
The plot demonstrates that orbital eccentricity variations are
not necessarily monotonic functions of $\epsilon$.  Consequently,
neither are the cross sections, as shown in Figs. \ref{cse0001}-\ref{cse001}.
}
\label{noniso1}
\end{figure}

\begin{figure}
\centerline{
\psfig{figure=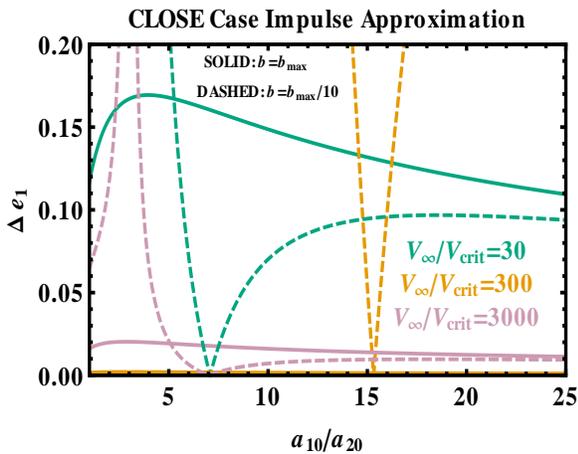,height=6cm,width=8cm} 
}
\caption{
The change in $e_1$ as a function of $\epsilon$ for a few different
curves of constant $\eta$ in the {\tt close} case.  The extrema seen 
here emphasize the complex
dependence of the orbital excitation on several variables.
}
\label{noniso2}
\end{figure}

We plot Eq. (\ref{epecc}) in Figs. \ref{noniso1} and \ref{noniso2} in a regime which 
showcases the nonmonotinicity of $e_{1f}^{(c)}\left(\epsilon\right)$.  Note in 
particular how the orange curves are higher than the black curves, just as in
the cross section plots.

Further, the cross section itself is dependent on $\epsilon$:

\begin{eqnarray}
\sigma &\propto& \frac{b_{\rm max}^2}{\left(a_{10} + a_{20}\right)^2}
\propto \left(\frac{b_{\rm max}}{a_{10}}\right)^2 \left(1 + \frac{1}{\epsilon} \right)^{-2}
\\
&\propto& \beta^2 + 2 \beta \left( \frac{V_{{\rm circ}, 0}}{V_{\infty}} \right)^2 \left(1 + \frac{1}{\epsilon} \right)^{-1}
\end{eqnarray}

\noindent{where} the constant of proportionality could itself be a complex function of $\epsilon$
given, for example, Eq. (\ref{epecc}).

We caution that these results are based on a single orbital configuration set and
assume that the impulse approximation holds, which becomes increasingly unlikely
as the inner planet's semimajor axis is decreased (Eq. \ref{impcri}).  Nevertheless,
they demonstrate how the dependence may be explained.

\subsection{Eccentricity Excitation Frequencies} \label{exfreq}

Having obtained cross sections, we
can now determine the frequency with which
planets' eccentricities are excited to particular values.
In particular, we are interested in the
{\it number of times over a main sequence lifetime
that $\left|\Delta e_1 \right| > \Upsilon$ occurs}.
Let us denote this number by $\mathcal{N}$, and the space density
of a patch of the Milky Way as $n$ and the main sequence
lifetime as $t_{\rm MS}$ (as in Section 1).  Then

\begin{eqnarray}
\mathcal{N} &=& \sigma(\left|\Delta e_1 \right| > \Upsilon,\epsilon,\eta) n V_{\infty} t_{\rm MS}
\nonumber
\\
&=& \sigma(\left|\Delta e_1 \right| > \Upsilon,\epsilon,\eta) n \eta t_{\rm MS} 
\sqrt{\frac{2 G M_{s1}}{a_{10}}}
\sqrt{\frac{\delta \left(1+\epsilon\right)}{1 + \delta}}
\nonumber
\\
&=& \xi \left(\frac{a_{10}}{1000 \ {\rm AU}}\right)^{\frac{3}{2}} \left(\frac{n}{0.5 {\rm pc}^{-3}}  \right)
        \left(\frac{t_{\rm MS}}{10^{10}{\rm yr} }  \right)
\label{fancyn}
\end{eqnarray}

\noindent{where}

\begin{equation}
\xi \equiv 0.016 \sigma_{\rm norm}(\left|\Delta e_1 \right| > \Upsilon,\epsilon,\eta) \eta \frac{\left(1 + \epsilon\right)^{\frac{5}{2}}}{\epsilon^2}
\end{equation}

\noindent{is} determined entirely from the numerical simulations, and is the only quantity
in Eq. (\ref{fancyn}) determined by the numerical simulations.  By setting $\mathcal{N} = \xi$, 
one could obtain a ``fiducial'' value of $\mathcal{N}$ for wide orbit planets in the field
with typical main sequence lifetimes.  Because $\mathcal{N} \propto a_{10}^{3/2}$, planets
on tight orbits are well-protected.  Nevertheless, a nonzero fraction of these planets will
be affected.  Thus, even if just a few percent of the $\sim 10^{11}$ Milky Way stars host planets, 
millions of tight-orbit planets may be affected.

Using our cross sections, we plot $\xi$, which represent fiducial values of $\mathcal{N}$, in
Figs. \ref{count0001}-\ref{counteje}.  On the right axes of these plots, we indicate what the
value of $\mathcal{N}$ would be for $a_{10} = 10$ AU, which is $10^{-3}$ less than the
$a_{10} = 1000$ AU case.  Further, we shade three regions on each plot, the horizontal extent
of which correspond to $10 {\rm km/s} \le V_{\infty} \le 100 {\rm km/s}$ for $a_{10} = 1000$ AU 
(left panels) and $a_{10} = 10$ AU (right panels).  The vertical extents have no physical meaning
and were chosen to be nonintrusive.  This velocity range corresponds to the typical range of
stellar velocities in the Galactic Disc.   Therefore, these figures allow us to read off directly
quantities of interest.  

Before analyzing the consequences of these plots, we first attempt to explain the
trends observed in these figures based on our analytics.

\subsubsection{Explanation for the Frequency Trends}

The features in Figs. \ref{count0001}-\ref{counteje} are highly dependent on the range of $b$ chosen.  
For example, if
one chose $b > (b_{{\rm stat,>}}^{(c)})$ (Eq. \ref{bstatgreater}) exclusively, then the maximum
eccentricity variation would be $e_{\rm ext,max}^{(c)}$ (Eq. \ref{eextmax}).
If instead $b$ was sampled only inside $b_{{\rm eje,2}}^{(c)}$, then the
{\it minimum} eccentricity variation would be given by $e_{2f}^{(f)}(b)$ (Eq. \ref{far2e}).

Although the numerical integrations model interactions at random orientations, we can use 
our two limiting analytical cases in order to help explain the bumps in Figs. \ref{count0001}-\ref{count1}.
In the {\tt far} case, the eccentricity excitation monotonically decreases with both $b$
and $V_{\infty}$ (Fig. \ref{aefar}).  Alternatively, in the {\tt close} case (Fig. \ref{aenear}), 
qualitative differences are more varied.  In that figure, we can superimpose horizontal lines 
which would be related to the values of $\Upsilon$ chosen in Figs. \ref{count0001}-\ref{count1}.
Then we can count the number of instances when the Fig. \ref{aenear} curves are above the horizontal
lines: this yields a subset of $\mathcal{N}$.

Low horizontal lines, under $e_{\rm ext,min}^{(c)}$, all attain the same contribution for 
$b < (b_{{\rm stat,>}}^{(c)})$.  However, for $b > (b_{{\rm stat,>}}^{(c)})$, the contribution steadily increases
as $V_{\infty}$ is increased.  This behaviour is seen in the blue points of Fig. \ref{count0001}.
When the blue points tail off, effectively $V_{\infty}$ has become high enough that the entire curve
to the right of $b_{{\rm stat,>}}^{(c)}$ is under the horizontal line.  The two rightmost
blue points correspond to a value of $V_{\infty}$ so high that the region
$b < b_{{\rm eje,2}}^{(c)}$ now becomes important.  In this region, the dipping Fig. \ref{aenear} curves dip low
enough so that $e_{\rm ext,min}^{(c)} < \Upsilon$, causing a large drop in $\mathcal{N}$.
This oscillatory behaviour is repeated in the blue curves of Figs. \ref{count001}-\ref{count1}
as $\Upsilon$, and hence the horizontal line in Fig. \ref{aenear}, steadily moves upward.
By Fig. \ref{count1}, $\Upsilon$ is so high that increasing $V_{\infty}$ serves only
to decrease $\mathcal{N}$.  Figure \ref{count01} shows the greatest detail, with a
clear upward and downward trend plus modulations.  These modulations likely naturally
result from the random orientations sampled, as it is important to recall that Fig. \ref{aenear}
models only a single, ideal configuration.  

Figure \ref{counteje} is based on small
number statistics and hence must be treated with caution.  In particular, the linear
upward trends of the black dots and red squares from the minima explained
in the last paragraph are based on single data point
statistics.  As evidenced by Fig. \ref{cseje}, these single data points all have
the same normalized cross section, meaning that $\mathcal{N} \propto \eta$ 
(Eq. \ref{fancyn}).  The figure does demonstrate that ejection is possible,
even with stellar velocities towards the upper end of typical Disc velocities.

Also, in all cases, despite the variations, $\mathcal{N} \rightarrow 0$ as 
$V_{\infty} \rightarrow \infty$ (Eqs. \ref{fancyn}, \ref{sig} and \ref{bmax}).

\section{Interpretation of Results}

Our results demonstrate that exoplanets are in constant danger of losing their primordial
eccentricities.  The extent to which these eccentricities vary over time may 
i) provide a link to formation theories from the currently observed middle-aged systems,
ii) identify the most dynamically excited regions of the Milky Way, and
iii) impose a significant lower bound for the eccentricity variation of wide-orbit planets.

\subsection{Links to Formation Theories}

Classical core accretion typically forms planets within several tens
of AU of the parent stars.  These planets may begin their lives on
nearly circular orbits, particularly if they are born in
isolation.  Without additional planets in the system, and provided that
the formed planet is far away enough from its parent star to avoid
tidal circularization, this planet can be perturbed only by external
forces.  

Hence, the nonzero eccentricities of isolated planets may arise
from planetary system flybys.  The cumulative effect of planetary system flybys over
a main sequence lifetime could eliminate the near-circular signature
of a formation pathway.  A few percent of planets on 
tight orbits could have their eccentricities perturbed by
$10^{-3}$ (Fig. \ref{count001}), which is comparable to the smallest
observational errors yet achieved on planetary eccentricity measurements
\citep{wolszczan1994,weletal2012}.  A nonzero fraction of tight-orbit
planets will experience greater perturbations, with $\sim 0.01\%$
receiving a kick of over $0.1$ (Fig. \ref{count1}).  Given that the current number of observed
exoplanets is $\sim 10^3$, we have not yet observed enough exoplanets, on 
average, to have detected a tight-orbit planet with such a large kick.
Nevertheless, at least millions of such planets should exist in the Milky Way,
given the recent total exoplanet population estimate \citep{casetal2012}.

In multiple planet systems, the effect of flybys may be more pronounced.
Two planets on the verge of dynamical instability could be driven to scatter off of one
another after a sufficiently strong nudge from a flyby.  Small, flyby-induced
changes to particular dynamical signatures of formation, such as the circulation of the apsidal angle 
between two planets, can propagate over several Gyr to obscure the
formative value.

If, however, core-accreted planets are not born in isolation, and instead are continually
perturbed in dense clusters, then these planets will attain a nonzero eccentricity.
A detailed comparison of the relative contributions to a planet's dynamical history from 
its birth cluster versus its middle-aged interactions in the Galactic Disc may be crucial
in determining the types of planetary orbits seen in different Galactic environments. 
This study is a first step towards such a comparison.  Subsequent studies could
focus on merging the two types of simulations, or at least consider fast interactions
with initial non-zero planetary eccentricities.

Nevertheless, we can provide a broad estimate here by computing an eccentricity
distribution similar to that of Fig. 9 in \cite{boletal2012}, which illustrates
a post-cluster planetary eccentricity distribution.  Assume a population of 
any number of Solar-mass stars each with a 10 Gyr Main Sequence lifetime and a 
space density of $n=0.5$ pc$^{-3}$.  Each star has a Jupiter-mass planet
orbiting on a circular orbit all of the same semimajor axis.  Then
we can use Eq. (\ref{fancyn}) with $\epsilon = 1$ and a given velocity
distribution of the stars to compute $\mathcal{N}$.  Because 
$\sigma_{\rm norm}(\left|\Delta e_1 \right| > \Upsilon,\epsilon,\eta)$
is a discrete function computed numerically, we create an interpolating
function based on those data points, for a given $\Upsilon$.  Then, 
for the velocity range of interest, we use the mean value theorem on
the interpolating function to compute an averaged value of $\mathcal{N}$ 
for a given $\Upsilon$.  If $\mathcal{N} \ge 1$, then we say that 100\% of 
that population suffered an eccentricity of at least $\Upsilon$.
Recall that because our numerical integrations modeled single encounters,
we do not know how additive the eccentricities are due to repeated perturbations
when $\mathcal{N} \ge 2$.

\begin{figure}
\centerline{
\psfig{figure=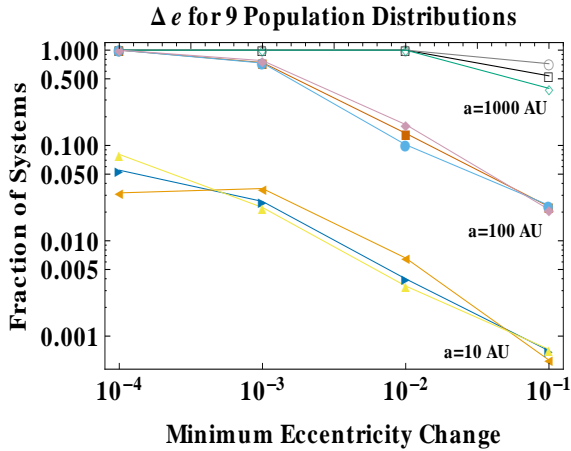,height=6cm,width=8cm} 
}
\caption{
The minimum eccentricities achieved by the fraction 
of systems in 9 different model populations.  All populations
assume a Jupiter-mass planet is orbiting a Solar-mass star
on a circular orbit at the same semimajor axis, a space
density of $n = 0.5$ pc$^{-3}$ and a main sequence lifetime
of 10 Gyr.  The symbols and lines correspond to: 
open black squares  ($V_{\infty} = 10-100$ km/s, $a = 1000$ AU),
filled vermilion squares  ($V_{\infty} = 10-100$ km/s, $a = 100$ AU),
filled right-pointing blue triangles  ($V_{\infty} = 10-100$ km/s, $a = 10$ AU),
open gray circles  ($V_{\infty} = 10-30$ km/s, $a = 1000$ AU),
filled aqua circles  ($V_{\infty} = 10-30$ km/s, $a = 100$ AU),
filled left-pointing orange triangles  ($V_{\infty} = 10-30$ km/s, $a = 10$ AU),
open green diamonds  ($V_{\infty} = 80-100$ km/s, $a = 1000$ AU),
filled purple diamonds  ($V_{\infty} = 80-100$ km/s, $a = 100$ AU), and
filled upward-pointing yellow triangles ($V_{\infty} = 80-100$ km/s, $a = 10$ AU).
The plot broadly suggests regions of phase space where
eccentricity excitation due to fast ``middle-aged'' 
Galactic Disc encounters may be comparable to or smaller 
than those achieved from
other eras of the planet's lifetime.
}
\label{aaron}
\end{figure}

We apply this procedure to 9 populations: for $a = 10, 100$, and 1000 AU,
and for flat distributions of $V_{\infty}$ in the ranges [10-100 km/s], [10-30 km/s] 
and [80-100 km/s].  The results are presented in Fig. \ref{aaron}.
A comparison with \cite{boletal2012} is difficult because of the different setups of the two
papers.  However, our Fig. \ref{aaron} does illustrate that a few percent
of the planets at $a = 100$ AU experience eccentricity changes of at
least $0.1$, which may be comparable to those achieved in birth clusters.
Further, for fast perturbers, the eccentricity change is a strong function 
of semimajor axis and a weak function of the velocity range chosen.  Additionally, the 
values in the plot are dependent on $n$ in a linear fashion,
such that for dense environments, the fractions may increase by a
factor of 2-3.  Overall, more detailed comparisons are necessary.

\subsection{Dynamically Excited Galactic Regions}

Given that $\mathcal{N} \propto n \eta \sigma(\eta)$ (Eq. \ref{fancyn}), the extent 
of the planetary orbital disruption may significantly depend on the 
Galactic environment of the host star.  Additionally, the migration history
of an exoplanet host star through regions of differing spatial density
and velocity dispersions will affect the resulting perturbations on
orbiting planets.  Unlike planetary motion, stellar orbits are typically
not closed, and can suddenly transition from, for example, being ensconced
in a dense tidal tail to traveling in the sparse region between
two spiral arms.  Even the region exterior to the Galactic Disc
is complex: the two broadly overlapping structural components of the Milky Way's halo 
feature distinct spatial density profiles \citep{caretal2007}.

Exoplanet host star velocities can vary by approximately two orders of 
magnitude. HIP 13044, which is thought to be of extragalactic origin, has a
measured systematic velocity of 300 km/s with respect to Sun \citep{setetal2010}.
Alternatively, \cite{heletal1999} suggest that relic debris streams from Milky
Way formation show internal velocity dispersions of just a few km/s.
In the Solar neighborhood (within 30 pc of the Sun), no single stellar 
velocity component exceeds 50 km/s \citep{nakmor2012}.

These results suggest that the value of $\mathcal{N}$ may vary by a few
orders of magnitude depending on the region studied.  This variation may
be important for characterizing the abundance and location of exoplanets
when assessing the Milky Way's global population.  In particular, for
dense enough environments, wide orbit planets may survive only for a small
fraction of the host's main sequence lifetime.  Conversely, sparse environments
would allow planetary systems to retain their formation signatures for several Gyr.
Generally, regions closer to the Galactic Centre are denser, and hence
perhaps harbor more dynamically excited exoplanets than in the Solar neighborhood.
Independently, this conclusion also arises from modelling the effect of galactic tides
on exoplanets.

\subsection{Consequences for Wide-orbit Planets}

At least three exoplanets have been detected orbiting their
parent stars at semimajor axes exceeding $10^3$ AU 
\citep{goletal2010,kuzetal2011,luhetal2011}.  Several others
are thought to orbit at separations of $10^2$ AU - $10^3$ AU.  The population
of wide orbit planets may be large, but remains difficult
to distinguish from the purportedly vast free-floating planet
population \citep{sumetal2011,benetal2012}.
Unlikely to have formed in their current locations via core accretion,
wide-orbit planets perhaps already represent the victims of
internal dynamical jostling \citep{veretal2009,boletal2012} or recaptured
free-floaters \citep{perkou2012}.  Regardless, at these distances,
these planets become even more susceptible to influence from
external flybys.

\begin{figure*}
\centerline{
\psfig{figure=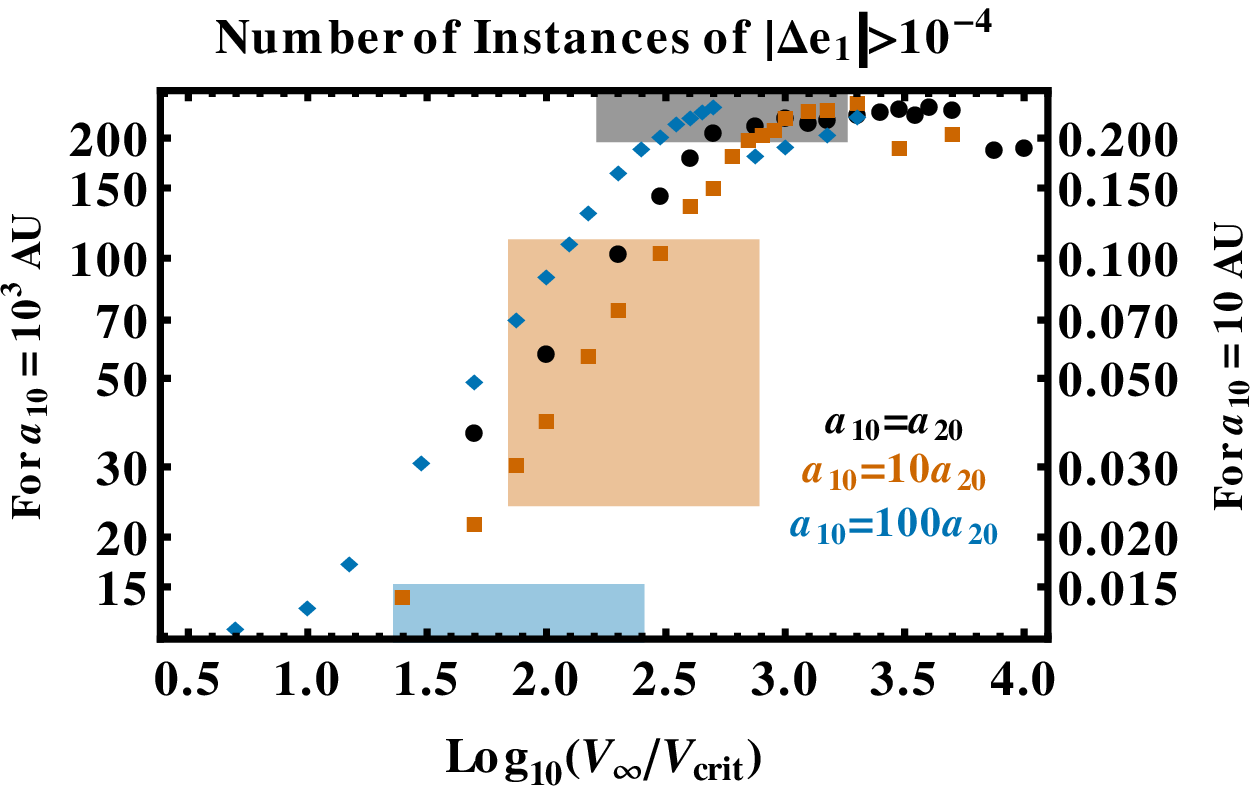,height=6cm,width=8cm} 
\
\
\
\
\psfig{figure=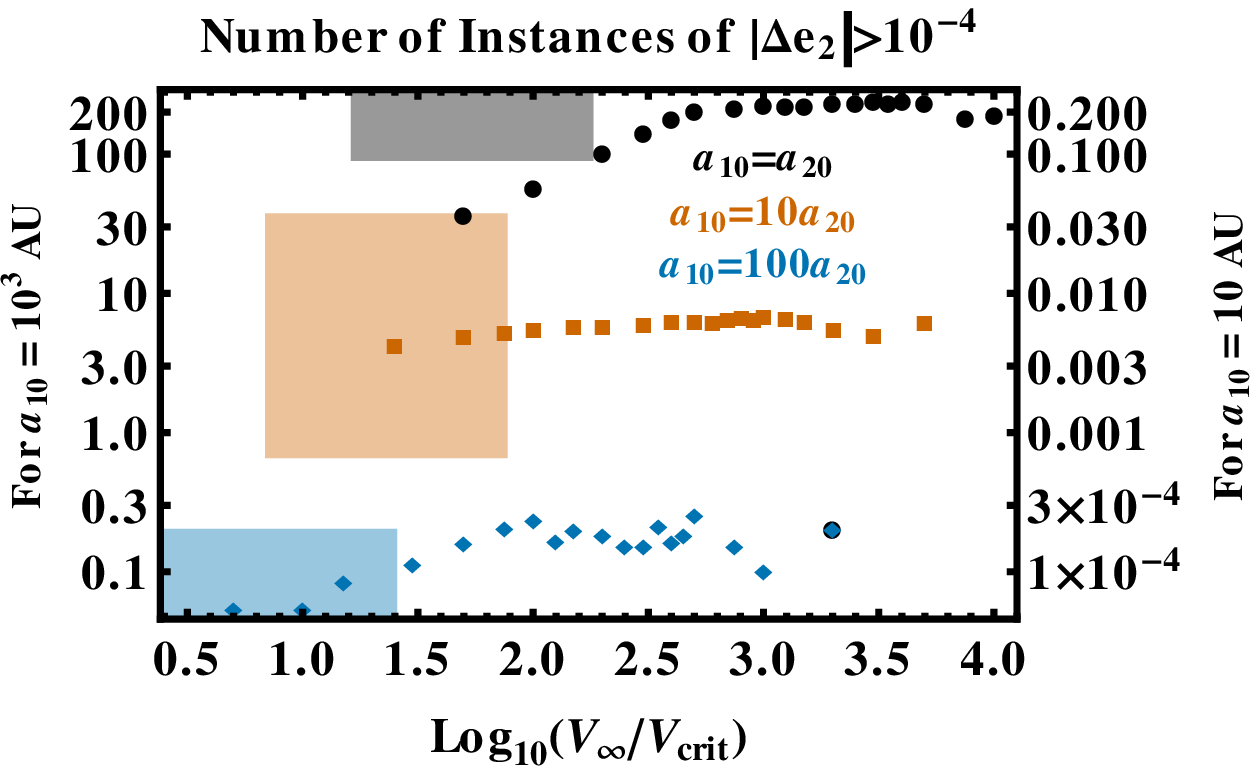,height=6cm,width=8cm}
}
\caption{
The number of times over a typical main sequence lifetime ($10$ Gyr)
in a Galactic region with space density of $0.5$ pc$^{-3}$ that 
$\left|\Delta e_1\right| \ge 10^{-4}$ occurs (left panel)
and $\left|\Delta e_2\right| \ge 10^{-4}$ occurs (right panel)
for $a_{10} = 10^3$ AU (left axes) and $a_{10} = 10$ AU (right axes).  
The shaded regions correspond
to the realistic Galactic field velocity range $10$km/s - 100km/s for
$a_{10} = 10^3$ AU (left panel) and $a_{10} = 10$ AU (right panel).
}
\label{count0001}
\end{figure*}

\begin{figure*}
\centerline{
\psfig{figure=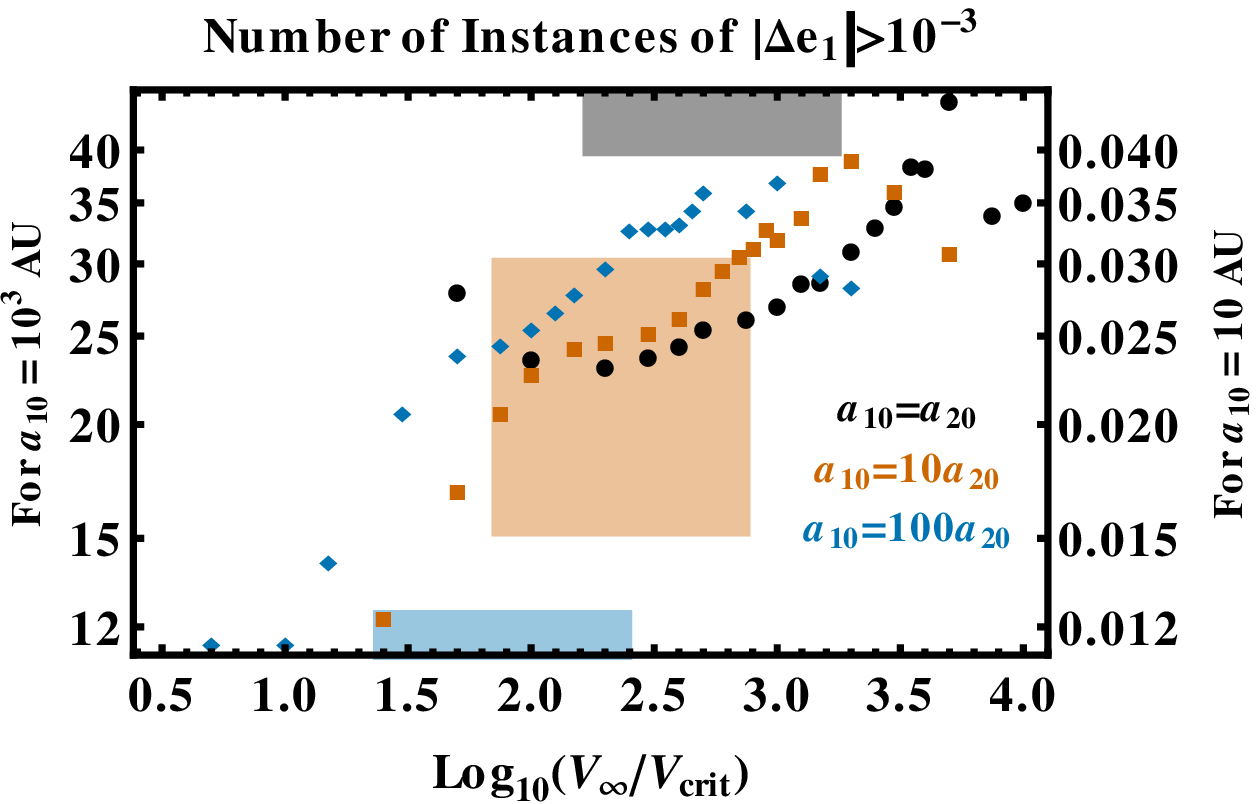,height=6cm,width=8cm} 
\
\
\
\
\psfig{figure=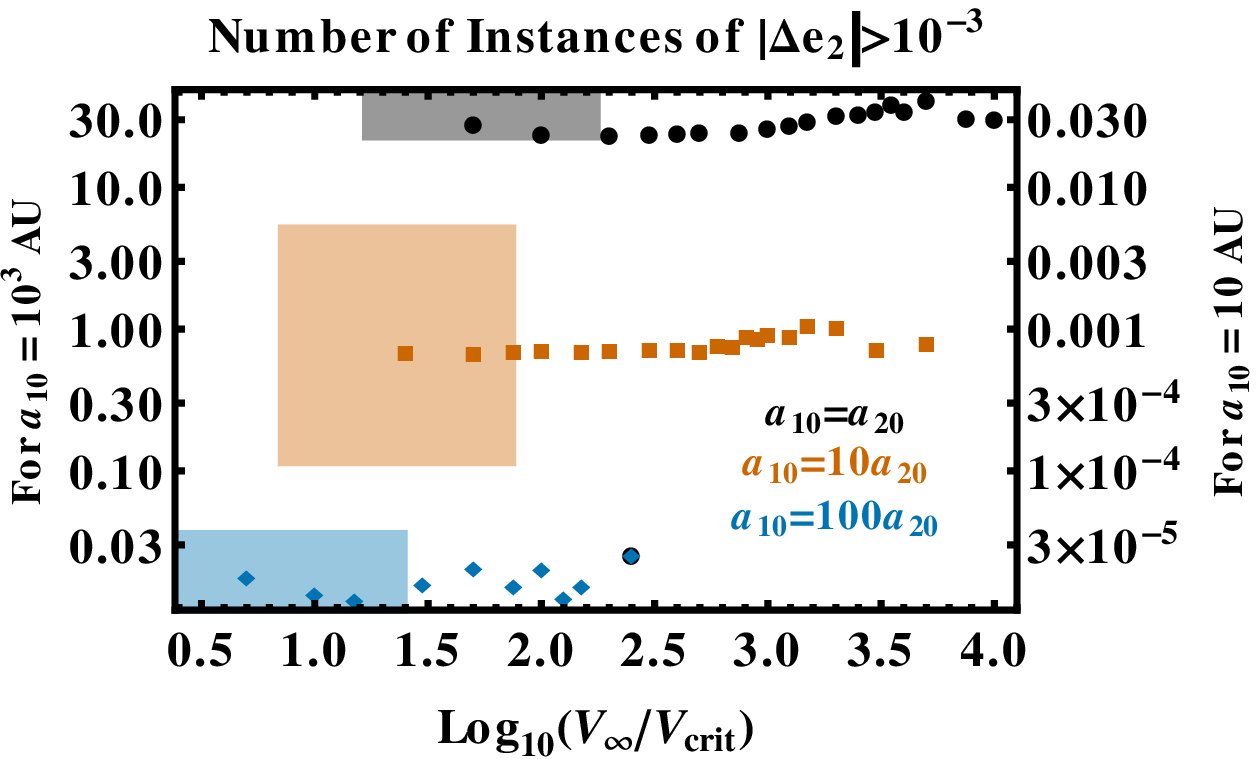,height=6cm,width=8cm}
}
\caption{
Same as Fig. \ref{count0001} but for $\Upsilon = 10^{-3}$.
}
\label{count001}
\end{figure*}

\begin{figure*}
\centerline{
\psfig{figure=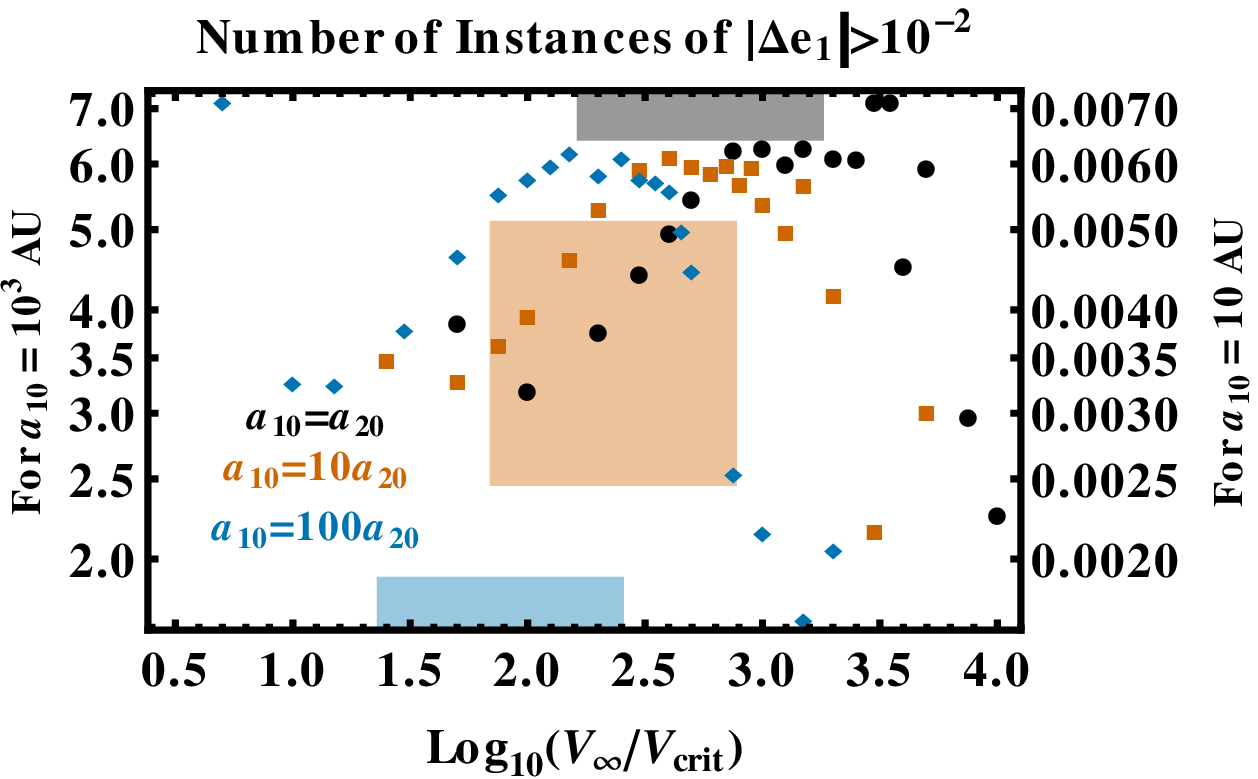,height=6cm,width=8cm} 
\
\
\
\
\psfig{figure=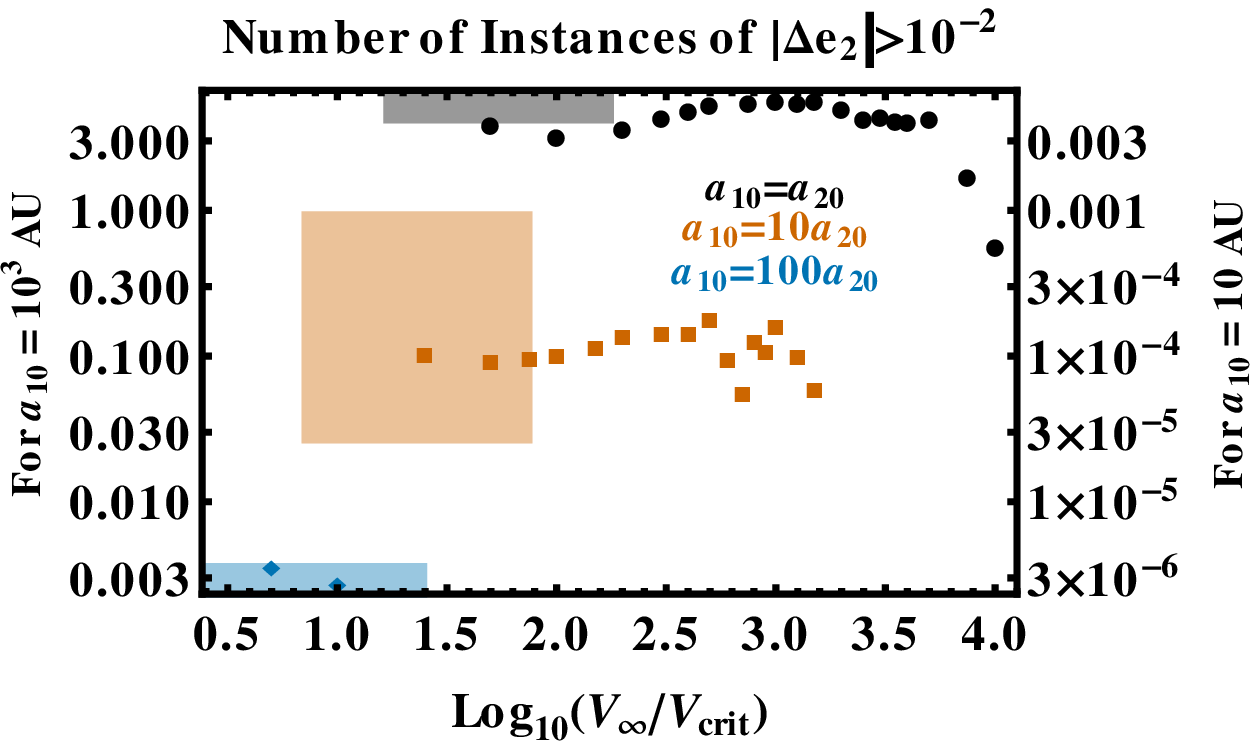,height=6cm,width=8cm}
}
\caption{
Same as Fig. \ref{count0001} but for $\Upsilon = 10^{-2}$.
}
\label{count01}
\end{figure*}

\begin{figure*}
\centerline{
\psfig{figure=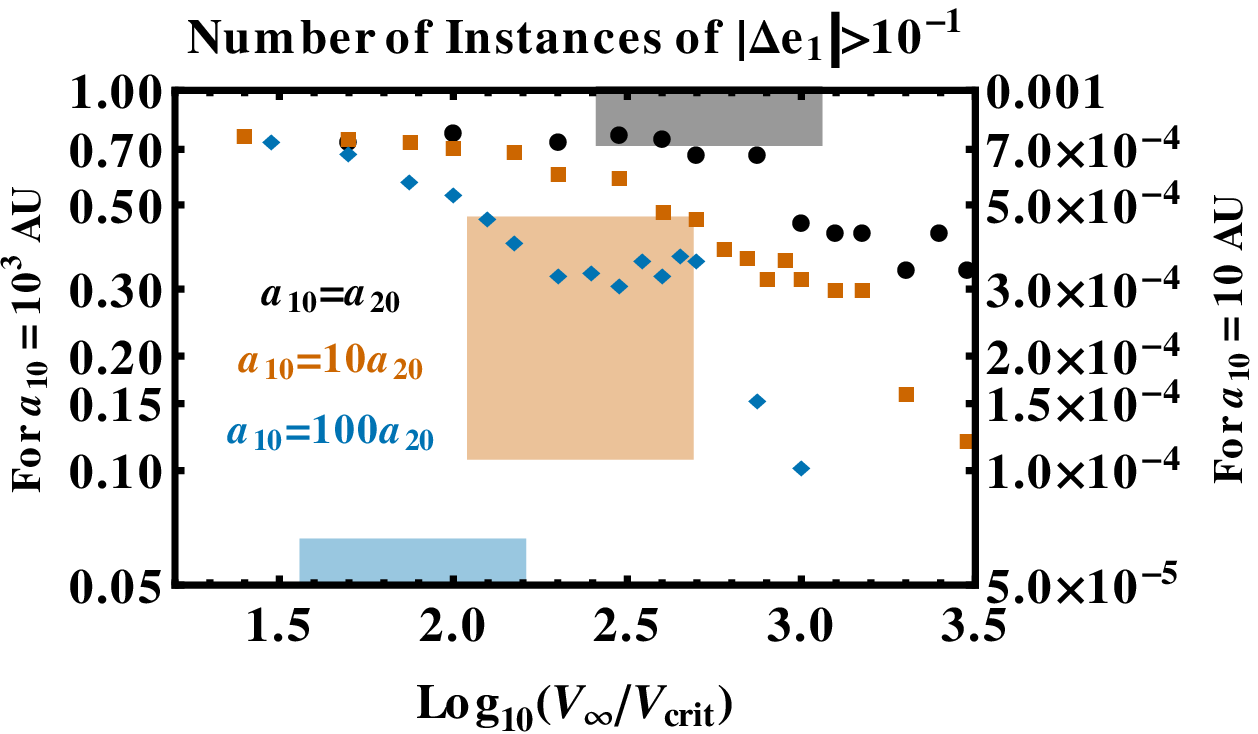,height=6cm,width=8cm}
\
\
\
\
\psfig{figure=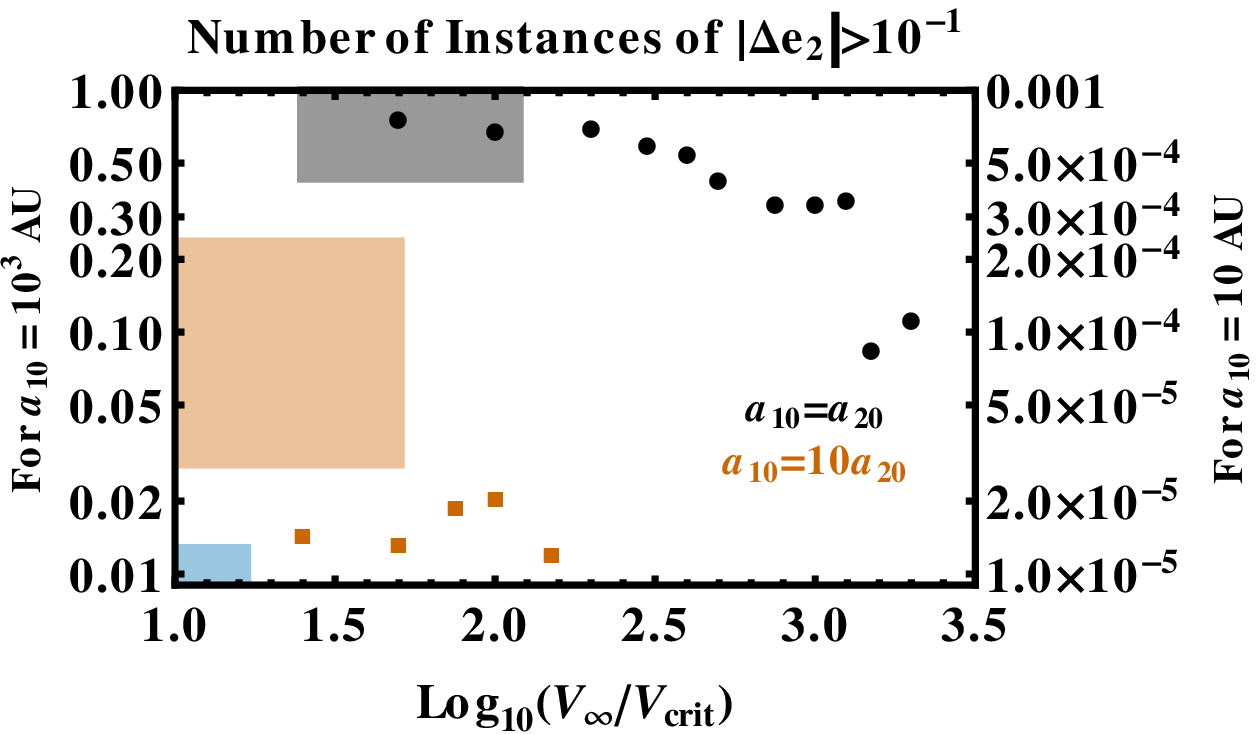,height=6cm,width=8cm}
}
\caption{
Same as Fig. \ref{count0001} but for $\Upsilon = 10^{-1}$.
}
\label{count1}
\end{figure*}

\begin{figure*}
\centerline{
\psfig{figure=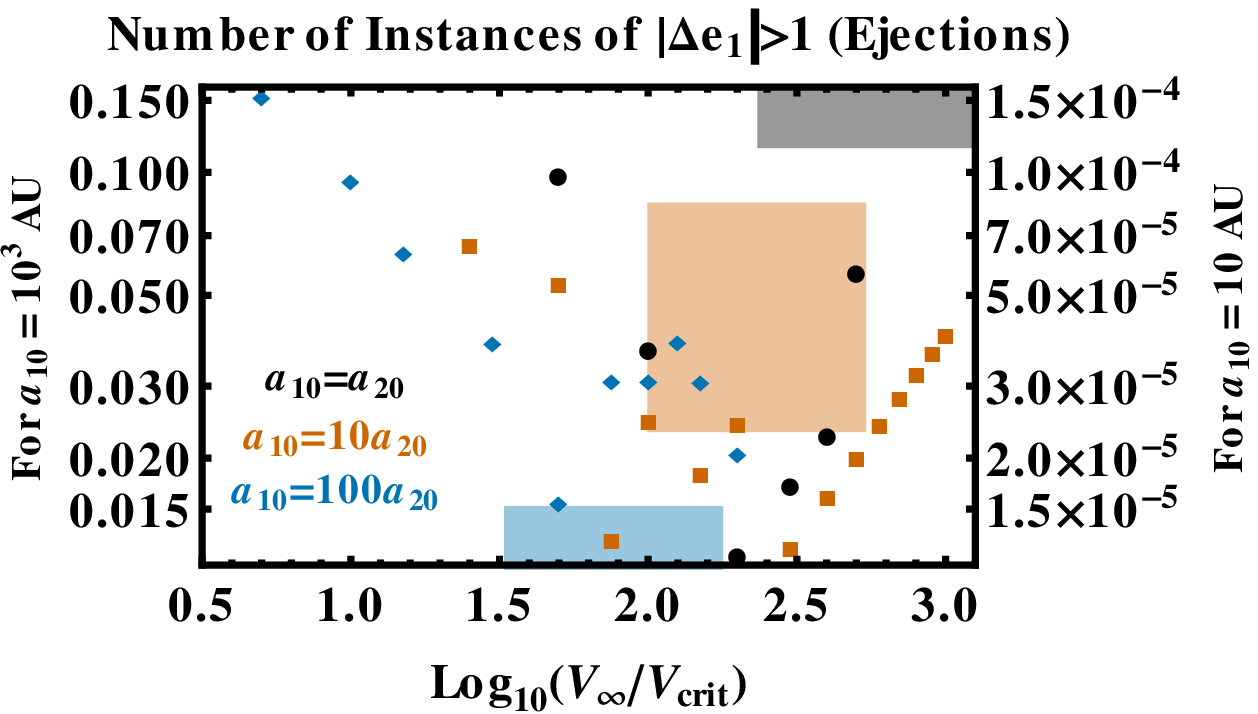,height=6cm,width=8cm} 
\
\
\
\
\psfig{figure=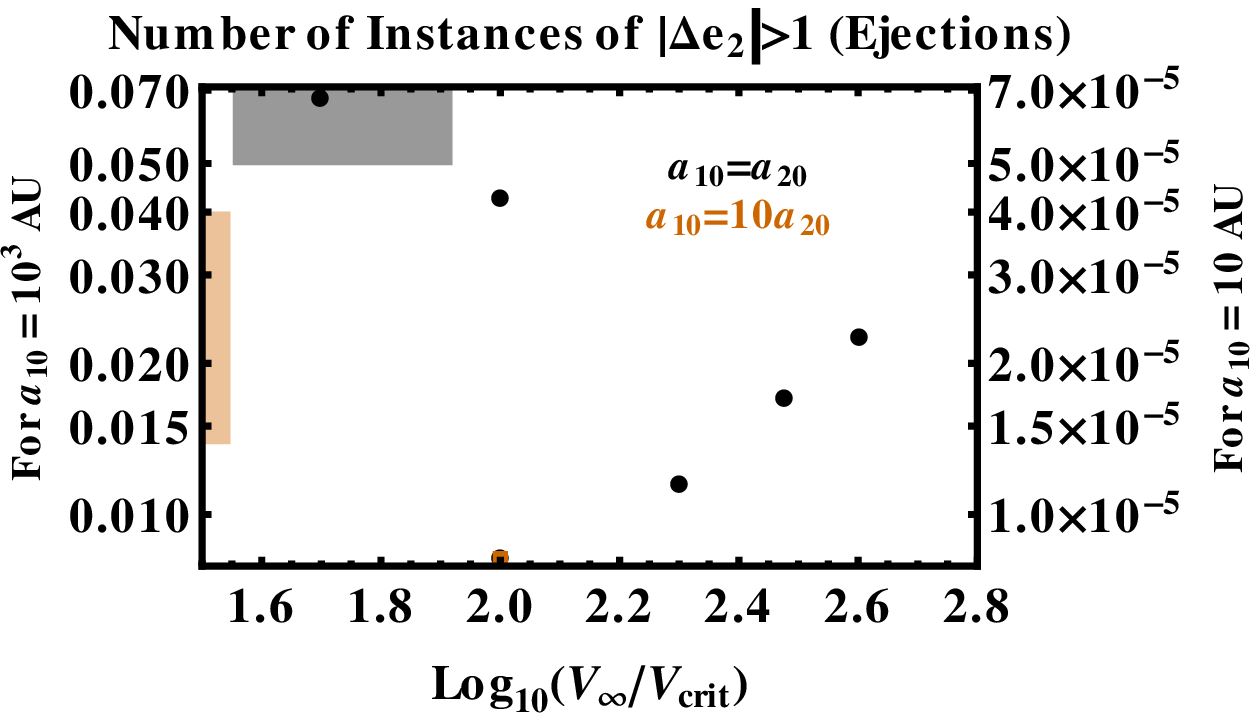,height=6cm,width=8cm}
}
\caption{
Same as Fig. \ref{count0001} but for ejections.
}
\label{counteje}
\end{figure*}

A wide orbit planet will typically have its eccentricity kicked
by at least $0.1$ roughly once over its host star's main sequence
lifetime (Fig. \ref{count1}).  Further, the planet is likely to experience hundreds
of kicks at the $10^{-4}$ level (Fig. \ref{count0001}).  The probability of ejection is
on the order of a few percent (Fig. \ref{counteje}).  These values can vary by a factor
of a few depending on the size of the intruder system's planetary
orbit ($\epsilon$).  Thus, wide orbit planets could represent an additional source
of the free-floating planet population, which cannot be explained
by planet-planet scattering alone \citep{verray2012}.  Further,
the significant eccentricity and semimajor axis kick given to wide orbit planets
during their parent star's main sequence could hasten escape during that star's post-main
sequence evolution \citep{veretal2011,vertou2012}.

\section{Discussion}

Here we consider a few potential extensions to this work.
First, we remove the assumption that the planetary masses
are equal and estimate how our results might change.  Second,
we discuss other related few-body interactions in the
Galactic Disc.

\subsection{Unequal Planetary Masses}

Here we briefly consider how orbital parameters might be perturbed when
the planetary masses are unequal.  The ratio of planetary
masses should be most important when the planets are near
each other during the close encounter of the two systems.
Therefore, let us consider the {\tt close} configuration,
and specifically focus on the region around the planet-planet
collision point (roughly bounded by  
$b_{{\rm stat},<}$ and $b_{{\rm stat},>}$)

Denote $\delta_1 \equiv M_{p1}/M_{s1}$ and 
$\delta_2 \equiv M_{p2}/M_{s2}$.  For 
simplicity, choose $M_{s1} = M_{s2}$
and $a_{10} = a_{20}$, as in Section \ref{impulse}.  If
we carry out the same analytic procedure in that
section and the Appendix, then we find that the eccentricity change
of Planet \#2 is similarly described by Eq. (\ref{e2close}),
except now:

\begin{eqnarray}
Z_4 &=& 2 G M_{s1} V_{\infty}^2 \left(6 a_{20} - b\left(3 - 2 \delta_2  \right)  \right)
\\  
Z_5 &=& V_{\infty}^4 \left(4 a_{20}^2 + b^2 \left(1 - 2 \delta_2 \right) - b a_{20} \left(4 + \delta_1 - 3\delta_2 \right)  \right)
\end{eqnarray}

The result is the critical points around the planet-planet collision region
become:

\begin{eqnarray}
b_{{\rm stat,<}}^{(c)} &\approx& a_{20} \left[ \frac{4 - 
                                    \sqrt{ \delta_{1}^2 + \delta_1 \left(8 - 6 \delta_2 \right) + \delta_2 \left(8 + 9 \delta_2 \right)  }}
                                   {2 - 4 \delta_2} \right]
\\
b_{{\rm stat,>}}^{(c)} &\approx& a_{20} \left[ \frac{4 + 
                                    \sqrt{ \delta_{1}^2 + \delta_1 \left(8 - 6 \delta_2 \right) + \delta_2 \left(8 + 9 \delta_2 \right)  }}
                                   {2 - 4 \delta_2} \right]
\end{eqnarray}

\noindent{which} is equivalent to Eqs. (\ref{bstatless})-(\ref{bstatgreater}) when $\delta_1 = \delta_2$.

\begin{figure}
\centerline{
\psfig{figure=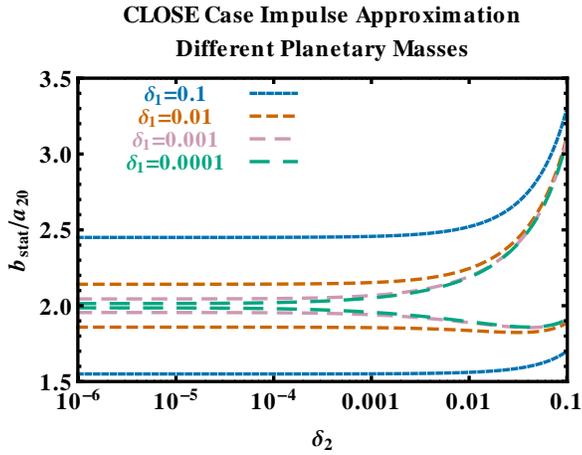,height=6cm,width=8cm}
}
\caption{
How the gravitational focusing region of both planets varies as
functions of both planetary masses.  Plotted are the two critical
points $b_{\rm stat}$ (Eqs. \ref{bstatless}-\ref{bstatgreater} and, 
e.g. Fig. \ref{aeneardelta}), which indicate the impact parameters at
which planetary eccentricity remains unchanged during a close encounter. 
}
\label{diffmass1}
\end{figure}

We plot these critical points as functions of the mass ratios in Fig. \ref{diffmass1}.
The plot demonstrates that given a Jupiter-mass Planet \#1, the region
of planet-planet gravitational influence changes by $\approx 0.1 a_{20}$ if Planet \#2's mass
is an Earth-mass versus a Jupiter-mass.  In the latter case, the region of influence
is greater.  Note also how the asymmetry of the two critical $b_{\rm stat}$ points
is enhanced when the planetary masses approach the stellar masses.

\subsection{Other System Configurations}

Scattering simulations for different hierarchical configurations of 4 bodies,
or for more than 4 bodies, would provide a more complete picture
of planetary orbital excitation from passing stars during the host
star's middle age.  However, the phase space to be explored is prohibitive.
Nevertheless, because the few-body problem admits few analytical solutions,
studies often must rely on numerical integrations.

Alternatively, in the Galactic Disc, the impulse formalism may 
be generalized to any number of bodies
in any orientations.  Although the resulting analytical formulas
are unlikely to be as compact as those presented here, they 
-- subject to the assumption in Eq. (\ref{impcri}) --  would
be able to sample the entire phase space.  Such a formalism
could be useful, for example, in modeling how secular or resonant
evolution of multi-planet systems might change naturally over 
time.  \cite{zaktre2004} consider secular eccentricity propagation.
For resonant systems, this same propagation might kick planets into a deeper
or shallower mean motion resonance, if not out of the resonance 
entirely.  Results from the {\it Kepler} mission illustrate
that there is an abundance of near-resonant planets \citep{lisetal2011,fabetal2012}.

In cases other than the {\tt close} and {\tt far} cases, impulses
would cause both a perpendicular and parallel kick.  The net
effect could be modelled as a single impulse.  If a planet
is on an eccentric orbit before the kick, then the true anomaly
of the planet must be taken into account.  In principle, one
could remove the error bars associated with Poisson counting
statistics in Figs. \ref{cse0001}-\ref{cseje} by generating
those figures analytically, and then generalizing the figures
with a given distribution of eccentricities.  In this way,
one can also quantify the preference of a planet's eccentricity
to increase versus decrease given an initial nonzero value.

This formalism should also work for hierarchical systems: those
with stars, planets and moons.  The Solar System demonstrates
that moons typically orbit planets within half of a Hill radius.
Further, a planet's Hill radius is proportional to its semimajor
axis.  Therefore, wide orbit planets with 
moons\footnote{Wide-orbit planets scattered out to their current
locations could have retained moons, whether the moons were formed
in the circumplanetary disc or were captured satellites.}
could feature a widely spaced moon orbit, one which extends
to several percent of the planet's semimajor axis.  At a distance of 1000 AU, such
an orbit would be comparable to the Neptune-Sun separation, and
hence could be disrupted by passing stars.

Also, given the possible vast population of free-floating giant planets
\citep{sumetal2011}, passing giant planets might be more common
than passing stars.  Then, the resulting binary-single interactions
with a passing free-floater and a planetary system could become
important \citep{varetal2012}.  If a giant free-floating planet of mass $M_p$ were
to pass by a system with a planet of mass $M_p$ orbiting a star of mass $M_s$,
then the critical velocity of this configuration is 
$\sqrt{(2\delta+1)/(2+\delta)} \approx 71\%$ of the critical velocity
of the traditional stellar flyby binary-single scattering configuration.
This reduction in $V_{\rm crit}$ is not enough to claim that the system will
be completely ionized; hence, this situation may be treated in a similar impulse situation
as this work.  In the perhaps more exotic situation of two pairs of
free-floating planet binaries suffering a close encounter, $V_{\rm crit}$ would
be reduced from the traditional four-star encounter by a factor of 
$1/\sqrt{\delta} \approx 32$.  This reduction is significant enough 
that ionization would be much more likely in that case
for typical Galactic field velocities.

\section{Conclusion}

We have modeled the close encounter of two single-planet exosystems
in the Galactic Disc, which mimicks a common occurence during
middle-aged planetary evolution.  We obtained analytical formulae and numerical cross
sections which may be useful for future population studies of exoplanets
in specific regions of the Milky Way.  The resulting change in orbital parameters for 
wide-orbit ($a \approx 100-1000$ AU) planets is significant (with a typical $\Delta e$ of 
several hundredths to over a tenth) and potentially measurable, suggesting 
that these planets are highly unlikely to retain a static orbit during main 
sequence evolution.  Although tight-orbit planets (with $a \lesssim 10$ AU)
are more resistant to orbital changes, millions in the Milky Way will be
affected, and lose their primordial orbital signatures.  The most 
dynamically excited Milky Way exoplanets are likely to reside in 
the densest Galactic regions.

\section*{Acknowledgments}

We thank the referee for a careful read of the manuscript and
astute and helpful suggestions.

\appendix

\section{Additional Analytics}

This Appendix expounds upon the analytical results in 
Section \ref{impulse}.  The resulting formulae may be applied
generally to a particular exosystem of study, and contribute
to our analytic understanding of the general four-body problem.

We can relate the effective impact parameters to the closest approach distances of
all of the planet-star combinations through geometry.  We obtain:

\begin{eqnarray}
b_{s1p2} &=& \left(q \pm a_{10}\right) \frac{b V_{\infty}^2}{\mu} 
\left[ 1 + \left( \frac{b V_{\infty}^2}{\mu} \right)^2  \right]^{-\frac{1}{2}}
\\
b_{p1s2} &=& \left(q \pm a_{20}\right) \frac{b V_{\infty}^2}{\mu}
\left[ 1 + \left( \frac{b V_{\infty}^2}{\mu} \right)^2  \right]^{-\frac{1}{2}}
\\
b_{p1p2} &=& \left(q \pm a_{10} \pm a_{20}\right) \frac{b V_{\infty}^2}{\mu}
\left[ 1 + \left( \frac{b V_{\infty}^2}{\mu} \right)^2  \right]^{-\frac{1}{2}}
\end{eqnarray}

\noindent{where} the upper and lower signs correspond to the {\tt far}
and {\tt close} cases, respectively.  In the {\tt close} case, $b_{\rm min}$ is
the value of $b$ which gives $b_{p1p2} = 0$.  Eq. (\ref{peri}) yields:

\begin{equation}
b_{\rm min} =  \sqrt{ \left(a_{10} + a_{20} \right) \left(a_{10} + a_{20} + \frac{2G\mu}{V_{\infty}^2}  \right)}
\end{equation}

When $b < b_{\rm min}$, then $b_{p1p2} < 0$ and the systems cross orbits.
For a low enough $b$ (when $b_{s1p2} < 0$ or $b_{p1s2} < 0$), the stars directly 
pass through the region in-between
the other system's planetary orbit.  

Now we impose the analytic simplification described
in Section \ref{ansimp} and apply the fiducial values
from Section \ref{fidsimp} to accompany the analytics. 

To gain physical insight into the following situations, consider how
the perpendicular impulses from Eqs. (\ref{stanb}) - (\ref{bp1p2})
tend toward zero for both $b \rightarrow 0$ and $b \rightarrow \infty$.
Therefore, each impulse is maximized for a particular finite value of $b$.
These values are given by:

\begin{eqnarray}
b_{{\rm crit},s1s2} &=& \frac{G \mu}{V_{\infty}^2}
\nonumber
\\
b_{{\rm crit},s1p2} &=& b_{{\rm crit},p1s2} 
\nonumber
\\
&=&  
\sqrt{
\left( \frac{  \left(1 + \delta\right) G M_{S1}}{V_{\infty}^2} \pm a_{20} \right)
 \left( \frac{ \left(5 + \delta\right) G M_{S1}}{V_{\infty}^2} \pm a_{20} \right)
}
\nonumber
\\
b_{{\rm crit},p1p2} &=&
2\sqrt{
\left( \frac{        \delta G M_{S1}}{V_{\infty}^2} \pm a_{20} \right)
 \left(\frac{\left(2+\delta\right) G M_{S1}}{V_{\infty}^2} \pm a_{20} \right)
}
\nonumber
\end{eqnarray}

\noindent{where} the upper and lower signs denote the {\tt far} 
and {\tt close} cases, respectively.   For planetary systems,
$\delta$ is small and hence the expressions for $b_{{\rm crit},s1p2}$ and 
$b_{{\rm crit},p1p2}$ may be shortened.
For the fiducial values
we adopted above, $b_{{\rm crit},s1s2} \approx 1.97$ AU,
very close to the collision point of the stars.
Further, in the {\tt far} case, 
$b_{{\rm crit},s1p2}^{(f)} \approx 997.04$ AU and 
$b_{{\rm crit},p1p2}^{(f)} \approx 1998.03$ AU,
which are both a few AU away from $a_{10} = a_{20}$.
In the {\tt close} case, 
$b_{{\rm crit},s1p2}^{(c)} \approx 1002.96$ AU and 
$b_{{\rm crit},p1p2}^{(c)} \approx 2001.97$ AU.
In this case, note further that when $\delta = 0$,
then $b_{{\rm crit},p1p2}^{(c)} = b_{\rm min}$; otherwise,
$b_{{\rm crit},p1p2}^{(c)}$ is slightly higher:

\begin{eqnarray}
\frac{b_{{\rm crit},p1p2}^{(c)}}
     {b_{\rm min}} 
&\approx&
1 + 2 \delta \left( \frac{a_{20}}{b_{\rm min}}\right)^2
             \left( \frac{V_{{\rm circ},0} }{V_{\infty}}\right)^2
\nonumber
\\
&\times& 
\left[
2 + \left(2 + \delta \right) \left( \frac{V_{{\rm circ},0}}{V_{\infty}}  \right)^2
\right]
\label{bp1p2min}
\end{eqnarray}

\noindent{For} our fiducial case, 
$b_{{\rm crit},p1p2}^{(c)} - b_{\rm min} \approx 0.0019$ AU $= 2.8 \times 10^5$ km, which
is smaller than the radius of the Sun but not of any of the Solar System planets.

These critical values interact with one another to produce the interesting dynamics
below.  We first consider the {\tt far} case.


In the limit of $b = 0$, the stars will collide and impart a large perpendicular
kick to the planets.  The kick will be greatest at $b = b_{{\rm crit},s1s2}$, which is
about only $0.2\%$ of $a_{20}$ for our fiducial case.  Nevertheless, reductions of 
Eqs. (\ref{af}) and (\ref{ef}) show that the planet will be ejected 
($a_{2f} \rightarrow \infty$, $e_{2f} \ge 1$) for all $b$ from 0 to $b_{{\rm eje}}^{(f)}$, where

\begin{equation}
b_{{\rm eje}}^{(f)} \approx \frac{a_{20}}{2 V_{\infty}}
\left[2 V_{{\rm circ},0} - V_{\infty} + \sqrt{V_{\infty} \left(V_{\infty} + 12 V_{{\rm circ},0}  \right)} \right]
\label{far1}
\end{equation}

\noindent{or}, at about $118$ AU for our fiducial case.  This
result shows how a passing star at $\sim 100$ AU can
rip a bound planet off of another star, even if the
passing star is at opposition with the other star's planet.




The eccentricity and semimajor axis perturbations approximated by 
Eqs. (\ref{far2e})-(\ref{far2a}) are independent 
of planetary mass because, in this case the planets are always ``far'' ($\ge 2 a_{20} = 2000$ AU 
$\gtrsim b_{{\rm crit},p1p2}^{(f)}$) 
from each other.\footnote{Equation (\ref{far2e}) does not reduce to
Eqs. (9) or (11) of \cite{hegras1996} because the assumptions used to derive those
latter two formulas are different: encounters are assumed to be slow, the encounter
trajectory is assumed to be parabolic, and the
impact parameter is assumed to be large relative to the planet-star semimajor
axes.  Further, the power-law dependence is reported in terms of pericenter distance.}  
Similarly, because $b_{{\rm eje}}^{(f)} + a_{20} > b_{{\rm crit},p1p2}^{(f)}$, we should
expect the eccentricity and semimajor axis distributions to be smooth functions
of $b > b_{{\rm eje}}^{(f)}$.

This formalism allows us to estimate the contribution to
Planet \#2's eccentricity variation from the potential of Planet \#1 alone:

\begin{eqnarray}
e_{\chi}^{(f)} &\equiv& 
\left|   
\frac
{|\Delta \vec{V}_{\bot}|_{p1s2} - |\Delta \vec{V}_{\bot}|_{p1p2}}
{|\Delta \vec{V}_{\bot}|_{p2}}
\right|
\\
&\approx&
\left|
\frac{\delta b^2}{2b\delta \left(b+a_{20}\right) - \left(b + 2 a_{20} \right)^2   }
\right|
\label{erat}
\end{eqnarray}

\noindent{Equation} (\ref{erat}) shows that the relative contribution from the planet
is an increasing function of $b$.  The maximum contribution is

\begin{equation}
e_{\chi_{\rm max}}^{(f)} = e_{\chi}^{(f)}(b \rightarrow \infty) =  \frac{\delta}{1-2\delta}
\label{eflimit}
\end{equation}

\noindent{showing} that Planet \#1 completely dominates the evolution when $\delta = 1/3$.
Similarly, we can quantify the change in Planet \#2's semimajor axis from Planet \#1 alone:

\begin{eqnarray}
a_\chi^{(f)} &\equiv&
\frac
{\left(a_{2f} - a_{20}\right)_{\rm Planet \ \#1 \ Only}}
{\left(a_{2f} - a_{20}\right)_{\rm Total}}
=
\frac
{e_{\chi}^{{(f)}^2} \left(1 - e_{2f}^{{(f)}^2}\right)}
{1 - e_{2f}^{{(f)}^2}  e_{\chi}^{{(f)}^2}}
\nonumber
\\
&=&
\delta^2 b^2
\left[
\frac
{\left(b Z_1\right)^2 - \left(Z_2\right)^2 }
{\left(Z_1 Z_3\right)^2 - \left(\delta b Z_2\right)^2   }
\right]
\label{diffsq}
\end{eqnarray}

\noindent{where}

\begin{eqnarray}
Z_1 &=& \frac{b + a_{20}}{a_{20}}
\\  
Z_2 &=& 2 \left(b + 2a_{20}\right) \frac{V_{{\rm circ},0}}{V_{\infty}} 
\\
Z_3 &=& \left(b + 2 a_{20} \right)^2 - 2 \delta b \left(b + a_{20}\right)
.
\end{eqnarray}

\noindent{Because} it is expressed as differences of squares, Eq. (\ref{diffsq})
readily reveals the conditions that will zero out the planetary
contribution.

Similar to the eccentricity, the relative contribution to
$a_{2f}$ from the planet is an increasing function of $b$.
In the limit $b \rightarrow \infty$,

\begin{equation}
a_{\chi_{\rm max}}^{(f)} =  \left(\frac{\delta}{1-2\delta}\right)^2
\end{equation}

\noindent{again showing} that Planet \#1 completely dominates the evolution when $\delta = 1/3$.
Comparing $a_{\chi_{\rm max}}^{(f)}$ and $e_{\chi_{\rm max}}^{(f)}$ suggests that intruding planets
have a greater capacity to alter other planets' eccentricities than their semimajor axes.

\begin{figure}
\centerline{
\psfig{figure=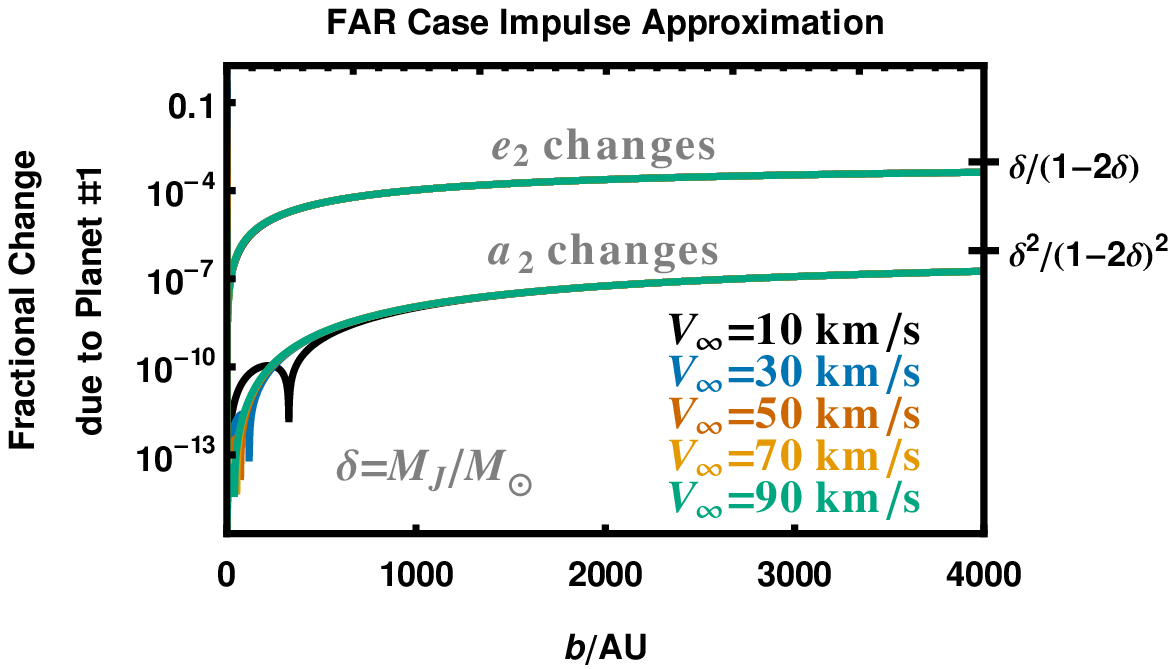,height=6cm,width=8cm} 
}
\centerline{}
\centerline{
\psfig{figure=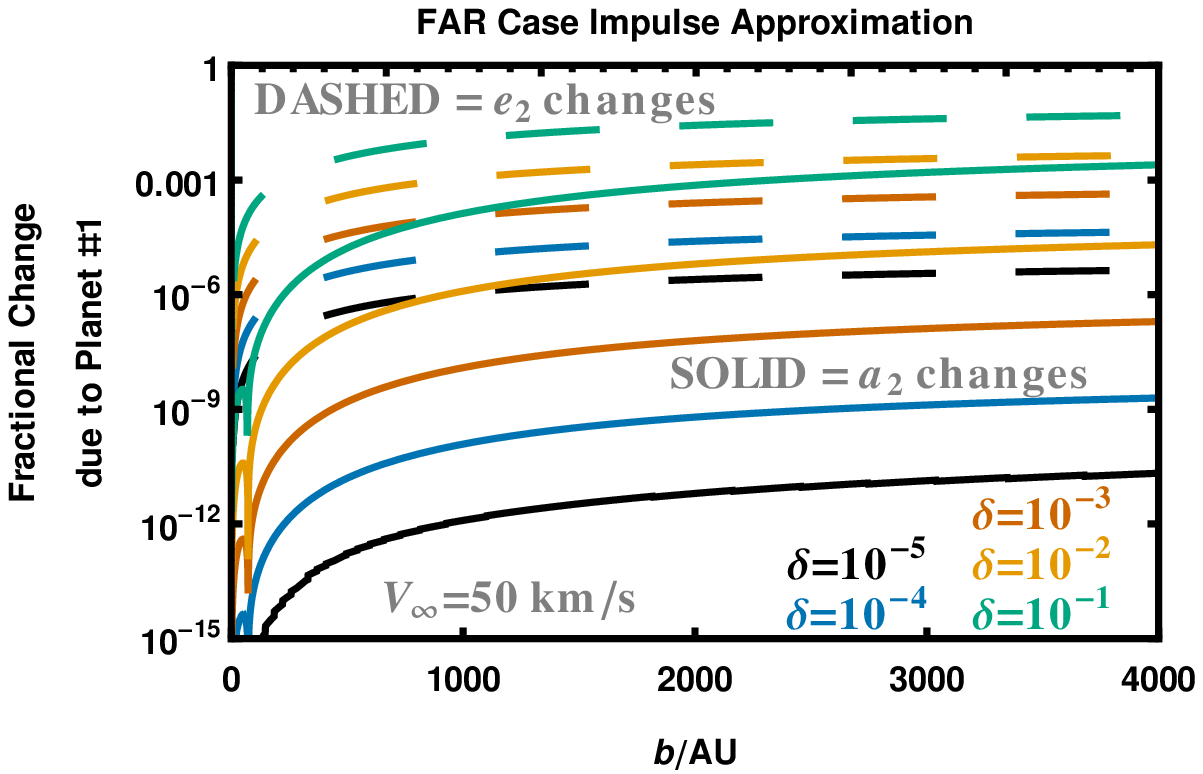,height=6cm,width=8cm}
}
\caption{
Eccentricity and semimajor axis variation of Planet \#2 
due to Planet \#1 alone in the {\tt far} case.  In the
upper panel, $V_{\infty}$ is varied; the resulting differences
are negligible.  In the lower panel, $\delta$ is varied,
illustrating the sensitive dependence of $e_\chi^{(f)}$
and $a_\chi^{(f)}$ on the planet-star mass ratio.  For distant
encounters of the most massive exoplanets and low mass host stars, the planetary
contribution can represent several percent of the overall contribution.
}
\label{aecont}
\end{figure}

Figure \ref{aecont} graphically illustrates Planet \#1's contribution to the evolution
of Planet \#2 in the {\tt far} case.  The plots demonstrate the contrastingly weak and 
strong dependencies of $e_{\chi}^{(f)}$ and $a_{\chi}^{(f)}$ on $V_{\infty}$ and
$\delta$, respectively.  Also, $e_{\chi}^{(f)} > a_{\chi}^{(f)}$ always.  For the most
massive-possible exoplanets ($\approx 11M_J-16M_J$; \citealt*{spietal2011}) and impact parameters
of a few thousand AU, the planetary contribution may reach $10\%$ of the overall contribution.

\begin{figure}
\leftline{
\psfig{figure=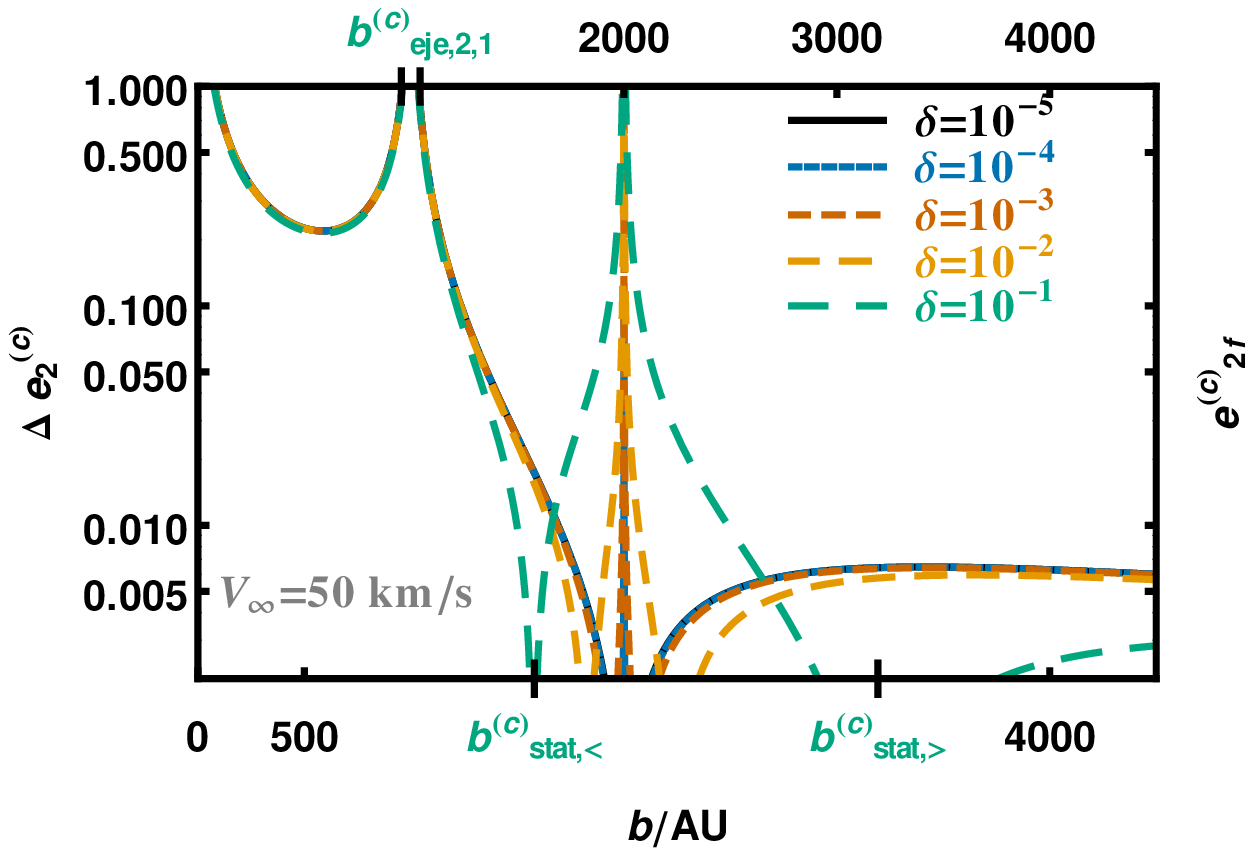,height=6cm,width=8cm} 
}
\centerline{}
\leftline{
\hspace{0.4mm}
\psfig{figure=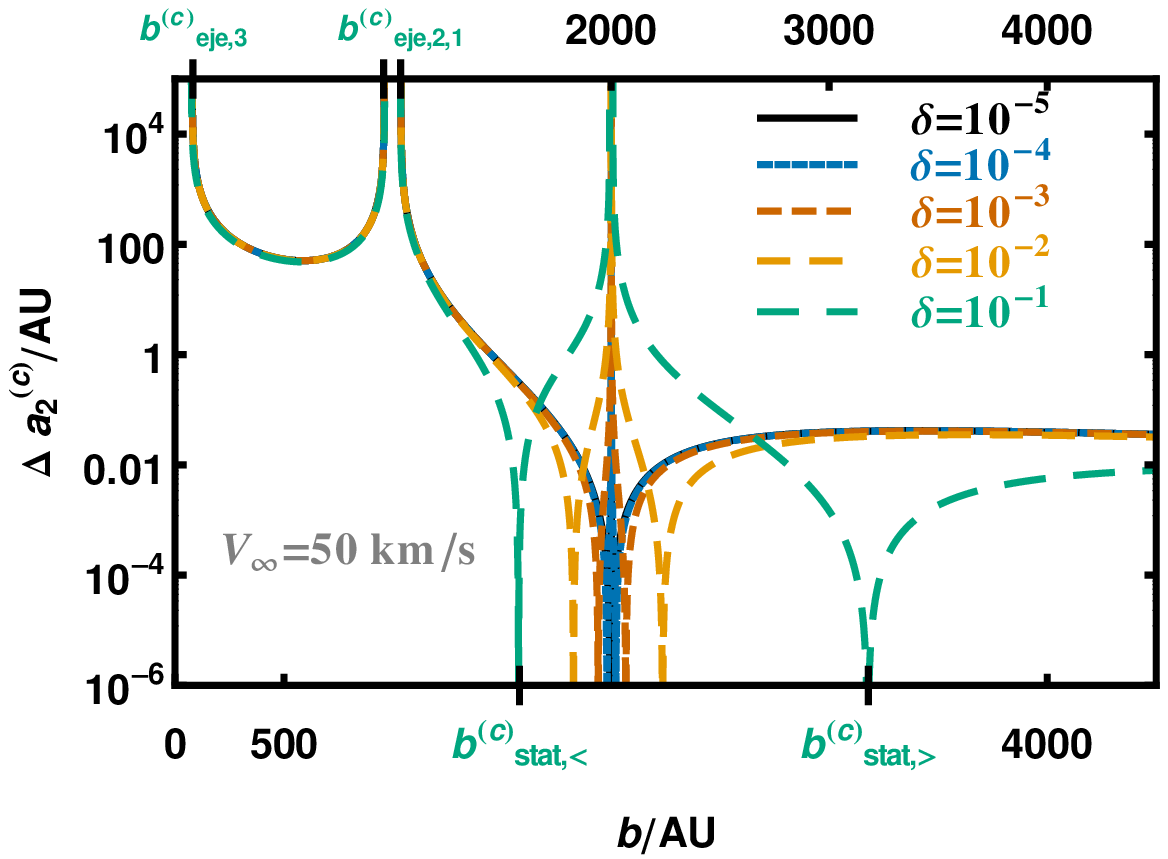,height=6cm,width=7.8cm}
}
\caption{
Same as Fig. \ref{aenear}, except for different curves
of $\delta$ instead of $V_{\infty}$.  Higher values of
$\delta$ have a marked effect on the region where both
planets suffer a close encounter with each other, in which the
contribution from the parent stars is negligible.  Some critical
points not marked in Fig. \ref{aenear} are marked here.
}
\label{aeneardelta}
\end{figure}

Now we perform a similar analysis for the {\tt close} case.
Here, where the position vectors
from each star to their orbiting planet point towards each other,
the resulting orbital parameter
evolution is a more complicated function of $b$.  Figures \ref{aenear} and 
\ref{aeneardelta} may be helpful guides for the following discussion.

In the limiting case of
$b = b_{\rm min}$, the planets collide with each other,
and the kicks on each other have no perpendicular component 
(as can be seen in Eq. \ref{bp1p2}).  Further, the perpendicular kicks from
both stars would cancel out.  Therefore, in this limit -- 
if the collision could be neglected -- the planets' orbital parameters 
would remain nearly unchanged.  However, even a slight nonzero
distance between the planets would produce strong perpendicular
kicks, much stronger than the kicks from the stars, and cause
the planets to escape.  This kick
is highest at $b_{{\rm crit},p1p2}^{(c)}$, which differs from 
$b_{\rm min}$ on the order of a giant planet radius (Eq. \ref{bp1p2min}).  Therefore,
around the vicinity of $b_{\rm min}$, the planets either collide
with each other or escape.

As $b$ deviates from $b_{\rm min}$, eventually the kick contributions from 
both Star \#1 and Planet \#1 will cancel out the back reaction
from Star \#1 on Star \#2.  The result is that planet's orbital
elements would remain unchanged.  This situation occurs at:

\begin{eqnarray}
b_{{\rm stat,<}}^{(c)} &\approx& a_{20} \left[ \frac{2 - \sqrt{\delta \left(4 + \delta \right)}}
                                   {1 - 2 \delta} \right]
\label{bstatless}
\\
b_{{\rm stat,>}}^{(c)} &\approx& a_{20} \left[ \frac{2 + \sqrt{\delta \left(4 + \delta \right)}}
                                   {1 - 2 \delta} \right]
\label{bstatgreater}
\end{eqnarray}

\noindent{where} the subscripts $<$ and $>$ indicate that $b_{{\rm stat}}$
is less than or greater than $b_{\rm min}$.  For our fiducial case,
$b_{{\rm stat,<}}^{(c)} \approx 1942$ AU and $b_{{\rm stat,>}}^{(c)} \approx 2066$ AU.

For $b < b_{{\rm stat,<}}$, the perturbation on Planet \#2 becomes
high as Star \#1 approaches.  In the vicinity of $b \approx a_{20}$, 
where Planet \#2 collides
with Star \#1, the planet is either ejected or destroyed for
$b_{{\rm eje,2}} \le b \le b_{{\rm eje,1}}$, where

\begin{equation}
b_{{\rm eje,2}}^{(c)}\approx\frac{a_{20}}{2 V_{\infty}}
\left[2 V_{{\rm circ},0} + V_{\infty} + \sqrt{V_{\infty} \left(V_{\infty} - 12 V_{{\rm circ},0}  \right)} \right]
,
\end{equation}

\begin{equation}
b_{{\rm eje,1}}^{(c)} \sim 2 a_{20} - b_{{\rm eje,2}}^{(c)}
.
\end{equation}

\noindent{For} our fiducial case, $b_{{\rm eje,1}}^{(c)} \sim 1074$ AU and
$b_{{\rm eje,2}}^{(c)} \approx 926$ AU.
For $b < b_{{\rm eje,2}}^{(c)}$, the perturbations on Planet \#2 stay
high as the two stars approach each other (bottom panel of Fig. \ref{cartoon}).  
The planet will escape for $b < b_{{\rm eje,3}}^{(c)}$, where

\begin{equation}
b_{{\rm eje,3}}^{(c)}\approx\frac{a_{20}}{2 V_{\infty}}
\left[2 V_{{\rm circ},0} + V_{\infty} - \sqrt{V_{\infty} \left(V_{\infty} - 12 V_{{\rm circ},0}  \right)} \right]
\end{equation}

\noindent{In} our fiducial case, $b_{{\rm eje,3}}^{(c)} \approx 137$ AU.
In between $b_{{\rm eje,2}}^{(c)}$ and $b_{{\rm eje,3}}^{(c)}$,
the eccentricity and semimajor axis variations are {\it minimized} at

\begin{equation}
b_{{\rm ext,min}}^{(c)}\approx 
\left(2 - \sqrt{2}\right) a_{20}
\left[1 + 2 \left(\frac{V_{{\rm circ},0}}{V_{\infty}}\right)^2 \right]
\end{equation}

\noindent{or}, a value of $\approx 587$ AU for our fiducial case.
The resulting minimum eccentricity and semimajor axes
values are

\begin{eqnarray}
e_{\rm ext,min}^{(c)} &\approx& \left(6 + 4\sqrt{2}\right) \left(\frac{V_{{\rm circ},0}}{V_{\infty}}\right)
,
\label{eextmin}
\\
a_{\rm ext,min}^{(c)} &\approx& a_{20}
\left[ 1 - \left(6 + 4\sqrt{2}\right)^2 \left(\frac{V_{{\rm circ},0}}{V_{\infty}}\right)^2 \right]^{-1}
\label{aextmin}
\end{eqnarray}

\noindent{or} $e_{\rm ext,min}^{(c)} = 0.366$ and $a_{\rm ext,min}^{(c)} = 1.155 a_{20}$.
Eqs. (\ref{eextmin}) and (\ref{aextmin}) imply that Planet \#2 cannot
remain bound at {\it any} $b < b_{{\rm eje,1}}^{(c)}$ if
$V_{\infty} \le  11.7 V_{{\rm circ},0} \approx 10.8 {\rm km/s}$. 
Therefore, Fig. \ref{aenear} does not feature
a black solid curve (which represents $V_{\infty} = 10 \ {\rm km/s}$) for 
$b \lesssim 1074$ AU in either panel.  However, when
using this critical relation, one should remember that
the impulse approximation starts to break down as $V_{\infty}$ decreases according
to Eq. (\ref{impcri}).

Now let us consider $b > b_{{\rm stat,>}}^{(c)}$.  In this regime, the planet-planet interaction
becomes negligible, and Planet \#2's evolution is dominated by
$|\Delta \vec{V}_{\bot}|_{s1s2}$ and $|\Delta \vec{V}_{\bot}|_{s1p2}$.
These impulses partially, but not completely, cancel each other out,
and admit the greatest net perturbation on Planet \#2 at 

\begin{equation}
b_{{\rm ext,max}}^{(c)}\approx 
\left(2 + \sqrt{2}\right) a_{20}
\left[1 + 2 \left(\frac{V_{{\rm circ},0}}{V_{\infty}}\right)^2 \right]
\end{equation}

\noindent{or} $\approx 3421$ AU, which correspondingly results in maximum
eccentricity and semimajor axes values of

\begin{figure}
\centerline{
\psfig{figure=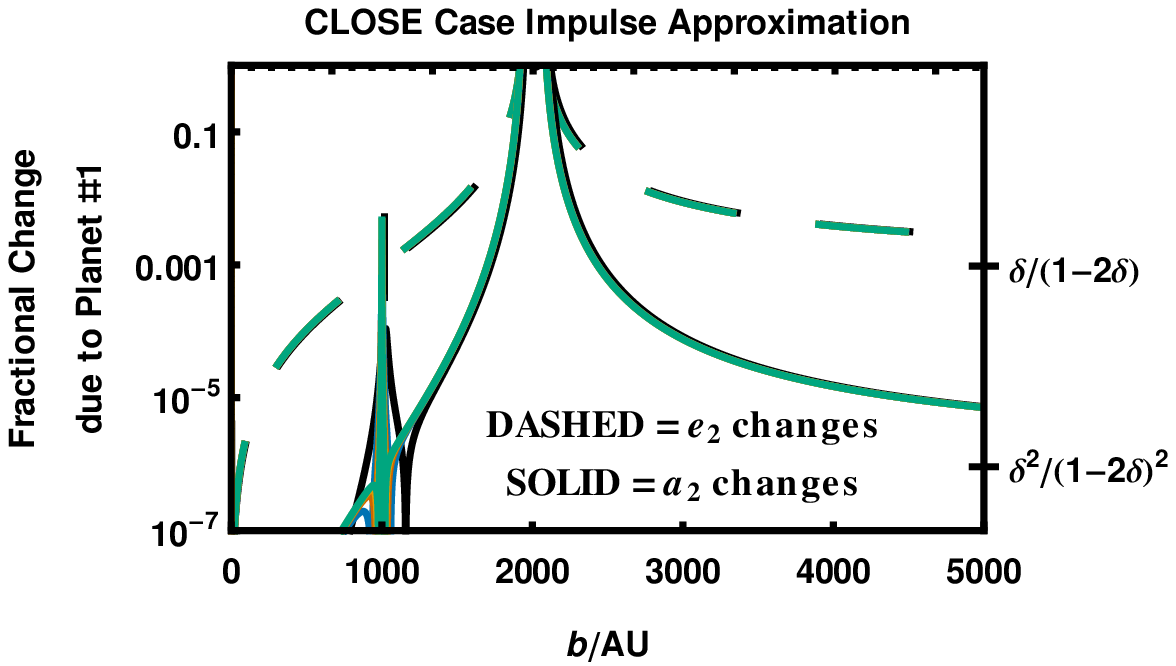,height=6cm,width=8cm} 
}
\centerline{}
\centerline{
\psfig{figure=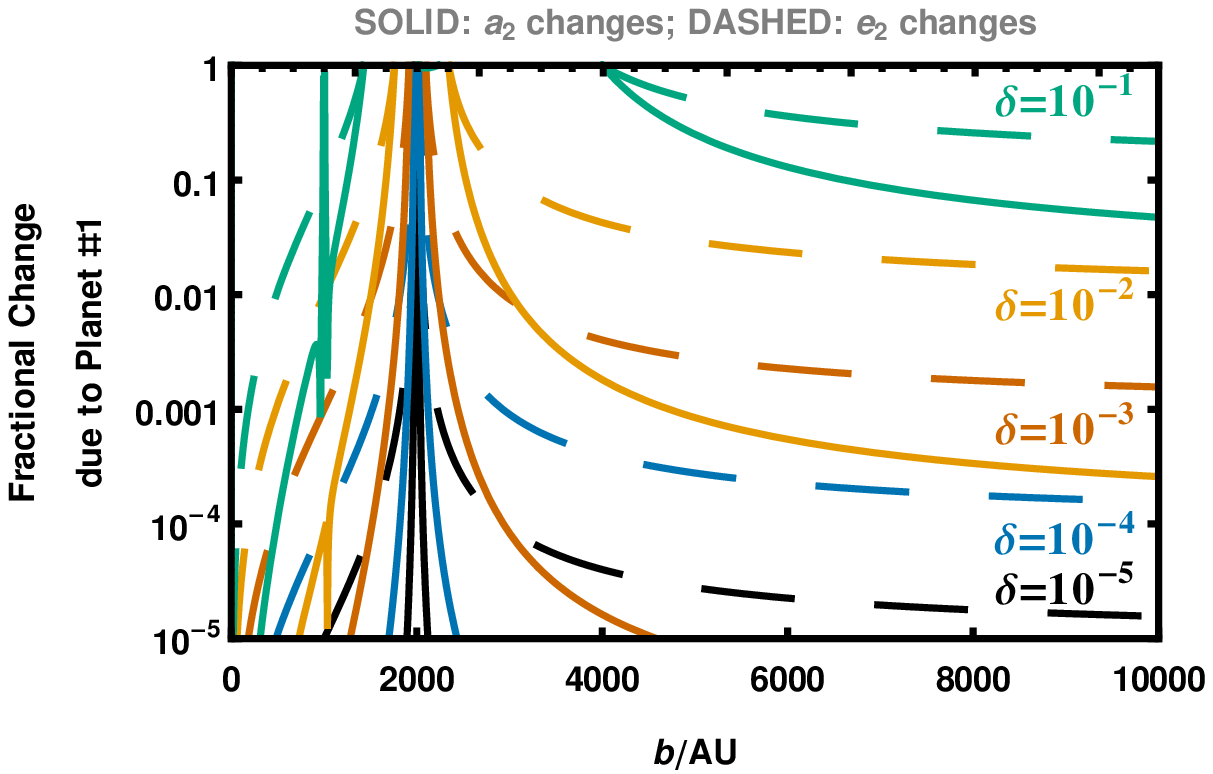,height=6cm,width=8cm}
}
\caption{
Eccentricity and semimajor axis variation of Planet \#2 
due to Planet \#1 alone in the {\tt close} case.  Unlike
in Fig. \ref{aecont}, Planet \#1 may dominate the evolution
over a large region of impact parameter phase space.
}
\label{aenearcont}
\end{figure}

\begin{eqnarray}
e_{\rm ext,max}^{(c)} &\approx& \left(6 - 4\sqrt{2}\right) \left(\frac{V_{{\rm circ},0}}{V_{\infty}}\right)
,
\label{eextmax}
\\
a_{\rm ext,max}^{(c)} &\approx& a_{20}
\left[ 1 - \left(6 - 4\sqrt{2}\right)^2 \left(\frac{V_{{\rm circ},0}}{V_{\infty}}\right)^2 \right]^{-1}
\label{aextmax}
\end{eqnarray}

\noindent{or} $e_{\rm ext,max}^{(c)} \approx 0.011$ and $a_{\rm ext,max}^{(c)} = 1.00012 a_{20}$.  
Equations (\ref{eextmax}) and (\ref{aextmax})
importantly let us explore if Planet \#2 could ever be ejected when $b > a_{10} + a_{20}$.
Ejection is possible only if $V_{\infty} \le  0.34 V_{{\rm circ},0} \approx 0.32 {\rm km/s}$.

Finally, to complete our 
exploration of the impact parameter phase space, 
as $b \rightarrow \infty$, the
orbital changes asymptotically approach zero.

Planet \#1 plays a much larger role in altering the orbit of Planet
\#2 in the {\tt close} case instead of the {\tt far} case.  This can
be seen by the dependence of $e_{2f}$ on $\delta$,  even though in route
to the derivation of $e_{2f}$, we neglected a higher order term due to $\delta$.
The fully general case (Eq. \ref{ef}) makes no assumptions whatsoever about $\delta$,
meaning that the formula is just as applicable to four stars as it is to two
stars and two planets.  Hence, we use Eq. (\ref{ef}) in order to plot
Fig. \ref{aeneardelta}, which illustrates the dependence on $\delta$.

The plot provides an effective region of planet-planet influence, perhaps
interpreted as the region where planet-planet gravitational focusing is important.
For $\delta = 0.1$, this region is nearly as large as $0 < b < b_{{\rm stat,<}}^{(c)}$.
Note however, that the orbital parameter variations for $b < b_{{\rm stat,<}}^{(c)}$
are nearly completely independent of $\delta$.  For $b > b_{{\rm stat,<}}^{(c)}$, greater
values of $\delta$ have an overall weaker effect, because of the locations at which the forces
partially cancel out one another.

Now we can estimate what fraction of Planet \#2's orbital
changes are due to Planet \#1 alone.  We find that $e_\chi^{(c)}$
is equal to

\begin{equation}
\frac
{\delta b^2 V_{\infty}^2 \left(b - a_{20} \right)  }
{\left[2GM_{s1} + V_{\infty}^2 \left(b - a_{20} \right)   \right] 
\left[ 4a_{20}^2 + b^2 \left(1 - 2 \delta \right) -
2 a_{20} b \left(2-\delta\right) \right]   }
\end{equation}

\noindent{which} takes the same limit of $\delta/(1-2\delta)$ for $b \rightarrow \infty$
as $e_\chi^{(f)}$ (Eq. \ref{eflimit}).  Similarly, the limit of $a_\chi^{(c)}$
as $b \rightarrow \infty$ is $\delta^2/(1-2\delta)^2$.  We plot
these contributions in Fig. \ref{aenearcont}.  The left and right
panels illustrate the dependencies on $V_{\infty}$ and $\delta$ respectively.
Like in the {\tt far} case, here $e_\chi^{(f)}$ and $a_\chi^{(f)}$ are
greatly sensitive to $\delta$.  Unlike in the {\tt far} case,
there is a region of impact parameter phase space where Planet \#1's contribution
dominates the evolution.  For $\delta \gtrsim 10^{-2}$, the width of this
region can extend beyond $10^3$ AU. Even for $b > 10^4$ AU, the contribution
due to massive exoplanets may still be a few percent of the overall contribution.


\label{lastpage}

\end{document}